\documentclass[a4paper,11pt]{article}
\pdfoutput=1 

\usepackage{jcappub} 
\usepackage[T1]{fontenc} 
\usepackage{amsmath}
\usepackage{amsfonts}
\usepackage{amsthm}
\usepackage{amssymb}
\usepackage{latexsym, array, multirow, verbatim, enumerate}
\usepackage{slashed}
\usepackage{booktabs}
\usepackage{tabularx}
\usepackage{cancel}
\usepackage{dcolumn}
\usepackage{bm}
\usepackage{graphicx, caption, subcaption} 
\usepackage{url}
\usepackage{pifont}
\usepackage{empheq}
\usepackage[utf8]{inputenc}
\usepackage{bm}
\usepackage{blkarray}
\usepackage[font=footnotesize,labelfont=sf]{caption}
\usepackage{cleveref}
\usepackage{float}
\usepackage{multirow}
\usepackage{physics}
\graphicspath{{/} }
\usepackage{hyperref}
\usepackage{xcolor}
\usepackage[normalem]{ulem}

\newcommand{\hquad}{\hspace{3em}}
\newcommand{\solarmass}{\textup{M}_\odot}

\title{\boldmath Exploring Faraday rotation signatures and population bounds for primordial magnetic black holes}

\author[a]{Arka Banerjee,}
\author[a]{Lalit Singh Bhandari,}
\author[b]{Ashwat Jain,}
\author[a]{and Arun M. Thalapillil}

\affiliation[a]{ Department of Physics,\\ Indian Institute of Science Education and Research Pune,\\ Pune 411008, India}
\affiliation[b]{ Wadham College,\\ University of Oxford,\\ Oxford, United Kingdom}

\emailAdd{arka@iiserpune.ac.in}
\emailAdd{bhandari.lalitsingh@students.iiserpune.ac.in}
\emailAdd{ashwat.jain@wadham.ox.ac.uk}
\emailAdd{thalapillil@iiserpune.ac.in}

\abstract{ Primordial black holes bearing magnetic charges may bypass the constraints imposed by Hawking radiation, thereby enabling reasonable present-day populations, even for masses below $10^{15}\,\mathrm{g}$—a range previously considered improbable. They could, therefore, conceivably contribute to a component of dark matter. We investigate novel Faraday rotation signatures exhibited by primordial magnetic black holes while also establishing new Parker-type bounds on their populations. For the latter, we bound the dark matter fraction from intergalactic magnetic fields in cosmic voids $\left(f_{\text{\tiny DM}} \lesssim 10^{-8}\right)$ and cosmic web filaments $\left(f_{\text{\tiny DM}} \lesssim 10^{-4}\right)$, notably eclipsing previous bounds. Exploring Faraday rotation effects, we discern a pronounced rotation of the polarization angle and the rotation measure values for extremal primordial magnetic black holes with masses $M^{\text{\tiny ex.}}_{\text{\tiny BH}}\gtrsim10^{-6}~ \solarmass$. This makes them potentially detectable in current observations. A comparative investigation finds that the effects are notably greater than for a neutron star, like a Magnetar, with a similar magnetic field at the surface. Moreover, the polarization angle maps for primordial magnetic black holes exhibit unique features, notably absent in other astrophysical magnetic configurations. In this context, we also introduce a simple integral measure, offering a quantitative measure for their discrimination in many scenarios. These traits potentially suggest a robust avenue for their observational detection and differentiation. 
}

\begin{document}
\maketitle
%%%%%%%%%%%%%%%%%%%%%%%%%%%%%%%%%%%%%%%
\section{Introduction}\label{sec:intro}
%%%%%%%%%%%%%%%%%%%%%%%%%%%%%%%%%%%%%%%
    The nature of dark matter is among the most intriguing problems in physics today (see, for instance,\,\cite{Rubin_1970, Bertone:2004pz, Bertone:2016nfn, Feng:2010gw, ParticleDataGroup:2022pth, Cooley:2022ufh} and references therein). There have been numerous investigations using a variety of different methods which have strongly supported its existence---for example, galactic rotation curves\,\cite{Rubin_1970}, observations of the bullet cluster\,\cite{Clowe_2006}, power spectrum of cosmic microwave background (CMB)\,\cite{Planck_2015}, gravitational lensing\,\cite{BARTELMANN_2001}, and the large-scale structure of the universe\,\cite{Blumenthal_1984}.  However, despite the extensive amount of observational data, the microphysical nature of dark matter remains elusive. A wide array of candidates for dark matter have been proposed, ranging from weakly interacting massive particles (WIMPs)\,\cite{Steigman_1984}, axions\,\cite{Preskill_1982}, fuzzy dark matter\,\cite{Hu_2000}, hidden sector particles\,\cite{POSPELOV_2008}, to primordial black holes (PBHs)\,\cite{Zel'dovich_1967, Carr_1974, Carr_1975, Carr_2010, Carr_2016, Bird:2022wvk}, to name a few.    
    
    In recent times, the last of these proposed candidates---PBHs---have been of significant interest\,\cite{Zel'dovich_1967, Kribs_1999, Ballesteros_2018, Carr_1974, Carr_1975, Carr_2010, Carr_2016, Carr_2021, Carr_2024nlv, Sasaki_2016, Sasaki_2018, Niikura_2019, Lu_2019, Griest_2011,Bai:2018bej, Tamta:2024pow, Raidal_2017, Raidal_2019, Vaskonen_2021, CLESSE_2018, Araya_2020, Bird:2022wvk, Saga_2020, Papanikolaou_2023, Estes_2023, Zhang_2023, Hooper:2023nnl, Hooper:2022cvr}. There have been keen efforts to investigate various aspects of their formation\,\cite{Zel'dovich_1967, Kribs_1999, Ballesteros_2018, Carr_2016,Heydari:2021gea,Heydari:2021qsr,Heydari:2023rmq,Heydari:2023xts,Heydari:2024bxj}, mass distribution\,\cite{Carr_1975, Carr_2016}, signatures from gravitational waves\,\cite{Sasaki_2016, Raidal_2017, Vaskonen_2021}, gravitational lensing\,\cite{Niikura_2019, Lu_2019, Griest_2011, Tamta:2024pow}, and other pertinent phenomena\,\cite{Papanikolaou_2023, Zhang_2023}. In the early universe, their formation may have been a consequence of the gravitational collapse of large overdensities\,\cite{Carr_1974, Kribs_1999, Ballesteros_2018} emerging during inflation, and their population may have subsequently dwindled due to evaporation by Hawking radiation\,\cite{Hawking_1975}. This is especially true for the smaller mass PBHs since the Hawking temperature (for Schwarzschild black holes) is inversely proportional to the mass. Consequently, smaller mass Schwarzschild PBHs with masses $M_{\text{\tiny BH}}\lesssim 10^{15}\,\text{g or } 10^{-18} \,\solarmass$\,\cite{Page_1976} would have mostly evaporated by now (see, for instance,\,\cite{Carr_2021} and references therein). 
    
    Interestingly, however, this terminal evaporation can be evaded if the PBHs are magnetically charged\,\cite{Hiscock:1983ag}. If so, their Hawking temperature is reduced, and as evaporation continues, the Hawking temperature ultimately becomes zero for an extremal magnetically charged black hole, for which the mass equals the magnetic charge. For instance, in a scenario where magnetic monopoles existed in the early universe, the absorption of $N$ monopoles by the PBHs could statistically result in the accumulation of monopole charges proportional to $\sqrt{N}$\,\cite{ Stojkovic:2004hz, Maldacena_2020,Araya_2020, Estes_2023}, leading to the formation of magnetically charged PBHs or Magnetic Black Holes (MBHs)\footnote{These PBHs can absorb a significant portion of the population of magnetic monopoles, converting them into MBHs\,\cite{ Stojkovic:2004hz}}. Following absorption, they persist while emitting Hawking radiation, consequently losing mass and gradually approaching the extremal limit, eventually leading to zero Hawking temperature. This leads to a cessation of radiation, and hence even PBHs\footnote{ Rapidly spinning PBHs may result in extremal Kerr black holes having spin $J=GM_{\text{\tiny BH}}^2/c$, which also have vanishing Hawking temperatures\,\cite{Taylor:2024fvf,deFreitasPacheco:2020wdg}. Nevertheless, angular momentum is rapidly degraded in typical astrophysical environments, making these scenarios less potent compared to MBHs. Additionally, for the case of electrically charged PBHs electric charges are more readily neutralised in astrophysical environments, making them less relevant. } with masses $M_{\text{\tiny BH}}\lesssim 10^{15}\,\text{g or } 10^{-18} \,\solarmass$ can potentially survive, allowing for a relatively significant population of them to persist until the current epoch. There can be alternative mechanisms for the production of MBHs. One possibility involves the direct production of MBHs during an electroweak phase transition\,\cite{Cho:2016npz,Cho:2024vyq}. Another scenario considers the pair production of MBHs from cosmic strings in the presence of a background magnetic field\,\cite{Emparan:1995je,Ashoorioon:2020mos}.  
    We will primarily focus our study on non-spinning MBHs. Note also in passing that non-extremal MBHs having masses $M_{\text{\tiny BH}}\gtrsim 10^{15}\,\text{g}$ are metastable anyway due to their low Hawking temperatures compared to smaller mass MBHs. Consequently, they will avoid terminal Hawking evaporation even without achieving the extremal limit, allowing them to persist.
    
     MBHs have garnered significant theoretical interest in the past\,\cite{Cho:1975uz,Bais:1975gu,Galtsov:1989ip,Lee_1991, Lee_1991_2, Lee_1994, Andersson_1996,Gibbons:1990um}. In more recent years, there has also been a growing interest in their phenomenological aspects \,\cite{Estes_2023, Diamond_2021, Bai_2020, Ghosh_2020, Kim_2020, Hooper:2023nnl, Hooper:2022cvr, Liu_2020, Liu:2020bag, Liu:2022wtq, Estes:2022buj, DeFelice:2023rra, Turimov:2022evw, Maldacena_2020,  Dyson:2023ujk,Pereniguez:2023wxf,Pereniguez:2024fkn}. These include, for instance, investigations into gravitational and electromagnetic radiation arising from a binary system of dyonic black holes \,\cite{Liu_2020, Liu:2020bag, Liu:2022wtq}, as well as gravitational waves from neutron star (NS) and MBH mergers\,\cite{Estes:2022buj}. Studies have also attempted to discern between MBHs and their electric counterparts by studying their quasinormal modes\,\cite{DeFelice:2023rra} and considerations of the motion of charged test particles surrounding them\,\cite{Turimov:2022evw}. Other studies have looked into the restoration of electroweak symmetry\,\cite{Ambjorn:1989bd,Ambjorn:1989sz,Ambjorn:1992ca} around MBHs, leading to the formation of an electroweak corona near them\,\cite{Maldacena_2020}, and its phenomenological implications\,\cite{Bai_2020}. Accretion of charged particles by rotating MBHs has been another aspect of interest\,\cite{Dyson:2023ujk}, along with explorations on the implications of a topologically induced black hole electric charge\,\cite{Kim_2020}, dark extremal PBHs\,\cite{Bai:2019zcd}, and solutions for MBHs in nonlinear electrodynamics---for instance, with a focus on their implications for black hole shadows\,\cite{Nomura_2021, Kumaran:2022soh, Allahyari:2019jqz}, gravitational lensing\,\cite{Kumaran:2022soh}, and MBH quasinormal modes\,\cite{Chakrabarty:2018skk}.
    
     As MBHs carry monopole charges, they simultaneously serve as sources of extreme gravitational fields and unique monopolar magnetic fields, potentially giving rise to uncommon astrophysical effects. For instance, interactions with magnetic fields in galaxies, galaxy clusters, and cosmic regions may result in the acceleration of MBHs. This, in turn, could lead to a significant depletion of these magnetic fields. It is then evident that the survival of these fields until the present epoch may, in many cases, put a bound on the population of the MBHs. Parker\,\cite{Parker_1970, Turner_1982} originally introduced this idea in the context of pure magnetic monopoles, thereby establishing limits on the flux of such monopoles based on the survival of galactic magnetic fields.\footnote{These constraints for pure magnetic monopoles have since been expanded through assessments of the survival of galactic seed fields\,\cite{Adams_1993, Lewis_1999}, intracluster magnetic fields\,\cite{Rephaeli_1982}, and primordial magnetic fields\,\cite{Kobayashi_2022, Kobayashi_2023}.}  Recently,\,\cite{Perri:2023ncd} has investigated the acceleration of magnetic monopoles to large velocities by intergalactic magnetic fields. These, in turn, lead to modification of the galactic Parker bounds on the flux of non-virialized, pure magnetic monopoles\,\cite{Perri:2023ncd}.  Additionally, in the mass range $M \lesssim \mathcal{O}(10^{19})\text{ GeV}$, pure monopole fluxes have been estimated for which the inter-galactic fields may be significantly affected\,\cite{Perri:2023ncd}. 
     
     For MBHs with masses  $M_{\text{\tiny BH}}\gtrsim10^{-6}\,\text{Kg}\sim \mathcal{O}(10^{20})\,\text{GeV}$\,\cite{Carr_1974,Carr_2021},
    constraints on their flux and dark matter fraction, specifically from intergalactic magnetic fields (IGMFs) in cosmic voids and cosmic web filaments, are as yet unexplored. We explore new bounds coming from these systems in this study. These bounds for extremal MBHs also differ from those applicable to pure monopoles due to the distinctive relationship between charge and mass (i.e. $Q_{\text{\tiny BH}}=\sqrt{4 \pi G/\mu_0 }M_{\text{\tiny BH}})$. In addition, IGMFs may have different generation mechanisms, for instance, galactic flux leakage\,\cite{Bertone:2006mr} or primordial magnetogenesis\,\cite{Kandus:2010nw}, which can result in distinctive magnetic field characteristics and, therefore, different final bounds.  It is worth mentioning that bounds on the flux of MBHs based on the survival of galactic magnetic fields and primordial magnetic fields during radiation-dominated and reheating eras have already been explored\,\cite{Kobayashi_2023}. Furthermore,\,\cite{Ghosh_2020,Bai_2020} have placed galactic Parker bounds, based on the field in the Andromeda galaxy, on the fraction of dark matter in the form of MBHs---$f_{\text{\tiny DM}} \lesssim \mathcal{O}(10^{-3})$. As mentioned, among the aims of this work is to study and derive novel bounds on the flux and the dark matter fraction in the form of MBHs. Specifically, we will apply considerations of the survival of IGMFs in cosmic web filaments and cosmic voids with different generation mechanisms to set strong bounds. For low mass MBHs with $M_{\text{\tiny BH}}\lesssim 10^{15}\,\text{g}$, we will put Parker-type bounds only for extremal MBHs. However, as discussed, for masses $M_{\text{\tiny BH}}\gtrsim 10^{15}\,\text{g}$, we also present bounds applicable to non-extremal MBHs along with extremal MBHs.

    Apart from the Parker bounds on MBHs, we are also interested in exploring the effects of MBHs on polarized radio sources---for instance, synchrotron emissions from pulsars or galaxies. The basic idea is that the plane of polarization of linearly polarized electromagnetic waves rotate when they travel through an ionic medium in the presence of a magnetic field. This birefringence phenomenon is known as Faraday rotation or the Faraday effect. Observations of Faraday rotation from point sources have been widely employed to probe the structure of magnetic fields along the line of sight \cite{Manchester_1972, Thomson_1980, Hamilton_1985, Hamilton_1987}.\footnote{ Recently, few works have investigated the intrinsic Faraday effect in pulsar radio emissions due to NS magnetospheres\,\cite{Wang_2011} and also explored the quantum electrodynamic effects on birefringence in the intermediate frequency regime\,\cite{Shannon_2006}, in similar contexts. In other related works, investigations have also been conducted on the birefringence effects of axion-like particles in NSs\,\cite{Fortin:2023jlg, Poddar_2020}.}  Of particular relevance to us are works exploring the characteristics of the interstellar medium---through observations of diffuse polarized emissions and their Faraday rotation imprints\,\cite{Eck_2019, Hill_2017, Sun_2015, Lenc_2016}. This includes observation of polarization angle map has been observed at both cluster scales for arcsecond scales for clusters\,\cite{Bonafede_2010,Govoni2006} and sub-micro-arcsecond scales for compact objects\,\cite{Johnson_2015,EventHorizonTelescope:2021bee,EventHorizonTelescope:2024hpu}. We will specifically leverage similar aspects to scrutinize the presence of MBHs.

  In this study, we investigate the Parker bounds on magnetic black holes (MBHs) derived from the survival of cosmological magnetic fields. Additionally, we explore the novel Faraday rotation signatures that arise from these exotic MBHs, which may serve as unique observational probes. The key findings of our research are summarized below, highlighting the implications of cosmological magnetic fields on the population of MBHs and its observable Faraday rotation signature.
    \begin{itemize}
      \item We have first explored Parker bounds on extremal MBHs estimating constraints on their fraction as a dark matter component using IGMFs in cosmic voids ($f_{\text{\tiny DM}} \lesssim 10^{-8}$, see Eqs.\,\eqref{eq:fDM_primordial_IGMF} and \,\eqref{eq:FMBH_primordial_IGMF}) and cosmic web filaments ($f_{\text{\tiny DM}} \lesssim 10^{-4}$, see Eq.\,\eqref{eq:fDM_cos_fil}). These limits arise from the unclustered MBHs exhibiting typical velocities of $v \sim 10^{-3}~ c$ in cosmic voids and filaments\,\cite{Padilla:2005ea,rost_2021}.  For non-extremal MBHs, these constraints depend on the ratio of their magnetic charge to their mass. These constraints are outlined in  Eqs.\;\eqref{eq:fDM_primordial_IGMF_non} and \eqref{eq:fDM_cos_fil_non}.  Our findings indicate that these constraints are considerably more stringent than those derived from galactic magnetic fields alone\,\cite{Kobayashi_2022,Bai_2020,Ghosh_2020} and are presently the most robust constraints on the population of such objects. It is important to note that these bounds are contingent upon the existence of cosmic structures with magnetic fields; specifically, there should be no cosmic voids or web filaments with depleted magnetic fields.

      \item We have also investigated in detail unique Faraday rotation effects induced by these extremal MBHs.
      \begin{itemize}
        \item Assuming a semi-realistic constant plasma density profile and a more realistic galactic plasma density profile, we compute the change in polarization angle and rotation measure (RM) values for a linearly polarized wave due to an extremal MBH located within the Milky Way. We find that these values are plausibly sufficient for current observations to detect for the extremal MBH magnetic charges $Q^{\text{\tiny ex.}}_{\text{\tiny BH}}\gtrsim10^{22}~ \text{A-m}$, equivalently, $M^{\text{\tiny ex.}}_{\text{\tiny BH}}\gtrsim10^{-6}~ \solarmass$ (see Figs.\,\ref{fig:gal_MW_RM_ex_QM_plot_mp}-\ref{fig:gal_MW_PA_gauss_NS_MBH_comp_v2} and Tab.\,\ref{tab:fr_NS_MBH}).
        \item  We have done a comparative analysis of the change in polarization angle and RM value due to an MBH relative to an NS.  Considering comparable magnetic field strengths at the surfaces of an NS and the outer horizon of an MBH, we find that the values obtained for an MBH are markedly greater ($\mathcal{O}(10^{8})$) in magnitude compared to those for an NS with this matched characteristic (see Fig.\,\ref{fig:gal_MW_PA_gauss_NS_MBH_comp_v2} and Tab.\,\ref{tab:fr_NS_MBH}). 
        \item We have also established a simple quantitative measure to globally distinguish between the polarization angle maps of an MBH and an NS in many scenarios (see Eqs.\,\eqref{eq:fr_ns_cp_cont_ineq} and \eqref{eq:fr_mbh_cp_cont_ineq}). 
      \end{itemize} 
    \end{itemize}
       We also point out that due to the uncommon monopolar nature of the MBH's magnetic field, other astrophysical magnetic configurations, which are always non-monopolar in nature, will not be able to mimic these MBH signatures very readily. 

   The paper is organized as follows. In Sec.\,\ref{sec:park_bound_pmbh}, we concisely revisit the Parker bound calculations and then adapt these calculations to comprehensively estimate limits on the MBH dark matter fraction from the IGMFs in the cosmic voids and the cosmic web filaments. In Sec.\,\ref{sec:farad_effec}, we then briefly present the standard Faraday rotation calculation in the presence of a non-uniform magnetic field and then employ these expressions to carefully investigate the unique Faraday rotation effects caused by an MBH. In this context, we also compare these effects with those produced by an NS. Our main findings and conclusions are finally summarised in Sec.\,\ref{sec:sum_and_conc}.

%%%%%%%%%%%%%%%%%%%%%%%%%%%%%%%%%%%%%%%
\section{Parker-type bounds on primordial magnetic black holes}
%%%%%%%%%%%%%%%%%%%%%%%%%%%%%%%%%%%%%%%
\label{sec:park_bound_pmbh}
   First, let us explore Parker-type bounds on MBHs due to galactic and galaxy cluster magnetic fields, and IGMFs in cosmic voids and cosmic web filaments. For our purposes, these magnetic fields may be characterized mainly by three parameters---the field strength $B$, the regeneration time $t_{\text{\tiny reg}}$ (the time scale over which the field may be regenerated), and the coherence length $l_{\text{\tiny c}}$ (the length scale over which the field remains relatively constant). It is useful to initially examine the characteristics of these field parameters, as they will prove instrumental in forthcoming subsections when we calculate the limits.

    %%%%%%%%%%%%%%%%%%%%%%%%%%%%%%%%%%%%%%%
    \subsection{Characteristics of astrophysical and cosmic magnetic fields}
    \label{subsec:mag_field}
    %%%%%%%%%%%%%%%%%%%%%%%%%%%%%%%%%%%%%%%
    Let us start by examining the magnetic fields in galaxies and galaxy clusters. 
    The galactic magnetic fields in many cases exhibit equipartition between their mean and turbulent components, with each having field strength $B\sim\mathcal{O}(10) \,\mu\text{G}\,$\,(see, for instance,\,\cite{Kandaswamy2008,Beck:2013bxa,Beck:2000dc,Widrow:2002ud} and references therein). This equipartition is a common characteristic of fields generated through turbulent dynamo processes\,\cite{equipartition, cho2009}. As anticipated, the mean component of the field demonstrates greater coherence, with a coherence length of approximately $\mathcal{O}(1)\,\text{kpc}$, while the turbulent component exhibits coherence on scales roughly 10 times smaller\,\cite{Kandaswamy2008,Beck:2013bxa}. As turbulent motions are the driving force behind the dynamo process, the latter can be regenerated swiftly in around $\mathcal{O}(10^{-1})\,\text{Gyr}$\,\cite{Kandaswamy2008,Beck:2013bxa}, whereas the mean field undergoes regeneration only on scales of approximately $ \mathcal{O}(1)\,\text{Gyr}$\,\cite{Kandaswamy2008,Beck:2013bxa}. We therefore broadly characterise the conventional features of the mean and turbulent components of galactic magnetic fields\,\cite{Kandaswamy2008,Beck:2013bxa} as 
    %%%%%%%%%%%%%%%
    \begin{eqnarray}\label{eq:B_gal}
        B_{\text{\tiny mean}}^{\text{\tiny gal}} \sim\mathcal{O}(10)\,\mu\text{G}\,, \quad l_\text{\tiny c, mean}^{\text{\tiny gal}} \sim \mathcal{O}(1)\,\text{kpc}\,,&& \quad t_{\text {\tiny reg, mean}}^{\text{\tiny gal}}   \sim \mathcal{O}(1)\,\text{Gyr}  \;, \nonumber \\
        B_{\text{\tiny turb}}^{\text{\tiny gal}} \sim \mathcal{O}(10)\,\mu\text{G}\,, \quad l_\text{\tiny c, turb}^{\text{\tiny gal}} \sim \mathcal{O}(10^{-1})\,\text{kpc}\,,&& \quad t_{\text {\tiny reg, turb}}^{\text{\tiny gal}}   \sim \mathcal{O}(10^{-1})\,\text{Gyr}  \;.
    \end{eqnarray}
    %%%%%%%%%%%%%%%

    In galaxy clusters, the magnetic fields still maintain equipartition with field strengths around $ \mathcal{O}(1)\,\mu \text{G}$\,(see \cite{Kandaswamy2008,Carilli:2001hj,Widrow:2002ud} and references therein). However, the fields exhibit greater coherence in most cases, with the coherence lengths of the mean fields expected to be around $l_\text{\tiny c, mean}^{\text{\tiny clus/iso}} \sim \mathcal{O}(10)\,$ kpc\, and that of the turbulent fields around $l_\text{\tiny c, turb}^{\text{\tiny clus/iso}} \sim \mathcal{O}(1)\,$ kpc\,\cite{Kandaswamy2008,Carilli:2001hj}. Nevertheless, the regeneration time for galaxy cluster fields are expected to be comparable to those observed in galactic fields---$t_{\text {\tiny reg, mean}}^{\text{\tiny clus/iso}} \simeq \frac{1}{3} t_{\text {\tiny reg, turb}}^{\text{\tiny clus/iso}} \sim \mathcal{O}(10^{-1})\,$ Gyr\,\cite{Kandaswamy2008,Carilli:2001hj}. 

    In addition to magnetic fields within galaxies and galaxy clusters, there are also cosmic magnetic fields. These fields\,\cite{Widrow:2002ud,Neronov:2009gh,NeronovCorr,DurrerNeronov,Ackermann,FermiLAT-HESS,alves,Bertone:2006mr,Chen2015,Dermer,Tashiro2014,Tavecchio,Pshirkov,Planck:2015zrl}---traditionally referred to as intergalactic magnetic fields (IGMFs), represent magnetic fields that pervade large-scale structures in the universe; manifesting themselves, for instance, in the voids and filaments of the cosmic web. IGMFs have been an area of strong interest due to their elusive nature and the experimental difficulties in observing such low-strength fields. Recently, the Fermi-LAT/H.E.S.S collaboration\,\cite{FermiLAT-HESS} have explored the strength of IGMFs through the electromagnetic cascades induced by Blazar-emitted gamma rays. They have established a conservative lower limit of approximately $ \mathcal{O}(10^{-15})\,\text{G}$\,\cite{FermiLAT-HESS} on the strength of IGMFs for coherence lengths $l_{\text{\tiny{c}}}\gtrsim \mathcal{O}(1)\,\text{Mpc}$.  This is further supported by earlier works\,\cite{Ackermann,Tavecchio,Dermer,NeronovVovk,Neronov:2009gh}, which bound the IGMF strengths to $B \gtrsim \mathcal{O}(10^{-15})-\mathcal{O}(10^{-16})\,\text{G}$ for coherence length $l_{\text{\tiny{c}}}\gtrsim \mathcal{O}(1)\,\text{Mpc} $. Moreover, based on helicity considerations of cosmic magnetic fields (drawing from Fermi satellite gamma-ray observations), the IGMF has been estimated to have a strength of approximately $B \gtrsim \mathcal{O}(10^{-14})\,$G \,\cite{Tashiro2014,Chen2015}, for larger coherence lengths of  $\mathcal{O}(10)\,$Mpc.   Simultaneously, MHD arguments assuming a disordered cosmic magnetic field at recombination have established upper limits of $B \lesssim \mathcal{O}( 10^{-9})\,\text{G}$\,\cite{Wasserman}. 
    
     The origins of these IGMFs is still not understood fully, and there are several proposed mechanisms. One of these posits that IGMFs are produced primordially, arising in the early universe\,(see, for example,\,\cite{DurrerNeronov,Widrow:2002ud} and references therein). In this case, it is anticipated that the IGMFs would possess a strength of $\mathcal{O}(10^{-15})\,G\lesssim B \lesssim \mathcal{O}(10^{-9})\,\text{G}$\,\cite{FermiLAT-HESS,Planck:2015zrl}, exhibiting coherence on scales $\mathcal{O}(1)\,\text{Mpc}$ or greater\,\cite{FermiLAT-HESS,Planck:2015zrl}.  Alternatively,\,\cite{Bertone:2006mr} has suggested that seeding by galactic fields via flux leakage, may lead to field strengths of $\mathcal{O}(10^{-12})\,\mathrm{G}\lesssim B  \lesssim \mathcal{O}(10^{-8})\,\mathrm{G}$\,\cite{Bertone:2006mr} over a time scale of $\mathcal{O}(10)$ Gyr\,\cite{Bertone:2006mr}. If the field is indeed primordial and lacks a mechanism to amplify the strength (for example, a mean-field dynamo), the regeneration time must be at least as long as the Hubble time $\mathcal{O}(10)\,\text{Gyr}$ for the field to persist until the present day.  
     
     It has also been speculated that the source of IGMFs in cosmic voids may be due to magnetization by cosmic rays from void galaxies\,\cite{BeckHanasz}. These cosmic rays, generated and accelerated by supernova events within galaxies residing in voids, are energetic enough to surpass the galaxies' virial velocities and escape into the surrounding void. As they traverse the void, these cosmic rays carry electric currents, contributing to the magnetization of the void region at an approximate rate of about $10^{-16}\,$G/Gyr\,\cite{BeckHanasz,Miniati:2010ne}. Remarkably, the estimate in\,\cite{BeckHanasz} aligns with the recent observations by Fermi-LAT/H.E.S.S collaboration\,\cite{FermiLAT-HESS}, suggesting strongly the presence of magnetic field strengths of the order of $10^{-15}\,$G in cosmic voids.
     
     IGMFs in cosmic web filaments have also been inferred recently through rotational measure maps\,\cite{Carretti:2022fqk,Carretti:2022tbj,Amaral:2021mly,Akahori:2010ym} and synchrotron radio emission studies\,\cite{OSullivan:2018shr,Vernstrom:2021hru} of extragalactic sources.
     These magnetic fields have typical strengths of $\mathcal{O}(10^{-9})\,\text{G}$\,\cite{Carretti:2022fqk,Carretti:2022tbj,Amaral:2021mly,Akahori:2010ym,OSullivan:2018shr,Vernstrom:2021hru} with coherence lengths $\mathcal{O}(1)\,\text{Mpc}$\,\cite{Carretti:2022fqk,Carretti:2022tbj,Amaral:2021mly,Akahori:2010ym}. This seems to be in accordance with the expected IGMF characteristics as discussed earlier. Again, the actual origin of such magnetic field strengths is unclear at the moment, but the amplification of a seed primordial field by a dynamo mechanism and galactic outflows are two of the plausible explanations.  Conjecturing such origins, we therefore assume the regeneration time to be $\mathcal{O}(10)\,\text{Gyr}$.

    Based on the above discussions, for our subsequent analyses, we classify the IGMFs depending on their location in the cosmic regions---IGMFs in cosmic web filaments and those in cosmic voids. We note that independent of the origin of the IGMFs in cosmic web filaments, one can conservatively assume the following field characteristics\,\cite{Carretti:2022fqk,Carretti:2022tbj,Amaral:2021mly,Akahori:2010ym,OSullivan:2018shr,Vernstrom:2021hru}
    %%%%%%%%%%%%%%%
    \begin{equation}\label{eq:fil}
       B^{\text{\tiny fil}} \sim  \mathcal{O}(10^{-9})\,\text{G}\,, \quad l_\text{\tiny c}^{\text{\tiny fil}} \sim \mathcal{O}(1)\,\text{Mpc}\,, \quad t_{\text {\tiny reg}}^{\text{\tiny fil}}   \sim \mathcal{O}(10)\,\text{Gyr}  \;.
    \end{equation}
    %%%%%%%%%%%%%%%
    The origin of IGMFs in cosmic voids may be due to primordial fields, galactic outflows, or magnetization by void galaxies. Unlike their filamentary counterparts, the characteristics of the magnetic fields in cosmic voids are thought to be more dependent upon their actual origins. For the case when the magnetic fields are produced due to galactic outflows\,\cite{Bertone:2006mr}, we have
     %%%%%%%%%%%%%%%
    \begin{equation}\label{eq:IGMF_gal}
       \mathcal{O}(10^{-12})\,\text{G}\lesssim B^{\text{\tiny void, outflow}} \lesssim  \mathcal{O}(10^{-8})\,\text{G}\,, \quad l_\text{\tiny c}^{\text{\tiny void, outflow}} \sim \mathcal{O}(1)\,\text{Mpc}\,, \quad t_{\text {\tiny reg}}^{\text{\tiny void, outflow}}   \sim \mathcal{O}(10)\,\text{Gyr}  \;.
    \end{equation}
    %%%%%%%%%%%%%%%
    However, for the fields with a primordial origin, we have typically\,\cite{FermiLAT-HESS}
    %%%%%%%%%%%%%%%
    \begin{equation}\label{eq:B_IGMF_prim}
        B^{\text{\tiny void, prim}} \gtrsim \mathcal{O}(10^{-15})\,\text{G}\,, \quad l_\text{\tiny c}^{\text{\tiny void, prim}} \sim \mathcal{O}(1-10)\,\text{Mpc}\,, \quad t_{\text {\tiny reg}}^{\text{\tiny void, prim}}   \sim \mathcal{O}(10)\,\text{Gyr}  \;.
    \end{equation}
    %%%%%%%%%%%%%%%
    Finally, for fields generated via magnetization by void galaxies, we have\,\cite{BeckHanasz}
    %%%%%%%%%%%%%%%
    \begin{equation}\label{eq:B_IGMF_void_gal}
        B^{\text{\tiny void, gal}} \gtrsim \mathcal{O}(10^{-15})\,\text{G}\,, \quad l_\text{\tiny c}^{\text{\tiny void, gal}} \sim \mathcal{O}(1)\,\text{Mpc}\,, \quad t_{\text {\tiny reg}}^{\text{\tiny void, gal}}   \sim \mathcal{O}(10)\,\text{Gyr}  \;.
    \end{equation}
    %%%%%%%%%%%%%%%
    Here, we have estimated the regeneration time $t_{\text {\tiny reg}}^{\text{\tiny void, gal}}$ using the field strength $B^{\text{\tiny void, gal}}$ and assuming a magnetization rate $10^{-16}\,$G/Gyr\,\cite{BeckHanasz,Miniati:2010ne}. 

    Understanding these magnetic field characteristics is crucial in estimating the Parker-type bounds on the flux $F_{\text {\tiny BH}}$ and fraction $f_{\text {\tiny DM}}$ of Dark Matter that is made up of MBHs. The original idea\,\cite{Parker_1970} was that magnetic monopoles traversing galactic magnetic fields would undergo acceleration and deplete the field's energy. To ensure the field's persistence until the present era, it must be getting regenerated at a rate at least equal to the depletion rate. Considering these two effects, therefore, provides a constraint on the allowed monopole fluxes. We extend this idea in the next section to scenarios which include large-scale cosmic magnetic fields.

\subsection{Theoretical framework for Parker-type bounds}
\label{sec:theory}
     Having discussed the characteristics of galactic, galaxy cluster, and cosmic magnetic fields, let us now derive the theoretical expressions for the constraints on MBH flux and MBH dark matter fraction from magnetic field survival considerations. We will suitably adapt methodologies from\,\cite{Parker_1970,Turner_1982} to the cases of interest.

     Let us consider an MBH with mass $M_{\text{\tiny BH}}$ and magnetic charge $Q_{\text{\tiny BH}}$, moving with a non-relativistic velocity $\vec{v}$  through a magnetic field of strength $\vec{B}_{\text{\tiny c}}$ and coherence length   $l_{\text{\tiny c}}$. The MBH will experience an acceleration\footnote{ The relativistic correction to the velocity is of order  $\sim\mathcal{O}(v^2/c^2)$. For the magnetic field of interest, these corrections to the velocity are conservatively estimated to be $\sim\mathcal{O}(10^{-6})\%$. }
    %%%%%%%%%%%%%%%
    \begin{equation}\label{eq:vdot}
        \frac{d\vec{v}}{dt} = \frac{Q_{\text{\tiny BH}} \vec{B}_{\text{\tiny c}}}{M_{\text{\tiny BH}}}\;.
    \end{equation}
    %%%%%%%%%%%%%%%
     In the above equation, we have neglected the influence of free electric fields due to their relative paucity in scenarios we are considering and consequently negligible effects\,\cite{BasicMHD}.
     
     We may define a velocity scale $v_{\text{\tiny mag}}$\,\cite{Turner_1982}, termed the magnetic velocity, which denotes the speed acquired by an MBH starting from rest within a single coherence length $l_{\text{\tiny c}}$. Using Eq.\,\eqref{eq:vdot}, we have,
    %%%%%%%%%%%%%%%
    \begin{equation}\label{eq:vmag_def}
        v_{\text{\tiny mag}} = \sqrt{\frac{2 Q_{\text{\tiny BH}} l_{\text{\tiny c}} B_{\text{\tiny c}}}{M_{\text{\tiny BH}}}}\;,
    \end{equation}
    %%%%%%%%%%%%%%%
    where, $B_{\text{\tiny c}}=|\vec{B}_{\text{\tiny c}}|$. Although $v_{\text{\tiny mag}}$ explicitly depends upon the charge $Q_{\text{\tiny BH}}$ and mass $M_{\text{\tiny BH}}$, in the extremal case with $Q_{\text{\tiny BH}}=\sqrt{4 \pi G/\mu_0 }M_{\text{\tiny BH}}$, these dependences completely cancel. As we mentioned, a primordial BH that is magnetically charged undergoes Hawking evaporation and naturally evolves to the extremal limit, where the Hawking temperature vanishes, and the evaporation then subsides. In this case of physical interest, therefore, the magnetic field characteristics $B_{\text{\tiny c}}$ and $ l_{\text{\tiny c}}$ completely determine $v_{\text{\tiny mag}}$.
    
     Based on whether the initial velocity $v_{\text{\tiny in}}$ of the MBHs is much larger or smaller than $v_{\text{\tiny mag}}$, as discussed in\,\cite{Turner_1982}, we may classify the MBHs into two categories -- fast MBHs, for which $v_{\text{\tiny in}} \gg v_{\text{\tiny mag}}$ and slow MBHs, for which $v_{\text{\tiny in}} \ll v_{\text{\tiny mag}}$. For fast MBHs, since $v_{\text{\tiny in}} \gg v_{\text{\tiny mag}}$, the change in velocity when an MBH traverses a coherence length is small, meaning $\Delta v \sim v_{\text{\tiny mag}}  \ll v_{\text{\tiny in}}$. On the contrary, slow MBHs ($v_{\text{\tiny in}} \ll v_{\text{\tiny mag}}$) undergo acceleration to a velocity comparable to $v_{\text{\tiny mag}}$, over a coherence length, and the change in velocity is then significant, i.e. $\Delta v \sim v_{\text{\tiny mag}}  \gg v_{\text{\tiny in}}$. As we will demonstrate later, these two cases will be relevant in different scenarios, resulting in distinct bounds. MBHs may be further divided into two subcategories -- clustered MBHs and unclustered MBHs, i.e., which are virialized or infalling towards a gravitational source, respectively. As we shall see later, MBHs belonging to these subcategories will have different velocity distributions, potentially resulting in different bounds.
    
    Let us first re-derive a few relevant theoretical expressions and results. To this end, consider non-relativistic MBHs all having the same charge $Q_{\text{\tiny BH}}$ and mass $M_{\text{\tiny BH}}$ moving through a magnetic field $\vec{B}_{\text{\tiny c}}$, which is assumed to be coherent over a distance $l_{\text{\tiny c}}$. Using Eq.\,\eqref{eq:vdot}, we can express the rate of change in kinetic energy of MBHs as
    %%%%%%%%%%%%%%%
    \begin{equation}\label{eq:KEdotBasic}
        \frac{\text{d}\mathcal{E}_\text{\tiny k}}{\text{d}t} = Q_{\text{\tiny BH}} \vec{B}_{\text{\tiny c}} \cdot \vec{v}\;.
    \end{equation}
    %%%%%%%%%%%%%%%
    Similarly, the second derivative of the kinetic energy can be written as,
    %%%%%%%%%%%%%%%
    \begin{equation}\label{eq:KEdoubledotNew}
        \frac{\text{d}^2\mathcal{E}_\text{\tiny k}}{\text{d}t^2}  = \frac{Q_{\text{\tiny BH}}^2 B_{\text{\tiny c}}^2}{M_{\text{\tiny BH}}} \;.
    \end{equation}
    %%%%%%%%%%%%%%%
    Notice that the right-hand side of the above equation is constant, and thus, all higher derivatives of the kinetic energy will vanish. Now, if $n$ number of MBHs enter the coherent magnetic field at time $t$ and traverse $l_{\text{\tiny c}}$ in a time $\Delta t$, we can express the average gain in their kinetic energy as
    %%%%%%%%%%%%%%%
    \begin{equation}\label{eq:finite_diff}
     \Delta \mathcal{E}_\text{\tiny k} \equiv  \left\langle \mathcal{E}_\text{\tiny k}(t+\Delta t)\right\rangle -\left\langle \mathcal{E}_\text{\tiny k}(t)\right\rangle=\Delta t\left\langle\frac{d \mathcal{E}_\text{\tiny k}}{d t}\right\rangle+\frac{1}{2}(\Delta t)^{2}\left\langle\frac{d^{2} \mathcal{E}_\text{\tiny k}}{d t^{2}}\right\rangle \;.
    \end{equation}
    Notice that Eq.\,\eqref{eq:finite_diff} is obtained by expanding $\mathcal{E}_\text{\tiny k}(t+\Delta t)$, and subsequently taking an average\footnote{Here, we define averege energy of MBHs as $\left\langle \mathcal{E}_\text{\tiny k}\right\rangle\equiv \frac{1}{n}\sum_{i}^{n}\mathcal{E}_\text{\tiny k,i}$, where $\mathcal{E}_\text{\tiny k,i}$ is the energy of $i$th MBH.} over n MBHs.  Additionally, since the higher derivatives of $\mathcal{E}_\text{\tiny k}$ are all zero, Eq.\,\eqref{eq:finite_diff} is an exact equation. Substituting Eqs.\,\eqref{eq:KEdotBasic} and \eqref{eq:KEdoubledotNew} in Eq.\,\eqref{eq:finite_diff}, we get
    %%%%%%%%%%%%%%%
    \begin{equation}\label{eq:deltaKENew_1}
        \Delta \mathcal{E}_\text{\tiny k} = Q_{\text{\tiny BH}} \langle \vec{B}_{\text{\tiny c}} \cdot \vec{v}_{\text{\tiny in}}  \rangle \Delta t+ \frac{Q_{\text{\tiny BH}}^{2} B_{\text{\tiny c}}^{2} }{2 M_{\text{\tiny BH}} } \Delta t^{2}\;.
    \end{equation}
    %%%%%%%%%%%%%%%

      Evidently, the first term depends on the velocity distribution of MBHs. As we shall see later, this will lead to two physical scenarios---   clustered/virialized MBHs and unclustered/infalling MBHs.  One expects an isotropic velocity distribution for clustered MBHs (virialized case). In this scenario, the first term in Eq.\,\eqref{eq:deltaKENew_1} will vanish, as $\langle \vec{B}_{\text{\tiny c}} \cdot \vec{v}_{\text{\tiny in}}  \rangle=0$. However, as we shall see, for unclustered MBHs (infalling case), the contribution from the first term will usually lead to stronger bounds.
      
      The energy gained by  MBHs comes at the expense of the background magnetic field $\vec{B}_{\text{\tiny c}}$. As discussed by\,\cite{Turner_1982}, and more recently in the context of IGMFs by\,\cite{Kobayashi_2022, Kobayashi_2023}, MBHs may also return energy to the magnetic fields when MBHs with opposite magnetic charges oscillate with the magnetic field; similar to electrostatic plasma oscillations or Langmuir oscillations. However, for MBHs in the scenarios we consider, the time period of such oscillations is significantly longer than the regeneration time, making them less relevant. Assuming no other back reactions on the field due to the MBHs, this leads to a depletion of the fields by the MBHs and places bounds on their fluxes based on magnetic field persistence till the current era. This concept forms the essence of Parker bounds and has been widely employed in previous studies\,\cite{Parker_1970, Turner_1982, Ghosh_2020, Adams_1993, Bai, Lewis_1999}.
       
    Now, summing over all the contributions, the field loses energy at a rate 
    %%%%%%%%%%%%%%%
    \begin{equation}\label{eq:dEdt}
        \left | \frac{\text{d}\mathcal{E}_\text{\tiny field}}{\text{d}t} \right | = 4 \pi l_{\text{\tiny c}}^{2}\Delta \mathcal{E}_\text{\tiny k} F_{\text{\tiny BH}}\;.
    \end{equation}
    %%%%%%%%%%%%%%%
    Here, $F_{\text{\tiny BH}}$ is the number of MBHs passing per unit area per unit time through the magnetic field\,\footnote{In some literature\,\cite{Turner_1982, Parker_1970}, flux is defined as number/(area $\cdot$ second $\cdot$ solid angle). Consequently, a flux $\tilde{F}$ defined in this way is related to the flux $F$ defined above by $\tilde{F} = F/\pi$. }.  The depletion time scale of the magnetic energy can be expressed as the ratio of the total enclosed magnetic energy stored in the field in a coherence region (taken to be a sphere of radius $l_{\text{\tiny c}}$) to the rate of energy loss and is given by
     %%%%%%%%%%%%%%%
     \begin{equation}\label{eq:t_dep}
        t_{\text{\tiny dep}} \simeq\frac{\mathcal{E}_\text{\tiny field}}{\left | \frac{\text{d}\mathcal{E}_\text{\tiny field}}{\text{d}t} \right |}= \frac{  B_{\text{\tiny c}}^{2}l_{\text{\tiny c}}}{6 \mu_{0}\Delta \mathcal{E}_\text{\tiny k} F_{\text{\tiny BH}} }\;.
    \end{equation}
    %%%%%%%%%%%%%%%
    Here, we have taken $\mathcal{E}_\text{\tiny field}=\left(\frac{1}{2 \mu_{0}} B_{\text{\tiny c}}^{2}\right)\left(\frac{4}{3} \pi l_{\text{\tiny c}}^{3}\right)$.  To ensure the field's persistence till the present day, the magnetic fields must regenerate faster than they are being depleted. Equivalently, the regeneration time must be smaller than the depletion time. Hence, we arrive at the condition for the survival of magnetic fields
    %%%%%%%%%%%%%%%
    \begin{equation}\label{eq:regeneration}
        t_{\text{\tiny reg}} \leq t_{\text{\tiny dep}}\simeq \frac{  B_{\text{\tiny c}}^{2}l_{\text{\tiny c}}}{6 \mu_{0}\Delta \mathcal{E}_\text{\tiny k} F_{\text{\tiny BH}} }\;.
    \end{equation}
    %%%%%%%%%%%%%%%
    Here, $t_{\text{\tiny reg}}$ is the regeneration time. Eq.\,\eqref{eq:regeneration} in turn puts a limit on the flux of MBHs
    %%%%%%%%%%%%%%%
    \begin{equation}\label{eq:fluxNew_gen}
        F_{\text{\tiny BH}} \leq \frac{  B_{\text{\tiny c}}^{2}l_{\text{\tiny c}}}{6 \mu_{0}\Delta \mathcal{E}_\text{\tiny k} t_{\text{\tiny reg}}  }\;.
    \end{equation}
    %%%%%%%%%%%%%%%
    
    In addition, using this bound on the flux of MBHs, we may now further establish an upper limit on the fraction $f_{\text{\tiny DM}}$ of dark matter (DM) in the form of extremal MBHs. The flux of MBHs moving with speed $v$ is related to the mass density $(f_{\text{\tiny DM}} \rho_{\text{\tiny DM}})$ of MBHs as
    %%%%%%%%%%%%%%%
    \begin{equation}\label{eq:FMBH_FDM_formal}
        F_{\text{\tiny BH}} = \frac{v \rho_{\text{\tiny DM}} f_{\text{\tiny DM}}}{M_{\text{\tiny BH}}}\;,
    \end{equation}
    %%%%%%%%%%%%%%%
    where, $\rho_{\text{\tiny DM}}$ is the total dark matter density. Using Eqs.\,\eqref{eq:fluxNew_gen} and \eqref{eq:FMBH_fDM_formal}, we may then express the limit on the dark matter fraction in the form of MBHs as
     %%%%%%%%%%%%%%%
    \begin{equation}\label{eq:FMBH_fDM_formal}
        f_{\text{\tiny DM}} \leq \frac{  B_{\text{\tiny c}}^{2}l_{\text{\tiny c}}M_{\text{\tiny BH}}}{6 \mu_{0}\Delta \mathcal{E}_\text{\tiny k} t_{\text{\tiny reg}}  v \rho_{\text{\tiny DM}}} \;.
    \end{equation}
    %%%%%%%%%%%%%%%
   
   Depending upon whether the MBHs are fast or slow, we may estimate the time $\Delta t$ taken to cover $l_{\text{\tiny c}}$, and using Eqs.\,\eqref{eq:deltaKENew_1}, \eqref{eq:FMBH_FDM_formal} and \eqref{eq:FMBH_fDM_formal}, we may place specific bounds on MBHs in various cases of interest. As discussed earlier, MBHs may be clustered (virialized) or unclustered (infalling towards a gravitational source), thereby having distinct velocity distributions. Clearly, from Eq.\,\eqref{eq:deltaKENew_1}, different velocity distributions will lead to different gains in kinetic energy by MBHs. This may, as a consequence, lead to different Parker-type bounds. Therefore, we will divide our analysis into two parts -- fast MBHs, which have $v_{\text{\tiny in}} \gg v_{\text{\tiny mag}}$, and slow MBHs, with $v_{\text{\tiny in}} \ll v_{\text{\tiny mag}}$, and then within each of these categories we will study clustered (virialized) and unclustered (infalling) MBHs.

    %%%%%%%%%%%%%%%%%%%%%%%%%%%%%%%%%
    \subsubsection{Fast MBHs ($v\sim v_{\text{\tiny in}} \gg v_{\text{\tiny mag}}$)}\label{subsec:fast_MBHs}
    Fast MBHs undergo a minimal change in their velocities when they traverse $l_{\text{\tiny c}}$. This is because their initial velocity, $v_{\text{\tiny in}}$, significantly exceeds the change in velocity induced by the magnetic field over $l_{\text{\tiny c}}$  (i.e., $\Delta v \sim v_{\text{\tiny mag}} \ll v_{\text{\tiny in}}$). Therefore, we may write $v\sim v_{\text{\tiny in}}$  and the time taken to cross  $l_{\text{\tiny c}}$ may simply be taken as $\Delta t\sim l_{\text{\tiny c}}/v_{\text{\tiny in}}$. Using Eq.\,\eqref{eq:deltaKENew_1}, we may then estimate the change in the kinetic energy of fast MBHs as
    %%%%%%%%%%%%%%%
    \begin{equation}\label{eq:deltaKENew_fast}
        \Delta \mathcal{E}_\text{\tiny k}^{\text{\tiny fast}} \simeq Q_{\text{\tiny BH}} l_\text{\tiny c}\langle \vec{B}_{\text{\tiny c}} \cdot \hat{v}_{\text{\tiny in}} \rangle + \frac{Q_{\text{\tiny BH}}^{2} B_{\text{\tiny c}}^{2} l_\text{\tiny c}^{2}}{2 M_{\text{\tiny BH}} v_{\text{\tiny in}}^{2}} \;,
    \end{equation}
    %%%%%%%%%%%%%%%
    where, $\hat{v}_{\text{\tiny in}}=\vec{v}_{\text{\tiny in}}/v_{\text{\tiny in}}$. 
    
    Notice that for the two relevant terms, one is linear and one is quadratic in $B_{\text{\tiny c}}$. The linear term depends upon the velocity distribution of MBHs, and we will have two physical scenarios.  MBHs may be clustered (virialized), leading to an isotropic velocity distribution, for which the term linear in $B_{\text{\tiny c}}$ vanishes. In this case, then, the dominant contribution to the energy gain is from the quadratic term.  On the other hand, when the MBHs are infalling towards a gravitational source, a directional flow is expected. In this case, both the linear and quadratic terms will contribute to the energy gain. We will calculate the limits on MBHs in these two specific cases in the next sections.

    For the case where the MBHs pass through multiple coherence length cells, uncorrelated field directions imply that the first term in Eq.\,\eqref{eq:deltaKENew_fast} vanishes. The second term will then scale as $N$, the number of coherence length cells that the MBHs traverse. However, we will see later that for cosmic voids and filaments, the very large coherence lengths imply that the number of cells traversed is $N \sim \mathcal{O}(1)$. We thus cannot neglect the contribution from the first term of Eq.\,\eqref{eq:deltaKENew_fast} in those cases.
    %%%%%%%%%%%%%%%%%%%%%%%%%%%%%%%%%%%
    \subsubsection*{Clustered MBHs}
    %%%%%%%%%%%%%%%%%%%%%%%%%%%%%%%%%%%
    As mentioned earlier, when MBHs are clustered or virialized, their velocity distribution is largely isotropic, and therefore the term $\langle \vec{B}_{\text{\tiny c}} \cdot \hat{v}_{\text{\tiny in}} \rangle=0$ over $l_{\text{\tiny c}}$ (since $\vec{B}_{\text{\tiny c}}$ is approximately constant). Substituting this in Eq.\,\eqref{eq:deltaKENew_fast}, we obtain the gain in the kinetic energy of MBHs
    %%%%%%%%%%%%%%%
    \begin{equation}\label{eq:deltaKENewUnif_fast_clustered}
        \Delta \mathcal{E}_\text{\tiny k}^{\text{\tiny fast,clust}} \simeq \frac{Q_{\text{\tiny BH}}^{2} B_{\text{\tiny c}}^{2} l_\text{\tiny c}^{2}}{2 M_{\text{\tiny BH}} v_{\text{\tiny in}}^{2}} \;.
    \end{equation}
    %%%%%%%%%%%%%%%
    Using Eqs.\,\eqref{eq:fluxNew_gen} and \eqref{eq:deltaKENewUnif_fast_clustered}, we then get the bounds on the flux of fast, clustered MBHs
      %%%%%%%%%%%%%%%
    \begin{equation}\label{eq:FBH_fast_clustered}
        F_{\text{\tiny BH}}^{\text{\tiny fast,clust}} \lesssim \frac{M_{\text{\tiny BH}} v_{\text{\tiny in}}^{2}}{3 \mu_{0} Q_{\text{\tiny BH}}^{2} l_{\text{\tiny c}} t_{\text{\tiny reg}}} \;.
    \end{equation}
    %%%%%%%%%%%%%%%
      Finally, combining this with Eq.\,\eqref{eq:FMBH_fDM_formal}, we find the expression for the DM fraction in this case
    %%%%%%%%%%%%%%%
    \begin{equation}\label{eq:fDM_fast_clustered}
        f_{\text{DM}}^{\text{\tiny fast,clust}}\lesssim \frac{M_{\text{\tiny BH}}^2 v_{\text{\tiny in}}}{{3 \mu_{0} Q_{\text{\tiny BH}}^{2} l_{\text{\tiny c}} t_{\text{\tiny reg}} \rho_{\text{\tiny DM}}}}\;.
    \end{equation}
     %%%%%%%%%%%%%%%
     
    Notice here that the bounds on $F_{\text{\tiny BH}}^{\text{\tiny fast,clust}}$ and $f_{\text{DM}}^{\text{\tiny fast,clust}}$ are independent of the magnetic field strength $\vec{B}_{\text{\tiny c}}$ and are inversely proportional to the other magnetic field characteristics i.e.  $l_{\text{\tiny c}} $ and  $t_{\text{\tiny reg}}$. Additionally, for extremal MBHs since $Q_{\text{\tiny BH}}=\sqrt{4 \pi G/\mu_0 }M_{\text{\tiny BH}}$, 
    the limit on $F_{\text{\tiny BH}}^{\text{\tiny fast,clust}}$ is inversely proportional to the mass $M_{\text{\tiny BH}}$ or magnetic charge $Q_{\text{\tiny BH}}$, while the limit on $f_{\text{DM}}^{\text{\tiny fast,clust}}$ is independent of the MBH mass or charge.
    %%%%%%%%%%%%%%%%%%%%%%%%%%%%%%%%%%%
    \subsubsection*{Unclustered MBHs}
    %%%%%%%%%%%%%%%%%%%%%%%%%%%%%%%%%%%
    Relevant scenarios with unclustered MBHs may arise, for instance, when they are infalling into a gravitational well before entering the coherence length region. In such circumstances, one may expect a uniform velocity distribution for the MBHs to be the apropos assumption. 
    
    Once they enter the coherence length region, these unclustered MBHs are accelerated uniformly in the direction of $B_{\text{\tiny c}}$. However, since the MBHs are fast, the change in velocity is small again. Therefore, we may write $\vec{v}\sim\vec{v}_{\text{\tiny in}}$ and $\langle \vec{B}_{\text{\tiny c}} \cdot \hat{v} \rangle\sim \langle \vec{B}_{\text{\tiny c}} \cdot \hat{v}_{\text{\tiny in}} \rangle =B_{\text{\tiny c}}v_{\text{\tiny in}}\cos\alpha$, where $\alpha $ is the angle between the velocity $\vec{v}_{\text{\tiny in}}$ and magnetic field $\vec{B}_{\text{\tiny c}}$.  Using Eq.\,\eqref{eq:deltaKENew_fast}, we then get
    %%%%%%%%%%%%%%%
    \begin{equation}\label{eq:deltaKENew_fast_unclustered}
        \Delta \mathcal{E}_\text{\tiny k}^{\text{\tiny fast,unclust}} \simeq Q_{\text{\tiny BH}} B_{\text{\tiny c}}l_\text{\tiny c}\cos\alpha + \frac{Q_{\text{\tiny BH}}^{2} B_{\text{\tiny c}}^{2} l_\text{\tiny c}^{2}}{2 M_{\text{\tiny BH}} v_{\text{\tiny in}}^{2}}=\frac{M_{\text{\tiny BH}}v_{\text{\tiny mag}}^2}{2}\left(\cos\alpha+\frac{v_{\text{\tiny mag}}^2}{4v_{\text{\tiny in}}^2}\right) \;.
    \end{equation}
    %%%%%%%%%%%%%%%
    When $\alpha < \cos^{-1}\left(-{\frac{v_{\text{\tiny mag}}^2}{4v_{\text{\tiny in}}^2}}\right)\sim \pi/2 $ (i.e., when MBHs have a component of velocity along the magnetic field strength), the MBHs will gain energy, and there will be a reduction in magnetic field strength. Otherwise, when they have a component opposite to the field, the MBHs will slow down and return energy to the field. To establish the Parker-type bounds, we focus our attention on cases where the magnetic fields are being attenuated. In this case, we may use Eqs.\,\eqref{eq:fluxNew_gen} and \eqref{eq:deltaKENew_fast_unclustered} to put bounds on the flux
     %%%%%%%%%%%%%%%
    \begin{equation}\label{eq:FBH_fast_unclustered}
        F_{\text{\tiny BH}}^{\text{\tiny fast,unclust}} \lesssim \frac{M_{\text{\tiny BH}} v_{\text{\tiny in}}^{2}}{3 \mu_{0} Q_{\text{\tiny BH}}^{2} l_{\text{\tiny c}} t_{\text{\tiny reg}}} \frac{1}{\left(1 + 4\frac{v_{\text{\tiny in}}^2}{v^2_{\text{\tiny mag}} }\cos \alpha\right)}\;.
    \end{equation}
    %%%%%%%%%%%%%%%
    Using Eq.\,\eqref{eq:FMBH_fDM_formal}, we may, as before, place bounds on the dark matter fraction
     %%%%%%%%%%%%%%%
    \begin{equation}\label{eq:fdm_fast_unclustered}
        f_{\text{DM}}^{\text{\tiny fast,unclust}} \lesssim \frac{M_{\text{\tiny BH}}^2 v_{\text{\tiny in}}}{{3 \mu_{0} Q_{\text{\tiny BH}}^{2} l_{\text{\tiny c}} t_{\text{\tiny reg}} \rho_{\text{\tiny DM}}}} \frac{1}{\left(1 + 4\frac{v_{\text{\tiny in}}^2}{v^2_{\text{\tiny mag}} }\cos \alpha\right)}\;.
    \end{equation}
    %%%%%%%%%%%%%%%
    
    Since $v_{\text{\tiny mag}}$ depends upon the magnetic field characteristics, unlike the clustered case, the bounds on $F_{\text{\tiny BH}}^{\text{\tiny fast,unclust}}$ and $f_{\text{DM}}^{\text{\tiny fast,unclust}}$ here are dependent on the magnetic field strength ${B}_{\text{\tiny c}}$. However, similar to the clustered case, for extremal MBHs with $Q_{\text{\tiny BH}}=\sqrt{4 \pi G/\mu_0 }M_{\text{\tiny BH}}$, 
    the limit on $F_{\text{\tiny BH}}^{\text{\tiny fast,unclust}}$ is inversely proportional to the mass $M_{\text{\tiny BH}}$ (or, equivalently, magnetic charge $Q_{\text{\tiny BH}}$) and the $f_{\text{DM}}^{\text{\tiny fast,unclust}}$ limit is independent of these. Also, in the limit when $4v^2_{\text{\tiny in}}\cos\alpha\gg v^2_{\text{\tiny mag}}$, i.e. when the component of the initial velocity of MBHs along the magnetic field is significantly greater than the magnetic velocity, we may write
    %%%%%%%%%%%%%%%
    \begin{equation}\label{eq:FBH_fast_unclustered_lim}
        F_{\text{\tiny BH}}^{\text{\tiny fast,unclust}} \lesssim \frac{{B}_{\text{\tiny c}} }{6 \mu_{0} Q_{\text{\tiny BH}}  t_{\text{\tiny reg}}\cos \alpha} \;,
    \end{equation}
    %%%%%%%%%%%%%%%
    and
    %%%%%%%%%%%%%%%
    \begin{equation}\label{eq:fdm_fast_unclustered_lim}
        f_{\text{DM}}^{\text{\tiny fast,unclust}} \lesssim \frac{{B}_{\text{\tiny c}}M_{\text{\tiny BH}}  }{6 \mu_{0} Q_{\text{\tiny BH}}  t_{\text{\tiny reg}}v_{\text{\tiny in}}\rho_{\text{\tiny DM}}\cos \alpha}\;.
    \end{equation}
    In this limit, the bounds on $F_{\text{\tiny BH}}^{\text{\tiny fast,unclust}}$ and $f_{\text{DM}}^{\text{\tiny fast,unclust}}$ are proportional to ${B}_{\text{\tiny c}}$, but independent of ${l}_{\text{\tiny c}}$. 
  
    We will use Eqs.\,\eqref{eq:FBH_fast_clustered}, \eqref{eq:fDM_fast_clustered}, \eqref{eq:FBH_fast_unclustered} and \eqref{eq:fdm_fast_unclustered} to estimate bounds on the flux and the dark matter fraction of extremal MBHs in the scenarios where they are fast, i.e. $v\sim v_{\text{\tiny in}} \gg v_{\text{\tiny mag}}$. In the next subsection, we will compute the corresponding expressions for the case of slow-moving MBHs with $v\sim v_{\text{\tiny in}}\ll v_{\text{\tiny mag}}$.
%%%%%%%%%%%%%%%%%%%%%%%%%%%%%%%%%
    \subsubsection{Slow MBHs ($v_{\text{\tiny in}} \ll v_{\text{\tiny mag}}$)}
    Let us now focus on slow-moving MBHs. Slow MBHs will experience an initial acceleration due to the magnetic fields and move in the direction of magnetic field lines. This makes them attain a velocity $v \simeq v_{\text{\tiny mag}}$ within the first coherence length they traverse. However, in subsequent coherence lengths, due to their low velocities, they may undergo substantial deflections, depending upon the magnetic field orientations, resulting in a random trajectory. Furthermore, unlike the fast MBH case, the time taken to cross the coherent length $l_{\text{\tiny c}}$ is no longer $ l_{\text{\tiny c}}/v_{\text{\tiny in}}$, and therefore Eq.\,\eqref{eq:deltaKENew_fast} doesn't hold. Irrespective of whether MBHs are clustered or unclustered, they will be accelerated in the direction of the magnetic field. And for initially clustered, slow MBHs, since the virial velocity $v_{\text{\tiny vir}}\equiv v_{\text{\tiny in}}\ll v_{\text{\tiny mag}}$, they will acquire a velocity $v\sim v_{\text{\tiny mag}}$, in the first coherence length region and become unclustered. Additionally, as we shall see later, in the galactic, cosmic void and cosmic filament systems we consider, the virial velocities typically satisfy $v_{\text{vir}}\gg v_{\text{mag}}$. Thus, the virialized MBHs are always typically categorised as fast for these systems.
    
    Using Eq.\,\eqref{eq:vdot}, we find that for $v_{\text{\tiny in}} \ll v_{\text{\tiny mag}}$ (and thus $\Delta v\simeq v_{\text{\tiny mag}}$), the time taken to cross $l_{\text{\tiny c}}$ is $\Delta t\simeq 2l_{\text{\tiny c}}/v_{\text{\tiny mag}}$. Substituting it in Eq.\,\eqref{eq:deltaKENew_1}, and taking $v_{\text{\tiny in}} \ll v_{\text{\tiny mag}}$, we get the energy acquired by the MBHs as
    %%%%%%%%%%%%%%%
    \begin{equation}\label{eq:deltaKENew_slow_1}
    \mathcal{E}^{\text{\tiny slow}}_{\text{\tiny k,1}}\equiv \Delta \mathcal{E}^{\text{\tiny slow}}_{\text{\tiny k,1}} = Q_{\text{\tiny BH}} B_{\text{\tiny c}} l_{\text{\tiny c}}\;.
    \end{equation}
    %%%%%%%%%%%%%%%
    Notice that the contribution of the first term in  Eq.\,\eqref{eq:deltaKENew_1} is negligible since $v_{\text{\tiny in}} \ll v_{\text{\tiny mag}}$. Also, unlike fast MBHs, the change in kinetic energy of slow MBHs doesn't scale as $N$ when they pass through $N$ multiple coherence lengths. We will now estimate this scaling and find the average change in energy when the MBHs pass a single coherence length. A similar treatment due to multiple coherence lengths has been previously investigated by \cite{Kobayashi_2023} for the change in monopole velocities.
    
    Let us take $v_{\text{\tiny N}}$ to be the velocity of a slow MBH when it crosses the $\text{N}^{\text{th}}$ coherence length region and $\vec{B}_{\text{\tiny c,N}}={B}_{\text{\tiny c}}\hat{B}_{\text{\tiny c,N}}$ to be the magnetic field there. Then, using Eq.\,\eqref{eq:deltaKENew_1}, one may write the change in its energy when it crosses the $\text{N}^{\text{th}}$ coherence length region as
    %%%%%%%%%%%%%%%
    \begin{equation}\label{eq:deltaKENew_slow}
        \Delta \mathcal{E}_\text{\tiny k,N}^{\text{\tiny slow}}\equiv\mathcal{E}_\text{\tiny k,N}^{\text{\tiny slow}}-\mathcal{E}_\text{\tiny k,N-1}^{\text{\tiny slow}} \simeq Q_{\text{\tiny BH}} \Delta t_{\text{\tiny N}}  \vec{B}_{\text{\tiny c,N}} \cdot \vec{v}_{\text{\tiny N}} + \frac{Q_{\text{\tiny BH}}^{2} B_{\text{\tiny c}}^{2} }{2 M_{\text{\tiny BH}} } \Delta t_{\text{\tiny N}}^{2}\quad \text{for N $\ge 2$}\;,
    \end{equation}
    %%%%%%%%%%%%%%%
    where $\mathcal{E}_\text{\tiny k,N}\equiv M_{\text{\tiny BH}}v^2_{\text{\tiny N}}/2$ is the energy of the MBH when  it crosses the $\text{N}^{\text{th}}$ cell and $\Delta t_{\text{\tiny N}}$ is the time the MBH takes to cross the $\text{N}^{\text{th}}$ cell. As mentioned earlier, $\Delta t_{\text{\tiny 1}} \simeq 2l_{\text{\tiny c}}/v_{\text{\tiny mag}}$, and we may assume that for $N\ge2$, $\Delta t_{\text{\tiny N}}\sim l_{\text{\tiny c}}/v_{\text{\tiny N-1}}$. Therefore, Eq.\,\eqref{eq:deltaKENew_slow} becomes,
    %%%%%%%%%%%%%%%
    \begin{equation}\label{eq:deltaKENew_slow_N1}
        \mathcal{E}_\text{\tiny k,N}^{\text{\tiny slow}}-\mathcal{E}_\text{\tiny k,N-1}^{\text{\tiny slow}} \simeq  \Delta \mathcal{E}^{\text{\tiny slow}}_{\text{\tiny 1}} \hat{B}_{\text{\tiny c,N}} \cdot \hat{v}_{\text{\tiny N-1}} + \frac{(\Delta \mathcal{E}^{\text{\tiny slow}}_{\text{\tiny 1}})^{2} }{4\mathcal{E}_\text{\tiny k,N-1}^{\text{\tiny slow}} }\quad \text{for N $\ge 2$} \;.
    \end{equation}
    %%%%%%%%%%%%%%%
    Here, we have used Eq.\,\eqref{eq:deltaKENew_slow_1} for the expression for  $\Delta \mathcal{E}^{\text{\tiny slow}}_{\text{\tiny 1}}$. Eq.\,\eqref{eq:deltaKENew_slow_N1} is a recursion relation that relates the energy of a slow MBH after it crosses the $\text{N}^{\text{th}}$ cell with the energy when it crosses the $(\text{N-1})^{\text{th}}$ cell. Since the magnetic fields in subsequent cells are assumed to be randomly oriented,  the first term in Eq.\,\eqref{eq:deltaKENew_slow_N1} doesn't contribute to the mean behaviour, and we may rewrite the above expression as
    %%%%%%%%%%%%%%%
    \begin{equation}\label{eq:deltaKENew_slow_N}
        \mathcal{E}_\text{\tiny k,N}^{\text{\tiny slow}}-\mathcal{E}_\text{\tiny k,N-1}^{\text{\tiny slow}} \simeq  \frac{(\Delta \mathcal{E}^{\text{\tiny slow}}_{\text{\tiny 1}})^{2} }{4\mathcal{E}_\text{\tiny k,N-1}^{\text{\tiny slow}} }\quad \text{for N $\ge 2$} \;,
    \end{equation}
    %%%%%%%%%%%%%%%
    
    The above recurrence relation has an asymptotic solution $\mathcal{E}_\text{\tiny k,N}^{\text{\tiny slow}} \simeq \sqrt{({N+1})/{2}}~\Delta \mathcal{E}^{\text{\tiny slow}}_{\text{\tiny 1}} $. Thus, the average gain in energy by an MBH while it passes a single cell is given by
    %%%%%%%%%%%%%%%
    \begin{equation}\label{eq:deltaKENew_slow_N}
     \Delta \mathcal{E}_\text{\tiny k}^{\text{\tiny slow}} \equiv \frac{\mathcal{E}_\text{\tiny k,N}^{\text{\tiny slow}}}{N}\sim \sqrt{\frac{N+1}{2N}} Q_{\text{\tiny BH}} B_{\text{\tiny c}} l_{\text{\tiny c}} \;,
    \end{equation}
    %%%%%%%%%%%%%%%   
    Using  Eqs.\,\eqref{eq:fluxNew_gen} and \eqref{eq:deltaKENew_slow_N}, we arrive at
    %%%%%%%%%%%%%%%
    \begin{equation}\label{eq:unclusteredSlowFMBH}
    F_{\text{\tiny BH}}^{\text{\tiny slow}} \lesssim \sqrt{\left(\frac{2N}{N+1}\right)}\frac{B_{\text{\tiny c}}}{6 \mu_0 Q_{\text{\tiny BH}} t_{\text{\tiny reg}}}\;.
    \end{equation}
    %%%%%%%%%%%%%%%
   
    Notice that, unlike fast clustered MBHs, the bound on the flux of slow MBHs is no longer independent of the magnetic field strength $B_{\text{\tiny c}}$ but varies linearly with it. Also, $N\sim d/l_{\text{\tiny c}}$, where $d$ is the total size of N coherence length regions. For $N=1$, when MBHs cover a single coherence length, this limit is, therefore, independent of $l_{\text{\tiny c}}$.
    For the slow MBH case, the average velocity after N cells can be determined using Eq.\,\eqref{eq:deltaKENew_slow_N} and is given by $v \sim \left({(N+1)}/(2N)\right)^{1/4} v_{\text{\tiny mag}}$. Therefore, 
    Eqs.\,\eqref{eq:FMBH_fDM_formal} and \eqref{eq:deltaKENew_slow_N} give us a constraint on the dark matter fraction
    %%%%%%%%%%%%%%%
    \begin{equation}\label{eq:unclusteredSlowfdm}
    f_{\text{\tiny DM}}^{\text{\tiny slow}} \lesssim \left(\frac{2N}{N+1}\right)^{3/4} \frac{B_{\text{\tiny c}} M_{\text{\tiny BH}}}{6  \mu_0 Q_{\text{\tiny BH}} t_{\text{\tiny reg}}v_{\text{\tiny mag}}\rho_{\text{\tiny DM}}} =\left(\frac{2N}{N+1}\right)^{3/4}\frac{B_{\text{\tiny c}}^{1/2} M_{\text{\tiny BH}}^{3/2}}{6\sqrt{2} \mu_0 Q_{\text{\tiny BH}} t_{\text{\tiny reg}}l_{\text{\tiny c}}^{1/2}\rho_{\text{\tiny DM}}} \;.
    \end{equation}
    %%%%%%%%%%%%%%%
    
    One must also consider the case when the MBHs are unable to cross a single coherence length in Hubble time $H_{\text{\tiny 0}}^{-1}$, where $H_{\text{\tiny 0}}$ is the Hubble constant. This happens when $v_{\text{\tiny mag}}\lesssim l_{\text{\tiny c}}H_{\text{\tiny 0}}$. Such a situation arises in the physical scenarios where $B_{\text{\tiny c}}\lesssim l_{\text{\tiny c}}M_{\text{\tiny BH}}H_{\text{\tiny 0}}^2/(2Q_{\text{\tiny BH}})$, i.e. small magnetic field strength $B_{\text{\tiny c}}$ or large coherence length $l_{\text{\tiny c}}$. We will see later that this situation is realized for cosmic voids and cosmic web filaments, where the number of cells traversed is always $N \sim \mathcal{O}(1)$. In such a case, the velocity acquired by the MBHs (from Eq.\,\eqref{eq:vdot}) in a Hubble time is given by $v_{\text{\tiny H0}}\sim Q_{\text{\tiny BH}}B_{\text{\tiny c}}/(M_{\text{\tiny BH}}H_{\text{\tiny 0}})$. Consequently, the energy gained is
     %%%%%%%%%%%%%%%
    \begin{equation}\label{eq:pb_slow_Echange_H0}
    \Delta \mathcal{E}_{\text{\tiny k,H0}}\equiv\frac{1}{2}M_{\text{\tiny BH}}v_{\text{\tiny H0}}^2  =  \frac{Q_{\text{\tiny BH}}^2 B_{\text{\tiny c}}^2}{2M_{\text{\tiny BH}} H_{\text{\tiny 0}}^2}\;.
    \end{equation}
    %%%%%%%%%%%%%%%
     Using  Eqs.\,\eqref{eq:fluxNew_gen} and \eqref{eq:pb_slow_Echange_H0}, we get the bounds on the flux of MBHs as,
    %%%%%%%%%%%%%%%
    \begin{equation}\label{eq:unclusteredSlowFMBH_H0}
    F_{\text{\tiny BH}}^{\text{\tiny slow,H0}} \lesssim \frac{l_{\text{\tiny c}}M_{\text{\tiny BH}}H_{\text{\tiny 0}}^2}{3 \mu_0 Q^2_{\text{\tiny BH}}t_{\text{\tiny reg}}}\;.
    \end{equation}
    %%%%%%%%%%%%%%%
    Finally, using Eqs.\,\eqref{eq:FMBH_fDM_formal} and \eqref{eq:pb_slow_Echange_H0} and taking $v\sim v_{\text{\tiny H0}}$ , we get the bound on the dark matter fraction as
    %%%%%%%%%%%%%%%
    \begin{equation}\label{eq:unclusteredSlowfdm_H0}
    f_{\text{\tiny DM}}^{\text{\tiny slow,H0}} \lesssim \frac{l_{\text{\tiny c}}M^3_{\text{\tiny BH}}H_{\text{\tiny 0}}^3}{3 \mu_0 B_{\text{\tiny c}}Q^3_{\text{\tiny BH}}\rho_{\text{\tiny DM}}t_{\text{\tiny reg}}} \;.
    \end{equation}
    %%%%%%%%%%%%%%%
    
   Notice that unlike previous cases, the bounds on $F_{\text{\tiny BH}}^{\text{\tiny slow,H0}}$ and $f_{\text{DM}}^{\text{\tiny slow,H0}}$  are directly proportional to the coherence length $l_{\text{\tiny c}}$. This happens because the gain in energy is independent of the coherence length. Moreover, the bound on $f_{\text{\tiny DM}}^{\text{\tiny slow,H0}}$ is inversely proportional to the magnetic field strength $B_{\text{\tiny c}}$.

    %%%%%%%%%%%%%%%
    	\begin{figure}[t]
 		\centering
 		\includegraphics[scale=1]{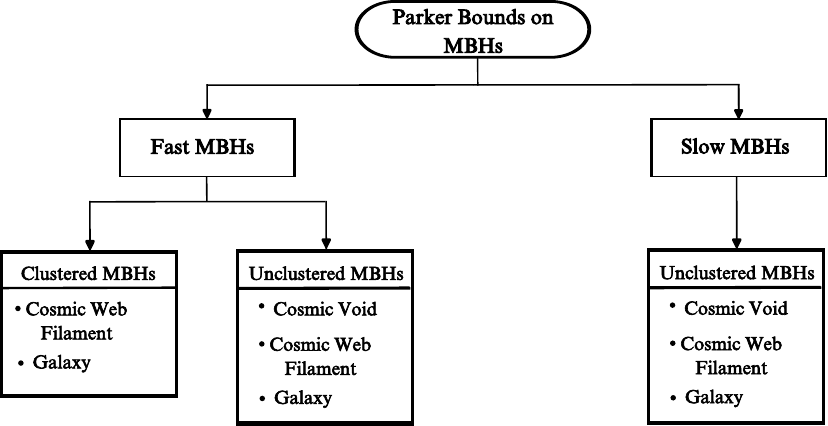}
 		\caption{Summary of the classification of MBH Parker bounds in this work. Please see the text for details. There is no clustered category for slow MBHs as in the systems under consideration the virial velocities typically satisfy $v_{\text{\tiny vir}} \gg v_{\text{\tiny mag}}$; always rendering them to be fast. }
 		\label{fig:Parker_bound_flowchart}
 	\end{figure}
    %%%%%%%%%%%%%%%
    
    Fig.\,\ref{fig:Parker_bound_flowchart} summarises the categories for Parker bounds on MBHs due to galactic, extra-galactic and cosmic magnetic fields. As we have calculated, depending on the initial velocity of the MBHs, they are divided into two categories---fast MBHs, which have $v_{\text{\tiny in}} \gg v_{\text{\tiny mag}}$, and slow MBHs, with $v_{\text{\tiny in}} \ll v_{\text{\tiny mag}}$. Additionally, depending upon whether the MBHs are clustered (virialized) or unclustered (infalling), bounds on MBHs are determined by Eqs.\,\eqref{eq:FBH_fast_clustered} and \eqref{eq:fDM_fast_clustered}, or Eqs.\,\eqref{eq:FBH_fast_unclustered} and \eqref{eq:fdm_fast_unclustered}. However, within cosmic voids, there exists a flow of MBHs directed toward gravitational sources, such as cosmic web nodes or filaments. In this case, if the MBHs are slow ($ v_{\text{\tiny in}} \ll v_{\text{\tiny mag}}$) and clustered, they will be accelerated in the direction of the magnetic field, and undergo a large change in velocity of the order of $ v_{\text{\tiny mag}}$. Due to this, they will have directional velocities, and hence, they will eventually become unclustered. Additionally, as we will see in the next section, the galactic, cosmic filament, and cosmic void systems under consideration have virial velocities $ v_{\text{\tiny vir}} \gg v_{\text{\tiny mag}}$. Therefore, slow clustered MBHs are unlikely to exist in these systems.   Furthermore, since IGMFs typically have large coherence lengths and small magnetic fields, the slow MBHs in cosmic voids and filaments may not cross even a single coherence length in Hubble time. In that case, the constraints will be determined by Eqs.\,\eqref{eq:unclusteredSlowFMBH_H0} and \eqref{eq:unclusteredSlowfdm_H0}. However, for the galactic case, the bounds on slow MBHs in the galaxy will be determined by  Eqs.\,\eqref{eq:unclusteredSlowFMBH} and \eqref{eq:unclusteredSlowfdm}.
    
%%%%%%%%%%%%%%%
\subsection{Bounds on primordial magnetic black holes}\label{subsec:numerical_bounds}
    We will now apply our analyses in Sec.\,\ref{sec:theory}  to the different magnetic field configurations discussed in Sec.\,\ref{subsec:mag_field}. Specifically, the constraints will be evaluated using Eqs.\,\,\eqref{eq:FBH_fast_clustered}, \eqref{eq:fDM_fast_clustered}, \eqref{eq:FBH_fast_unclustered}, \eqref{eq:fdm_fast_unclustered}, \eqref{eq:unclusteredSlowFMBH}, \,\eqref{eq:unclusteredSlowfdm},\,\eqref{eq:unclusteredSlowFMBH_H0} and \eqref{eq:unclusteredSlowfdm_H0}.
    As previously stated, we will explore constraints specific to different types of magnetic field configurations---IGMFs in cosmic filaments and cosmic voids, and galactic magnetic fields. For the cosmic filaments, we take the density of the dark matter as  $\rho_{\text{\tiny DM}}^{\text{\tiny fil}} \sim 10\,\rho_{\text{\tiny c}} \simeq 5.6 \times 10^{-5}\,\text{GeV/cm}^3$\,\cite{Cautun:2014fwa,Karachentsev:2018ysz}. However, for the cosmic voids, we take the dark matter density to be $\rho_{\text{\tiny DM}}^{\text{\tiny void}} \simeq 0.1\,\rho_{\text{\tiny c}} \sim 5.6 \times 10^{-7}\,\text{GeV/cm}^3$\,\cite{Cautun:2014fwa,Karachentsev:2018ysz}, where $\rho_{\text{\tiny c}}$ is the critical density of the universe\,\cite{Baumann}. As for the galactic magnetic fields, we assume that the DM density in the galaxies are of the same order of magnitude as the local DM density $\rho_{\text{\tiny DM}}^{\text{\tiny loc}}\simeq 0.3\,\text{GeV/cm}^3$\cite{Baushev:2012dm}. 
    
    First, we estimate the typical value of the magnetic velocity $v_{\text{\tiny mag}}$ using Eq.\,\eqref{eq:vmag_def}, which will determine whether the MBHs are fast or slow in each system.  In terms of magnetic field parameters, we may express this magnetic velocity as
    %%%%%%%%%%%%%%%
    \begin{equation}\label{eq:vmag_val}
        v_{\text{\tiny mag}} \simeq 2.6 \times 10^{-7} c \left ( \frac{Q_{\text{\tiny BH}}}{\text{A-m}} \right ) ^{1/2} \left ( \frac{\text{Kg}}{M_{\text{\tiny BH}}} \right ) ^{1/2}\left ( \frac{l_{\text{\tiny c}}}{1\,\text{Mpc}} \right ) ^{1/2} \left ( \frac{B}{ 10^{-15}\,\text{G}} \right ) ^{1/2}\;.
    \end{equation}
    %%%%%%%%%%%%%%%
    
    For an extremal MBH, using values presented in Sec.\,\ref{subsec:mag_field}, we observe that galactic magnetic fields possessing $B_{\text{\tiny c}}\sim \mathcal{O}(10)\,\mu\text{G} $ and $l_{\text{\tiny c}}\sim \mathcal{O}(1)\,\text{kpc}$ yield typically $v_{\text{\tiny mag}} \sim 10^{-4}\,\text{c}$ (where c is the speed of light). Similarly, for IGMFs in cosmic filaments with $B_{\text{\tiny c}}\sim  \mathcal{O}(10^{-9})\,\text{G} $ and $l_{\text{\tiny c}}\sim  \mathcal{O}(1)\,\text{Mpc}$, we get $v_{\text{\tiny mag}} \sim 10^{-5}c$. Now, since the virial velocities in galaxies and cosmic filaments are of order $10^{-3} c$\,\cite{sofue1987,Zhu:2017wug}, the clustered MBHs are fast.  Additionally, for cosmic void IGMFs generated primordially or from magnetizations due to void galaxies, we have    $B_{\text{\tiny c}}\simeq  \mathcal{O}(10^{-15})\,\text{G} $ and $l_{\text{\tiny c}}\gtrsim  \mathcal{O}(1-10)\,\text{Mpc}$, leading to $v_{\text{\tiny mag}} \sim 10^{-7}c-10^{-8}c$.  However, for a galactic outflow origin of IGMFs in cosmic voids with $\mathcal{O}(10^{-12})\,\text{G}\lesssim B_{\text{\tiny c}} \lesssim  \mathcal{O}(10^{-8})\,\text{G}$ and $l_{\text{\tiny c}}\sim  \mathcal{O}(1)\,\text{Mpc} $, we have $v_{\text{\tiny mag}} \sim 10^{-4}c-10^{-6}c$. For fast unclustered MBHs, we use the typical velocities of MBHs found in galaxies, filaments, and voids as $v_{\text{\tiny in}}\sim 10^{-3}c$\,\cite{Kashibadze:2020uux,rost_2021,Padilla:2005ea}.
    
    Notice that the magnetic velocity $v_{\text{\tiny mag}}$ for IGMFs is typically very small. In addition, due to the typically large coherence lengths ($\mathcal{O}(1)\,\text{Mpc}$), the time required by slow MBHs ($v\ll v_\text{\tiny mag}$) to cross a single coherence length is much greater than $10\, \text{Gyr}  $ (i.e. typically much greater than the Hubble time). Therefore, for the case of slow MBHs, the relevant expressions are Eqs.\,\eqref{eq:unclusteredSlowFMBH_H0} and \eqref{eq:unclusteredSlowfdm_H0}. 
    
    Putting all these together, let us discern the typical constraints on fast and slow MBHs. Under each of these cases, we will look at the various categories and limits of interest. In the fast unclustered (infalling) case, one may quantify the limits using Eqs.\,\eqref{eq:FBH_fast_unclustered}, and \eqref{eq:fdm_fast_unclustered}. One obtains, in this case 
    %%%%%%%%%%%%%%%
    \begin{eqnarray}\label{eq:fast_bound_unclustered_num_1}
       F_{\text{\tiny BH}}^{\text{\tiny fast,unclust}} &\lesssim& \frac{2.5 \times 10^{-23}}{1 + 0.6\times 10^8\cos\alpha \left(\frac{v_{\text{\tiny in}}}{ 10^{-3} c}\right)^{2} \left( \frac{ 10^{-15} \text{G}}{B_{\text{\tiny c}}}\right)\left(\frac{\text{Mpc}}{l_{\text{\tiny c}}}\right)\left(\frac{M_{\text{\tiny BH}}}{\text{kg}}\right)\left(\frac{\text{A-m}}{Q_{\text{\tiny BH}}}\right)}\left(\frac{M_{\text{\tiny BH}}}{\text{kg}}\right)\left(\frac{\text{A-m}}{Q_{\text{\tiny BH}}}\right)^2\;\nonumber  \\
         \quad &&\times\left(\frac{v_{\text{\tiny in}}}{ 10^{-3} c}\right)^{2}\left(\frac{\text{Mpc}}{l_{\text{\tiny c}}}\right)\left(\frac{\text{Gyr}}{t_{r e g}}\right)m^{-2} \text{s}^{-1}\;,\nonumber  \\ 
       f_{\text{DM}}^{\text{\tiny fast,unclust}} &\lesssim& \frac{0.8 \times 10^{-3}}{1 + 0.6\times 10^8\cos\alpha \left(\frac{v_{\text{\tiny in}}}{ 10^{-3} c}\right)^{2}\left( \frac{ 10^{-15} \text{G}}{B_{\text{\tiny c}}}\right)\left(\frac{\text{Mpc}}{l_{\text{\tiny c}}}\right)\left(\frac{M_{\text{\tiny BH}}}{\text{kg}}\right)\left(\frac{\text{A-m}}{Q_{\text{\tiny BH}}}\right)}\left(\frac{\rho_{\text{\tiny c}}}{\rho_{\text{\tiny DM}}}\right)\left(\frac{M_{\text{\tiny BH}}}{\text{kg}}\right)^2\;\nonumber  \\
         \quad &&\times\left(\frac{\text{A-m}}{Q_{\text{\tiny BH}}}\right)^2\left(\frac{v_{\text{\tiny in}}}{ 10^{-3} c}\right)\left(\frac{\text{Mpc}}{l_{\text{\tiny c}}}\right)\left(\frac{\text{Gyr}}{t_{r e g}}\right)\;.\nonumber  \\ 
    \end{eqnarray}
    %%%%%%%%%%%%%%%
    Specializing to the extremal case, this then reduces to
     %%%%%%%%%%%%%%%
    \begin{eqnarray}\label{eq:fast_bound_unclustered_num}
         F_{\text{\tiny BH}}^{\text{\tiny fast,unclust}} &\lesssim& \frac{3.7 \times 10^{-20}}{1 + 2.3\times 10^9\cos\alpha\left(\frac{v_{\text{\tiny in}}}{ 10^{-3} c}\right)^{2}\left( \frac{ 10^{-15} \text{G}}{B_{\text{\tiny c}}}\right)\left(\frac{\text{Mpc}}{l_{\text{\tiny c}}}\right)}\left(\frac{\text{kg}}{M_{\text{\tiny BH}}}\right)\;\nonumber  \\
         \quad &&\times\left(\frac{v_{\text{\tiny in}}}{ 10^{-3} c}\right)^{2}\left(\frac{\text{Mpc}}{l_{\text{\tiny c}}}\right)\left(\frac{\text{Gyr}}{t_{r e g}}\right)m^{-2} \text{s}^{-1}\;,\nonumber   \\ 
        f_{\text{DM}}^{\text{\tiny fast,unclust}} &\lesssim& \frac{12.2}{1 + 2.3\times 10^9\cos\alpha\left(\frac{v_{\text{\tiny in}}}{ 10^{-3} c}\right)^{2}\left( \frac{ 10^{-15} \text{G}}{B_{\text{\tiny c}}}\right)\left(\frac{\text{Mpc}}{l_{\text{\tiny c}}}\right)}\left(\frac{\rho_{\text{\tiny c}}}{\rho_{\text{\tiny DM}}}\right)\;\nonumber  \\
         \quad &&\times\left(\frac{v_{\text{\tiny in}}}{ 10^{-3} c}\right)\left(\frac{\text{Mpc}}{l_{\text{\tiny c}}}\right)\left(\frac{\text{Gyr}}{t_{r e g}}\right)\;.  
    \end{eqnarray}
    %%%%%%%%%%%%%%%
    However, if the MBHs are clustered or virialized, using Eqs.\,\eqref{eq:FBH_fast_clustered} and \eqref{eq:fDM_fast_clustered}, one obtains the constraints
    %%%%%%%%%%%%%%%
    \begin{eqnarray}\label{eq:fast_bound_clustered_num_1}
        F_{\text{\tiny BH}}^{\text{\tiny fast,clust}} &\lesssim& 2.5 \times 10^{-23}\left(\frac{M_{\text{\tiny BH}}}{\text{kg}}\right)\left(\frac{\text{A-m}}{Q_{\text{\tiny BH}}}\right)^2\left(\frac{v_{\text{\tiny in}}}{ 10^{-3} c}\right)^{2}\left(\frac{\text{Mpc}}{l_{\text{\tiny c}}}\right)\left(\frac{\text{Gyr}}{t_{r e g}}\right)m^{-2} \text{s}^{-1}\;,\nonumber  \\ 
        f_{\text{DM}}^{\text{\tiny fast,clust}}  &\lesssim& 0.8 \times 10^{-2}\left(\frac{\rho_{\text{\tiny c}}}{\rho_{\text{\tiny DM}}}\right)\left(\frac{M_{\text{\tiny BH}}}{\text{kg}}\right)^2\left(\frac{\text{A-m}}{Q_{\text{\tiny BH}}}\right)\left(\frac{v_{\text{\tiny in}}}{ 10^{-3} c}\right)\left(\frac{\text{Mpc}}{l_{\text{\tiny c}}}\right)\left(\frac{\text{Gyr}}{t_{r e g}}\right)\;.\nonumber  \\ 
    \end{eqnarray}
    %%%%%%%%%%%%%%%
    For the extremal MBHs, these simplify to
    %%%%%%%%%%%%%%%
    \begin{eqnarray}\label{eq:fast_bound_clustered_num}
         F_{\text{\tiny BH}}^{\text{\tiny fast,clust}} &\lesssim& 3.7 \times 10^{-20}\left(\frac{\text{kg}}{M_{\text{\tiny BH}}}\right)\left(\frac{v_{\text{\tiny in}}}{ 10^{-3} c}\right)^{2}\left(\frac{\text{Mpc}}{l_{\text{\tiny c}}}\right)\left(\frac{\text{Gyr}}{t_{r e g}}\right)m^{-2} \text{s}^{-1}\;,\nonumber  \\ 
        f_{\text{DM}}^{\text{\tiny fast,clust}}  &\lesssim& 12.2\left(\frac{\rho_{\text{\tiny c}}}{\rho_{\text{\tiny DM}}}\right)\left(\frac{v_{\text{\tiny in}}}{ 10^{-3} c}\right)\left(\frac{\text{Mpc}}{l_{\text{\tiny c}}}\right)\left(\frac{\text{Gyr}}{t_{r e g}}\right)\;.\nonumber  \\ 
    \end{eqnarray}
    %%%%%%%%%%%%%%%
    
    Let us now turn our attention to slow MBHs. When MBHs are moving slowly through galaxies, the bounds can be evaluated numerically using Eqs.\,\eqref{eq:unclusteredSlowFMBH}, and\,\eqref{eq:unclusteredSlowfdm}. We obtain, in this case
     %%%%%%%%%%%%%%%
    \begin{eqnarray}\label{eq:slow_bound_num_1}
         F_{\text{\tiny BH}}^{\text{\tiny slow}} &\lesssim&4.2 \times 10^{-21}\left(\frac{2N}{N+1}\right)^{1/2} \left(\frac{\text{A-m}}{Q_{\text{\tiny BH}}}\right)\left( \frac{B_{\text{\tiny c}}}{ 10\,\mu\text{G}}\right)\left(\frac{\text{Gyr}}{t_{\text{\tiny reg}}}\right)\left(\frac{\text{kpc}}{l_{\text{\tiny c}}}\right)^{1/2}m^{-2} \text{s}^{-1}\;, \\ 
        f_{\text{DM}}^{\text{\tiny slow}}  &\lesssim& 2.6 \times 10^{-4} \left(\frac{2N}{N+1}\right)^{3/4}\left(\frac{M_{\text{\tiny BH}}}{\text{kg}}\right)^{\frac{3}{2}}\left(\frac{\text{A-m}}{Q_{\text{\tiny BH}}}\right)^{\frac{3}{2}}\left(\frac{\rho_{\text{\tiny c}}}{\rho_{\text{\tiny DM}}}\right) \left( \frac{B_{\text{\tiny c}}}{ 10\,\mu\text{G}}\right)^{\frac{1}{2}}\left(\frac{\text{Gyr}}{t_{\text{\tiny reg}}}\right)\left(\frac{\text{kpc}}{l_{\text{\tiny c}}}\right)^{1/2}\;.\nonumber
    \end{eqnarray}
    %%%%%%%%%%%%%%%
     Here, $N\sim d_{\text{\tiny gal}}/l_\text{\tiny c, mean}^{\text{\tiny gal}}$ is an estimate of the number of coherence lengths that the slow MBHs may cross. For instance, with $d_{\text{\tiny gal}}\sim \mathcal{O}(10)\,\text{kpc}$ and $l_\text{\tiny c, mean}^{\text{\tiny gal}} \sim \mathcal{O}(1)\,\text{kpc}$ one has $N\sim \mathcal{O}(10) $. For extremal MBHs the above reduces to
    %%%%%%%%%%%%%%%
    \begin{eqnarray}\label{eq:slow_bound_num}
        F_{\text{\tiny BH}}^{\text{\tiny slow}} &\lesssim&1.6\times 10^{-19} \left(\frac{2N}{N+1}\right)^{1/2}\left(\frac{\text{kg}}{M_{\text{\tiny BH}}}\right)\left( \frac{B_{\text{\tiny c}}}{ 10\,\mu\text{G}}\right)\left(\frac{\text{Gyr}}{t_{\text{\tiny reg}}}\right)\left(\frac{\text{kpc}}{l_{\text{\tiny c}}}\right)^{1/2}m^{-2} \text{s}^{-1}\;, \\
        f_{\text{DM}}^{\text{\tiny slow}}  &\lesssim& 0.06\left(\frac{2N}{N+1}\right)^{3/4} \left(\frac{\rho_{\text{\tiny c}}}{\rho_{\text{\tiny DM}}}\right) \left( \frac{B_{\text{\tiny c}}}{ 10\,\mu\text{G}}\right)^{\frac{1}{2}}\left(\frac{\text{Gyr}}{t_{\text{\tiny reg}}}\right)\left(\frac{\text{kpc}}{l_{\text{\tiny c}}}\right)^{1/2}\;.\nonumber
    \end{eqnarray}
    %%%%%%%%%%%%%%%
    
  However, as discussed earlier, slow MBHs moving in IGMFs will not generally cross even a single coherence length in Hubble time. Hence, the bounds from IGMFs in cosmic web filaments and cosmic voids in the context of slow-moving MBHs will be determined by Eqs.\,\eqref{eq:unclusteredSlowFMBH_H0}, and\,\eqref{eq:unclusteredSlowfdm_H0}. From these, we have 
     %%%%%%%%%%%%%%%
    \begin{eqnarray}\label{eq:slow_bound_num_H0_1}
         F_{\text{\tiny BH}}^{\text{\tiny slow,H0}} &\lesssim&1.45 \times 10^{-25} \left(\frac{\text{A-m}}{Q_{\text{\tiny BH}}}\right)^2 \left(\frac{M_{\text{\tiny BH}}}{\text{kg}}\right)\left(\frac{l_{\text{\tiny c}}}{\text{Mpc}}\right)\left(\frac{\text{Gyr}}{t_{\text{\tiny reg}}}\right)m^{-2} \text{s}^{-1}\;, \\ 
        f_{\text{DM}}^{\text{\tiny slow,H0}}  &\lesssim& 343.33 \left(\frac{\text{A-m}}{Q_{\text{\tiny BH}}}\right)^3 \left(\frac{M_{\text{\tiny BH}}}{\text{kg}}\right)^3\left(\frac{l_{\text{\tiny c}}}{\text{Mpc}}\right)\left( \frac{ 10^{-15} \text{G}}{B_{\text{\tiny c}}}\right)\left(\frac{\text{Gyr}}{t_{\text{\tiny reg}}}\right)\;.\nonumber
    \end{eqnarray}
    %%%%%%%%%%%%%%%
    Again, the corresponding extremal MBH limits are
    %%%%%%%%%%%%%%%
    \begin{eqnarray}\label{eq:slow_bound_num_H0}
        F_{\text{\tiny BH}}^{\text{\tiny slow,H0}} &\lesssim&2.2 \times 10^{-22} \left(\frac{\text{kg}}{M_{\text{\tiny BH}}}\right)\left(\frac{l_{\text{\tiny c}}}{\text{Mpc}}\right)\left(\frac{\text{Gyr}}{t_{\text{\tiny reg}}}\right)m^{-2} \text{s}^{-1}\;, \\ 
        f_{\text{DM}}^{\text{\tiny slow,H0}}  &\lesssim& 2.0\times10^7 \left(\frac{l_{\text{\tiny c}}}{\text{Mpc}}\right)\left( \frac{ 10^{-15} \text{G}}{B_{\text{\tiny c}}}\right)\left(\frac{\text{Gyr}}{t_{\text{\tiny reg}}}\right)\;.\nonumber
    \end{eqnarray}
    %%%%%%%%%%%%%%%
   
%%%%%%%%%%%%%%%%%%%%%%%%%%%%%%%%%%%%%%%%%%%%%%%%%%%%%%%%%%%%%%%%%%%%%%%%
\subsubsection{Parker-type bounds from cosmic voids } 
    Parker-type bounds on MBHs due to cosmic void IGMFs are anticipated to result in stringent constraints. This expectation is motivated by the small field strengths and large coherence lengths in such cosmic structures. MBHs in cosmic voids will largely flow towards gravitational sources in the cosmic web nodes and filaments and, therefore, remain mostly unclustered.  Furthermore, the magnetic velocity in such systems is of the order of $v_{\text{\tiny mag}}\sim 10^{-7}c-10^{-8}c$ for fields generated primordially or due to void galaxies, and of the order of $v_{\text{\tiny mag}}\sim 10^{-4}c-10^{-6}c$ for fields generated from galactic outflow. Depending upon the velocity of MBHs, they may be fast ($v_{\text{\tiny in}}>v_{\text{\tiny mag}}$ ) or slow ($v_{\text{\tiny in}}<v_{\text{\tiny mag}} $ ), but unclustered. This will lead to different bounds. 
    Furthermore, depending on the production mechanism of the IGMFs in cosmic voids (see Sec.\,\ref{subsec:mag_field}), we have different field characteristics, which yield distinct bounds.
    
    In cosmic voids, IGMFs have field characteristics $B_{\text{\tiny c}} = \mathcal{O}(10^{-15})$ G, $l_\text{\tiny c} = \mathcal{O}(1-10)$ Mpc and $t_\text{\tiny reg} = \mathcal{O}(10)$ Gyr. Taking  $v_{\text{\tiny in}}\sim 10^{-3}c$\,\cite{Padilla:2005ea} in Eq.\,\eqref{eq:fast_bound_unclustered_num}, we can place very strong limits on unclustered extremal MBHs in cosmic voids
    %%%%%%%%%%%%%%%
    \begin{equation}\label{eq:fDM_primordial_IGMF}
       f_{\text{\tiny DM}}^{\text{\tiny fast,unclust}} \lesssim 10^{-8}\;.
    \end{equation}
    %%%%%%%%%%%%%%%
    This bound is based on the requirement that there must exist no cosmic void with a depleted magnetic field. If MBHs exist in the universe, the field depletion is most drastic when their flux is along the cosmic void field lines ($\alpha=0$). This is the scenario we assume to place the bound in Eq.\,\eqref{eq:fDM_primordial_IGMF}.

    Cosmic void IGMFs, if they have galactic flux leakage origins and characteristics (as quantified in Eq.\,\eqref{eq:IGMF_gal}) lead to less stringent bounds $f_{\text{\tiny DM}}^{\text{\tiny fast,unclust}} \lesssim 10^{-1}-10^{-5}$. 
    Note, in particular, that these bounds are independent of the mass of the extremal MBHs and are much stronger than previous bounds in the literature\,\cite{Ghosh_2020,Bai}. No significant limits on slow MBHs can be placed from cosmic voids, as seen trivially from Eq.\,\eqref{eq:slow_bound_num_H0}. 
    Furthermore, Eq.\,\eqref{eq:fast_bound_unclustered_num_1} limits the fraction of dark matter in the form of non-extremal MBHs, using IGMFs produced either primordially or via magnetizations due to void galaxies,
    %%%%%%%%%%%%%%%
    \begin{equation}\label{eq:fDM_primordial_IGMF_non}
       f_{\text{\tiny DM}}^{\text{\tiny fast,unclust}} \lesssim 10^{-9}\left(\frac{M_{\text{\tiny BH}}}{\text{kg}}\right)\left(\frac{\text{A-m}}{Q_{\text{\tiny BH}}}\right) \;.
    \end{equation}
    %%%%%%%%%%%%%%%
    Finally, if cosmic void IGMFs originate due to galactic flux leakage, we get weaker bounds on non-extremal MBHs as $f_{\text{\tiny DM}}^{\text{\tiny fast,unclust}} \lesssim 10^{-2}-10^{-6}\left({M_{\text{\tiny BH}}}/{\text{kg}}\right)\cdot\left({\text{A-m}}/{Q_{\text{\tiny BH}}}\right)$.  
    
    Now, for fast unclustered extremal MBHs in the context of primordially sourced or magnetization-induced cosmic void IGMFs, we obtain from Eq.\,\eqref{eq:fast_bound_unclustered_num} the limits
    %%%%%%%%%%%%%%%
    \begin{equation}\label{eq:FMBH_primordial_IGMF}
        F_{\text{\tiny BH}}^{\text{\tiny fast,unclust}} \lesssim 10^{-29}\left( \frac{\text{kg}}{M_{\text{\tiny BH}}}\right)\text{m}^{-2} \text{s}^{-1}\;.
    \end{equation}
    %%%%%%%%%%%%%%%
    Also,  from Eq.\,\eqref{eq:fast_bound_unclustered_num_1}, constraints on the flux of non-extremal MBHs are $F_{\text{\tiny BH}}^{\text{\tiny fast,unclust}}\lesssim 10^{-31}$ $\left( {\text{A-m}}/{Q_{\text{\tiny BH}}}\right)$ $\text{m}^{-2} \text{s}^{-1}$ for these IGMFs.
    
    Cosmic void IGMFs produced due to galactic flux leakage impose less stringent constraints on the flux of fast unclustered extremal MBHs,  $F_{\text{\tiny BH}}^{\text{\tiny fast,unclust}}\lesssim 10^{-22}-10^{-26}\left( {\text{kg}}/{M_{\text{\tiny BH}}}\right)$ $\text{m}^{-2} \text{s}^{-1}$, and the equivalent limits for non-extremal MBHs are  $F_{\text{\tiny BH}}^{\text{\tiny fast,unclust}}\lesssim 10^{-24}-10^{-28}$ $\left( {\text{A-m}}/{Q_{\text{\tiny BH}}}\right)$ $\text{m}^{-2} \text{s}^{-1}$. 

     However, for slow MBHs, it is clear from Eqs.\,\eqref{eq:slow_bound_num_H0_1} and \,\eqref{eq:slow_bound_num_H0}, that the bounds on MBH fluxes are independent of magnetic field strength.  Therefore, IGMFs in cosmic voids, independent of their origin, limit $ F_{\text{\tiny BH}}^{\text{\tiny slow,H0}} \lesssim  10^{-21} -10^{-22}( \,{\text{kg}}/{M_{\text{\tiny BH}}})\,\text{m}^{-2} \text{s}^{-1}$ for extremal MBHs, and  $ F_{\text{\tiny BH}}^{\text{\tiny slow,H0}} \lesssim  10^{-24} -10^{-25}({M_{\text{\tiny BH}}/{\text{kg}}})\cdot$ $( \,{\text{A-m}}/{Q_{\text{\tiny BH}}})^2\,\text{m}^{-2} \text{s}^{-1}$  for non-extremal MBHs.

%%%%%%%%%%%%%%%%%%%%%%%%%%%%%%%%%%%%%%%%%%%%%%%%%%%%%%%%%%%%%%%%%%%%
    \subsubsection{Parker-type bounds from cosmic filaments}
    Since the magnetic field strengths in cosmic web filaments are larger than in cosmic voids, we expect relatively weaker bounds on the flux and dark matter fraction of MBHs in this context. Again, the MBHs may be unclustered fast ($v>v_{\text{\tiny mag}}\simeq 10^{-5} $ c) or slow ($v<v_{\text{\tiny mag}}\simeq 10^{-5} $ c), leading to different bounds. Following Eq.\,\eqref{eq:fil}, we assume typical characteristics of magnetic fields in cosmic web filaments as $B_{\text{\tiny c}} = \mathcal{O}(10^{-9})$ G, $l_\text{\tiny c} = \mathcal{O}(1)$ Mpc and $t_\text{\tiny reg} = \mathcal{O}(10)$ Gyr. Taking  $v_{\text{\tiny in}}\sim 10^{-3}c$\,\cite{rost_2021}, we get bounds for the unclustered (infalling) extremal MBHs from Eq.\,\eqref{eq:fast_bound_unclustered_num} as
    %%%%%%%%%%%%%%%
    \begin{equation}\label{eq:fDM_cos_fil}
       f_{\text{\tiny DM}}^{\text{\tiny fast,unclust}} \lesssim 10^{-4}\,,
    \end{equation}
    %%%%%%%%%%%%%%%
    Again, this limit is based on the stipulation that no cosmic web filament must exist with a depleted magnetic field. Additionally, for unclustered non-extremal MBHs, Eq.\,\eqref{eq:fast_bound_unclustered_num_1} limits
    %%%%%%%%%%%%%%%
    \begin{equation}\label{eq:fDM_cos_fil_non}
       f_{\text{\tiny DM}}^{\text{\tiny fast,unclust}} \lesssim 10^{-5}\left(\frac{M_{\text{\tiny BH}}}{\text{kg}}\right)\left(\frac{\text{A-m}}{Q_{\text{\tiny BH}}}\right) \;.
    \end{equation}
    %%%%%%%%%%%%%%%
     For virialized/clustered extremal MBHs, we get a weaker bound $f_{\text{\tiny DM}}^{\text{\tiny fast,clust}} \lesssim\mathcal{O}(10^{-1})$ from Eq.\,\eqref{eq:fast_bound_clustered_num}. Using Eq.\,\eqref{eq:fast_bound_clustered_num_1}, we get bounds on clustered non-extremal MBHs as $ f_{\text{\tiny DM}}^{\text{\tiny fast,unclust}} \lesssim 10^{-4}\left({M_{\text{\tiny BH}}}/{\text{kg}}\right)^2$ $\left({\text{A-m}}/{Q_{\text{\tiny BH}}}\right)^2$. Again, for slow extremal MBHs we do not obtain any meaningful bounds on $ f_{\text{\tiny DM}}$ from cosmic web filament IGMFs. 
     
    For fast unclustered extremal MBHs, using Eqs.\,\eqref{eq:fast_bound_unclustered_num_1} and \eqref{eq:fast_bound_unclustered_num}, IGMFs in filaments limit  $F_{\text{\tiny BH}}^{\text{\tiny fast,unclust}} \lesssim 10^{-23} \,( {\text{kg}}/{M_{\text{\tiny BH}}})\,\text{m}^{-2} \text{s}^{-1},$ however for non-extremal MBHs the corresponding limit is $F_{\text{\tiny BH}}^{\text{\tiny fast,unclust}}\lesssim 10^{-25}$ $\left( {\text{A-m}}/{Q_{\text{\tiny BH}}}\right)$ $\text{m}^{-2} \text{s}^{-1}$. In case when fast MBHs are virialized, we get weaker constraints on their flux. Using Eq.\,\eqref{eq:fast_bound_clustered_num},  we obtain $F_{\text{\tiny BH}}^{\text{\tiny fast,clust}} \lesssim 10^{-21} \,( {\text{kg}}/{M_{\text{\tiny BH}}})\,\text{m}^{-2} \text{s}^{-1}$ for fast clustered extremal MBHs, whereas for the non-extremal MBHs, from Eq.\, \eqref{eq:fast_bound_clustered_num_1}, we get the limits $F_{\text{\tiny BH}}^{\text{\tiny fast,clust}} \lesssim 10^{-24} \,( {M_{\text{\tiny BH}}}/{\text{kg}})\cdot ( {\text{A-m}}/{Q_{\text{\tiny BH}}})^2\,\text{m}^{-2} \text{s}^{-1}$.

    As previously mentioned, since the bounds on the flux of slow MBHs due to IGMFs are independent of the magnetic field strength, we obtain similar bounds to those found in the cosmic void case. Therefore, using Eqs.\,\eqref{eq:slow_bound_num_H0_1} and \,\eqref{eq:slow_bound_num_H0}, we get $ F_{\text{\tiny BH}}^{\text{\tiny slow,H0}} \lesssim  10^{-22}( \,{\text{kg}}/{M_{\text{\tiny BH}}})\,\text{m}^{-2} \text{s}^{-1}$, for extremal MBHs, and  $ F_{\text{\tiny BH}}^{\text{\tiny slow,H0}} \lesssim  10^{-25}({M_{\text{\tiny BH}}/{\text{kg}}})\cdot( \,{\text{A-m}}/{Q_{\text{\tiny BH}}})^2\,\text{m}^{-2} \text{s}^{-1}$  for non-extremal MBHs.
     
%%%%%%%%%%%%%%%%%%%%%%%%%%%%%%%%%%%%%%%%%%%%%%%%%%%%%%%%%%%%%%%%%%%%
    \subsubsection{Parker-type bounds from galaxies}
    For galaxies, we can place bounds on extremal and non-extremal MBHs using Eqs.\,\eqref{eq:fast_bound_unclustered_num_1}-\eqref{eq:slow_bound_num}. We find that with the presently available data, the galaxies M31 and NGC4501 furnish the strongest galactic bounds on MBHs. The field characteristics of NGC4501 are\,\cite{NGC4501, 10.1093/mnras/sty3270}
     %%%%%%%%%%%%%%%
    \begin{equation}\label{eq:B_NGC4501_1}
        B^{\text{\tiny NGC4501}} \simeq  10^{-6}\,\text{G}\,, \quad l_\text{\tiny c}^{\text{\tiny NGC4501}} \gtrsim 100\,\text{kpc}\,, \quad t_{\text {\tiny reg}}^{\text{\tiny NGC4501}}   \simeq 2.5\,\text{Gyr}  \;.
    \end{equation}
    %%%%%%%%%%%%%%%
    From Eq.\,\eqref{eq:fast_bound_unclustered_num} this gives a bound for fast unclustered extremal MBHs
    %%%%%%%%%%%%%%%
    \begin{equation}\label{eq:bound_NGC4501_1}
       f_{\text{\tiny DM}}^{\text{\tiny fast,unclust}} \lesssim 10^{-3} \,. 
    \end{equation}
    %%%%%%%%%%%%%%%
    For slow extremal MBHs, comparable bounds are obtained as $f_{\text{\tiny DM}}^{\text{\tiny slow}} \lesssim 10^{-3} $. However, for virialized extremal MBHs we get a weaker limit of $f_{\text{\tiny DM}}^{\text{\tiny fast,clust}} \lesssim 10^{-2}$.  Corresponding limits for non-extremal MBHs due to NGC4501 can be derived using Eqs.\,\eqref{eq:fast_bound_unclustered_num_1}, \eqref{eq:fast_bound_clustered_num_1}, and  \eqref{eq:slow_bound_num_1}, which gives
    %%%%%%%%%%%%%%%
    \begin{equation}\label{eq:fDM_bound_NGC4501_1}
       f_{\text{\tiny DM}}^{\text{\tiny fast,unclust}} \lesssim \frac{ 10^{-6}}{1+0.6\cos\alpha\left(\frac{M_{\text{\tiny BH}}}{\text{kg}}\right)\left(\frac{\text{A-m}}{Q_{\text{\tiny BH}}}\right) }\left(\frac{M_{\text{\tiny BH}}}{\text{kg}}\right)^2\left(\frac{\text{A-m}}{Q_{\text{\tiny BH}}}\right)^2 \;.
    \end{equation}
    %%%%%%%%%%%%%%%
    Additionally, for virialized non-extremal MBHs, we get the limits $ f_{\text{\tiny DM}}^{\text{\tiny fast,clust}} \lesssim  10^{-6}\left({M_{\text{\tiny BH}}}/{\text{kg}}\right)^2\cdot\left({\text{A-m}}/{Q_{\text{\tiny BH}}}\right)^2$, while for slow non-extremal MBHs, we get  $ f_{\text{\tiny DM}}^{\text{\tiny slow}} \lesssim  10^{-6}\left({M_{\text{\tiny BH}}}/{\text{kg}}\right)^{3/2}\cdot\left({\text{A-m}}/{Q_{\text{\tiny BH}}}\right)^{3/2}$. 
    Similar bounds on clustered MBHs have been derived before, studying specific galaxies such as M31\,\cite{Bai_2020, Ghosh_2020}. 

    Moreover, NGC4501 puts strong constraints on the slow extremal MBH flux,
    %%%%%%%%%%%%%%%
    \begin{equation}\label{eq:FMBH_NGC4501_slow}
        F_{\text{\tiny BH}}^{\text{\tiny slow}} \lesssim 10^{-20}\left( \frac{\text{kg}}{M_{\text{\tiny BH}}}\right)\text{m}^{-2} \text{s}^{-1}\;.
    \end{equation}
    %%%%%%%%%%%%%%%
    Again, for fast unclustered extremal MBHs, we get similar limits $F_{\text{\tiny BH}}^{\text{\tiny fast,unclust}} \lesssim 10^{-20}\left( {\text{kg}}/{M_{\text{\tiny BH}}}\right)$ $\text{m}^{-2} \text{s}^{-1}$. However, for virialized MBHs we get weaker limits of $F_{\text{\tiny BH}}^{\text{\tiny fast,clust}} \lesssim 10^{-19}$ $\left( {\text{kg}}/{M_{\text{\tiny BH}}}\right)\text{m}^{-2} \text{s}^{-1}$. 
    The limits for the non-extremal MBH flux can be obtained from Eqs.\,\eqref{eq:fast_bound_unclustered_num_1}, \eqref{eq:fast_bound_clustered_num_1}, and  \eqref{eq:slow_bound_num_1}. For fast, non-extremal, unclustered MBHs, NGC4501 bounds 
    %%%%%%%%%%%%%%%
    \begin{equation}\label{eq:fDM_bound_NGC4501_1}
       F_{\text{\tiny BH}}^{\text{\tiny fast,unclust}} \lesssim \frac{ 10^{-22}}{1+0.6\cos\alpha\left(\frac{M_{\text{\tiny BH}}}{\text{kg}}\right)\left(\frac{\text{A-m}}{Q_{\text{\tiny BH}}}\right) }\left(\frac{M_{\text{\tiny BH}}}{\text{kg}}\right)\left(\frac{\text{A-m}}{Q_{\text{\tiny BH}}}\right)^2 \;.
    \end{equation}
    %%%%%%%%%%%%%%%
    For slow non-extremal MBHs, we get  $ F_{\text{\tiny BH}}^{\text{\tiny slow}} \lesssim  10^{-22}\left({\text{A-m}}/{Q_{\text{\tiny BH}}}\right)$, and for fast virialized non-extremal MBHs, we get $ F_{\text{\tiny BH}}^{\text{\tiny fast,clust}} \lesssim  10^{-22}\left({M_{\text{\tiny BH}}}/{\text{kg}}\right)\cdot \left({\text{A-m}}/{Q_{\text{\tiny BH}}}\right)^2$.
    
    %%%%%%%%%%%%%%%%%%%%%%%%%%%%%%%%%%%%%%%%%%%%%%%%%%%%%%%%%%
    
    In summary, in this section, we have predominantly explored bounds on MBHs in cosmic web filaments and cosmic voids. These bounds are both qualitatively and quantitatively different from previous bounds\,\cite{Rephaeli_1982} based on intracluster (IC) magnetic fields. While the effect of magnetic monopoles on IGMFs has been recently explored in\,\cite{Perri:2023ncd}, these bounds are relevant for monopoles of masses $M\lesssim\mathcal{O}(10^{19})\,\text{GeV}$. In contrast, the bounds on extremal MBHs are applicable for mass $M_{\text{\tiny BH}}\gtrsim10^{-6}\,\text{Kg}\sim \mathcal{O}(10^{20})\,\text{GeV}$\,\cite{Carr_1974,Carr_2021}. The present study is, therefore, complementary to these earlier studies.
    
    Having established the constraints on dark matter fraction constituted by MBHs, let us now investigate potential observational signatures of MBHs due to Faraday rotation effects. In the next section, we will explore and characterize the Faraday rotation properties of MBHs and juxtapose their signatures with magnetic NS. As we will demonstrate subsequently, their polarization angle maps exhibit significant differences. This characteristic will be an observational tool to distinguish between an MBH and an NS.

%%%%%%%%%%%%%%%%%%%%%%%%%%%%%%%%%%%%%%%
\section{Faraday rotations due to primordial magnetic black holes}
%%%%%%%%%%%%%%%%%%%%%%%%%%%%%%%%%%%%%%%
\label{sec:farad_effec}
        In the last section, we have explored the constraints on MBHs due to galactic, extra-galactic, and cosmic magnetic fields. These suggest that the population of such objects may be very small. We would, therefore, like to envision strategies to look for such rare objects in observational data. We now consider one possible signature of MBHs that may be unique to them and observationally relevant. 
        
        A germane question may be whether an MBH may be discerned from other astrophysical bodies that may mimic some of its characteristics, such as a neutron star. An intriguing effect arises when we consider the interaction of MBHs with light.  When a linearly polarized electromagnetic wave propagates through a magnetized medium, its plane of polarization rotates along its trajectory. This effect is commonly referred to as Faraday rotation. It occurs because the left and right circular components of the linearly polarized waves have distinct velocities in the medium, resulting in a rotation of the plane of polarization along the trajectory of the wave. In realistic astrophysical scenarios, like for an NS, the background magnetic field may be non-uniform\,\cite{Romani1990}. This may lead to a spatial dependence of the polarization angle in the plane perpendicular to the direction of motion. The spatial variation in polarization angle will depend on the characteristics of the magnetic field and, by implication, its specific astrophysical source.  
        
        To begin the analyses, we will first review, in some detail, the occurrence of Faraday rotation in a plasma threaded by a spatially non-uniform magnetic field. Then, we will investigate two scenarios for the source of the non-uniform magnetic field---an MBH and an NS. We will compare the Faraday effects in these two cases in detail. 
        
        First, let us explore Faraday rotation on an electromagnetic wave passing through an inhomogeneous plasma in an arbitrary non-uniform magnetic field. We follow the methodology outlined in\,\cite{Melrose_McPhedran_1991}, with the detailed calculations provided in Appendix\,\ref{appendix:Far_rot}. Consider an electromagnetic wave propagating through an inhomogeneous, weakly ionized hydrogen plasma\footnote{For a warm hydrogen plasma,\,\cite{Skilling_1971} has shown that away from resonances, the correction to the electron contribution to Faraday rotation is negligible compared to the cold plasma term.} along the $z$ axis. We can take the background magnetic field as       
       %%%%%%%%%%%%%%%
       \begin{equation}\label{eq:fr_fru_arb_mf}
	 	\vec{\mathcal{B}}_{\text{\tiny ext.}}(\vec r)=
	 	B_{\text{\tiny ext,x}}(\vec r)~\hat{x}+	B_{\text{\tiny ext,y}}(\vec r)~\hat{y}+	B_{\text{\tiny ext,z}}(\vec r)~\hat{z}\;,
        \end{equation}
        %%%%%%%%%%%%%%%        
       and, define $B_{\text{\tiny ext.}}(\vec r)\equiv\sqrt{B_{\text{\tiny ext,x}}^2+B_{\text{\tiny ext,y}}^2+B_{\text{\tiny ext,z}}^2}$. The electrons in the plasma experience a Lorentz force due to the propagating electromagnetic waves. The Lorentz force is given by, 
       %%%%%%%%%%%%%%%
        \begin{equation}\label{eq:fr_fru_arb_eom}
	 	m\text{\tiny e}\frac{d^2\vec{r}_{\text{\tiny e}}}{dt^2}=-e \left(\vec{E}+\frac{d \vec{r}_{\text{\tiny e}}}{d t}\times \vec{B}\right)\;,
	 \end{equation}
 	%%%%%%%%%%%%%%%  
    where $\vec{r}_{\text{\tiny e}}$, $m_{\text{\tiny e}}$ and $-e$ are the position, mass, and charge of electrons, respectively, and $\vec{E}$ and $\vec{B}$ are the total electric and magnetic fields. Since the protons are much heavier than electrons, their oscillations will be suppressed and, therefore, will have a negligible contribution to the Faraday rotation.
    
    In a steady state, we can now solve this Lorentz force equation by considering an oscillating perturbation around the background solution. This is given by
    %%%%%%%%%%%%%%%
	 \begin{eqnarray}\label{eq:fr_fru_arb_pert}
	 	\vec{r}_{\text{\tiny e}}&=&\vec{r}^{\,\text{(0)}}_{\text{\tiny e}}+\vec{r}^{\,\text{(1)}}_{\text{\tiny e}}e^{-i\omega t}\;, \nonumber\\
	 	\vec{E}(\vec{r})&=&0+\vec{E}^{\,\text{(1)}}(\vec{r})e^{-i\omega t}\;, \\
	 	\vec{B}(\vec{r})&=&\vec{\mathcal{B}}_{\text{\tiny ext.}}(\vec r)+\vec{B}^{\,\text{(1)}}(\vec{r})e^{-i\omega t}\;, \nonumber
	 \end{eqnarray}	 
	 %%%%%%%%%%%%%%%
     where $\vec{r}^{\,\text{(0)}}_{\text{\tiny e}}$ and $\vec{r}^{\,\text{(1)}}_{\text{\tiny e}}$ are the mean position and amplitude of displacement of the electron, respectively, and $\omega$, $\vec{E}^{\,\text{(1)}}$ and $\vec{B}^{\,\text{(1)}}$, are respectively the frequency, the electric field and the magnetic field of the traversing wave. Using Eqs.\,\eqref{eq:fr_fru_arb_eom}, and \eqref{eq:fr_fru_arb_pert}, we then get   
    %%%%%%%%%%%%%%%
	 \begin{equation}\label{eq:fr_fru_arb_eom_1}
	 	\frac{e \vec{E}^{\,\text{(1)}}}{m_{\text{\tiny e}}}=\omega^2\vec{r}^{\,\text{(1)}}_{\text{\tiny e}} 	+\frac{ie\omega}{m_{\text{\tiny e}}}\left(\vec{r}^{\,\text{(1)}}_{\text{\tiny e}}\times \vec{\mathcal{B}}_{\text{\tiny ext.}}(\vec r)\right)\;.
	 \end{equation}
     %%%%%%%%%%%%%%%
     
     The above equation can be solved to obtain the displacement $\vec{r}^{\,\text{(1)}}_{\text{\tiny e}}$ in terms of the electric field $\vec{E}^{\,\text{(1)}}$ of the passing wave.  Furthermore, these electron oscillations will produce a time-dependent dipole moment with an amplitude $\vec{P}\equiv- \mathcal{N}_{\text{\tiny e}} e \vec{r}^{\,\text{(1)}}_{\text{\tiny e}}$, where $\mathcal{N}_{\text{\tiny e}}$ is the number density of electrons. We can obtain the dielectric tensor using the relation between the induced dipole moment and the dielectric tensor, i.e., $\vec{P}=\epsilon_{0}({\bm \epsilon_{\text{\tiny r}}}-{\bm I})\cdot\vec{E}^{\,\text{(1)}}$, where $\epsilon_{\text{\tiny r}}$, $\epsilon_{0}$, and ${\bm I}$ are the dielectric tensor, the vacuum permittivity and the identity matrix respectively. Following Eqs.\,\eqref{eq:app_fr_fru_arb_eom_1}-\eqref{eq:app_fr_fru_arb_dir_dil_tens}, we obtain the dielectric tensor as
    %%%%%%%%%%%%%%%
    \begin{equation}\label{eq:fr_fru_arb_dir_dil_tens}
  		{\bm \epsilon_{\text{\tiny r}}}=\begin{pmatrix} 
  			1-\frac{\omega^2_{\text{\tiny p}}\left(\omega^2-\tilde{\omega}^2_{\text{\tiny x}} \right)}{\omega^2\left(\omega^2-\tilde{\omega}^2\right)} &  \frac{\omega^2_{\text{\tiny p}} \left(\tilde{\omega}_{\text{\tiny x}} \tilde{\omega}_{\text{\tiny y}}+i \omega  \tilde{\omega}_{\text{\tiny z}}\right)}{\omega^2\left(\omega^2-\tilde{\omega}^2\right)} & \frac{\omega^2_{\text{\tiny p}} \left(\tilde{\omega}_{\text{\tiny x}} \tilde{\omega}_{\text{\tiny z}}-i \omega  \tilde{\omega}_{\text{\tiny y}}\right)}{\omega^2\left(\omega^2-\tilde{\omega}^2\right)}   \\
  			\frac{\omega^2_{\text{\tiny p}} \left(\tilde{\omega}_{\text{\tiny x}} \tilde{\omega}_{\text{\tiny y}}-i \omega  \tilde{\omega}_{\text{\tiny z}}\right)}{\omega^2\left(\omega^2-\tilde{\omega}^2\right)} & 1-\frac{\omega^2_{\text{\tiny p}}\left(\omega^2-\tilde{\omega}^2_{\text{\tiny y}} \right)}{\omega^2\left(\omega^2-\tilde{\omega}^2\right)}&\frac{\omega^2_{\text{\tiny p}} \left(\tilde{\omega}_{\text{\tiny y}} \tilde{\omega}_{\text{\tiny z}}+i \omega  \tilde{\omega}_{\text{\tiny x}}\right)}{\omega^2\left(\omega^2-\tilde{\omega}^2\right)} \\ 
  			\frac{\omega^2_{\text{\tiny p}} \left(\tilde{\omega}_{\text{\tiny x}} \tilde{\omega}_{\text{\tiny z}}+i \omega  \tilde{\omega}_{\text{\tiny y}}\right)}{\omega^2\left(\omega^2-\tilde{\omega}^2\right)}    & \frac{\omega^2_{\text{\tiny p}} \left(\tilde{\omega}_{\text{\tiny y}} \tilde{\omega}_{\text{\tiny z}}-i \omega  \tilde{\omega}_{\text{\tiny x}}\right)}{\omega^2\left(\omega^2-\tilde{\omega}^2\right)} & 1-\frac{\omega^2_{\text{\tiny p}}\left(\omega^2-\tilde{\omega}^2_{\text{\tiny z}} \right)}{\omega^2\left(\omega^2-\tilde{\omega}^2\right)} \\
  		\end{pmatrix}\;.
  	\end{equation} 
   %%%%%%%%%%%%%%%
	 Here, $\tilde{\omega}_{\text{\tiny i}}\equiv e B_{\text{\tiny ext,i}}(\vec r)/m_{\text{\tiny e}}$ and $\tilde{\omega}\equiv e B_{\text{\tiny ext.}}(\vec r)/m_{\text{\tiny e}}$ are the component-wise and total electron cyclotron frequencies. $\omega_{\text{\tiny p}}^2\equiv e^2\mathcal{N}_{\text{\tiny e}}/(\epsilon_{0}m_{\text{\tiny e}})$ is the electron plasma frequency. 
      
    Evidently, due to the presence of a background magnetic field, the plasma is anisotropic in nature. In such cases, a medium has different refractive indices along different directions. To understand this behaviour, we will solve the Maxwell's equations to determine the characteristic modes and the corresponding refractive indices. 
    
    The Maxwell equations\,\cite{Constantinidis:2016wjo} in the present context may be expressed as
   %%%%%%%%%%%%%%%
	  \begin{eqnarray}
	 	\label{eq:fr_umf_arb_dir_ME_1}
	 	\nabla\cdot \vec{E}&=& \frac{e}{\epsilon_{0}}(\mathcal{N}_{\text{\tiny p}}-\mathcal{N}_{\text{\tiny e}})\;, \\
	 	\label{eq:fr_umf_arb_dir_ME_2}
	 	\nabla\cdot \vec{B}&=&\mu_{0}Q_{\text{\tiny BH}}\delta^3(\vec{r})\;, \\
	 	\label{eq:fr_umf_arb_dir_ME_3}
	 	\nabla\times\vec{E}+\frac{\partial \vec{B}}{\partial t}
	 	&=& 0\;, \\
	 	\label{eq:fr_umf_arb_dir_ME_4}
	 	\nabla\times\vec{B}-\frac{1}{c^2}\frac{\partial \vec{E}}{\partial t}&=&\mu_{0}\vec{J}=\mu_{0}e(\mathcal{N}_{\text{\tiny p}}\vec{v}_{\text{\tiny p}}-\mathcal{N}_{\text{\tiny e}}\vec{v}_{\text{\tiny e}})\;.
	 \end{eqnarray}
  %%%%%%%%%%%%%%%
    Here, $Q_{\text{\tiny BH}}$
  is the monopole magnetic charge, which may be one of the sources for the background magnetic field. Additionally, $\mathcal{N}_{\text{\tiny p}}$ and $\mathcal{N}_{\text{\tiny e}}$ represent the number densities of protons and electrons, while $\vec{v}_{\text{\tiny p}}$ and $\vec{v}_{\text{\tiny e}}$ denote the velocities of protons and electrons, respectively. Once again, we can neglect $\vec{v}_{\text{\tiny p}}$, since protons being heavier than electrons lead to  $v_{\text{\tiny p}}\ll v_{\text{\tiny e}}$.

   Eqs.\,\eqref{eq:fr_umf_arb_dir_ME_3}	 and \eqref{eq:fr_umf_arb_dir_ME_4} can be linearized using Eq.\,\eqref{eq:fr_fru_arb_pert} and they can be simplified to obtain (See Appendix\,\ref{appendix:Far_rot} for details)
        %%%%%%%%%%%%%%% 
	 \begin{equation}\label{eq:fr_umf_arb_dir_wave_eq}
	   \nabla\left(\nabla\cdot\vec{E}^{\,\text{(1)}}\right)-\nabla^2\vec{E}^{\,\text{(1)}}-\frac{\omega^2}{c^2}{\bm \epsilon_{\text{\tiny r}}}\cdot \vec{E}^{\,\text{(1)}}=0\;.
	 \end{equation}
	 %%%%%%%%%%%%%%%
  
    We can further write $\vec{E}^{\,\text{(1)}}(\vec{r})\propto e^{i\psi_{\text{\tiny ph.}}(\vec{r})}$ for a plane electromagnetic wave.  Here, $\psi_{\text{\tiny ph.}}$ is the phase of the electric field and is related to the wave vector as $\vec{k}=\nabla \psi_{\text{\tiny ph.}}(\vec{r})$.
    Putting it in Eq.\,\eqref{eq:fr_umf_arb_dir_wave_eq} and taking the eikonal limit, i.e. $\vert\nabla{k}\vert/k^2\ll 1$, we get
	  %%%%%%%%%%%%%%% 
	 \begin{equation}\label{eq:fr_umf_arb_dir_wave_eq_2}
	   \left(\vec{k}\cdot\vec{E}^{\,\text{(1)}}\right)\vec{k}-k^2\vec{E}^{\,\text{(1)}}+\frac{\omega^2}{c^2}{\bm \epsilon_{\text{\tiny r}}}\cdot \vec{E}^{\,\text{(1)}}\approx0\;.
	 \end{equation}
	 %%%%%%%%%%%%%%%
    Here, $k\equiv \abs{\vec{k}}$ is the magnitude of the wave vector, and note that it is also a function of $\vec{r}$. For a wave propagating along the $z$ direction, the wave vector takes the form $\vec{k}=(0,0,d \psi_{\text{\tiny ph.}}(\vec{r})/dz)$ and the phase can be written as 
     %%%%%%%%%%%%%%% 
	 \begin{equation}\label{eq:fr_umf_arb_dir_phase}
	  \psi_{\text{\tiny ph.}}(\vec{r})=\int dz~ k(\vec{r})=\frac{c}{\omega}\int dz~ n(\vec{r})\;.
	 \end{equation}
	 %%%%%%%%%%%%%%%
  Here, $n(\vec{r})\equiv c k(\vec{r})/\omega $ is the refractive index of the wave. Using Eqs.\,\eqref{eq:fr_fru_arb_dir_dil_tens} and \eqref{eq:fr_umf_arb_dir_wave_eq_2}, we get the refractive indices as 
     %%%%%%%%%%%%%%%
    \begin{equation}\label{eq:fr_umf_arb_dir_ref_ind}
 		n_{{ (\pm)}}(\vec{r})=\left(1-\frac{\omega^2_{\text{\tiny p}} \left(\omega^2-\omega^2_{\text{\tiny p}}\right)}{\omega^2\left(\omega^2-\omega^2_{\text{\tiny p}}-\frac{1}{2}\left(\tilde{\omega}^2_{\text{\tiny x}}+\tilde{\omega}^2_{\text{\tiny y}}\right)\pm\left(\frac{1}{4}\left(\tilde{\omega}^2_{\text{\tiny x}}+\tilde{\omega}^2_{\text{\tiny y}}\right)^2+(\omega^2-\omega^2_{\text{\tiny p}})^2\frac{\tilde{\omega}^2_{\text{\tiny z}}}{\omega^2}\right)^{1/2}\right)}\right)^{1/2}\;.
 	\end{equation}
    %%%%%%%%%%%%%%%
    
    Here, $n_{{ (\pm)}}$ are the refractive indices for the two characteristic modes. Notice that the refractive index is symmetric about the exchange of $\tilde{\omega}_{\text{\tiny x}}$ and $\tilde{\omega}_{\text{\tiny y}}$. Furthermore, its dependence on $\tilde{\omega}_{\text{\tiny z}}$ is different compared to $\tilde{\omega}_{\text{\tiny x}}$ and $\tilde{\omega}_{\text{\tiny y}}$, this is due to our choice of the direction of propagation being along the $z$-axis. 
    Additionally, as we will see later, these two characteristic modes will be further divided into branches, with our focus being solely on the propagating branches.
    
     %%%%%%%%%%%%%%%
    	\begin{figure}[t]
 		\centering
 		\includegraphics[scale=0.7]{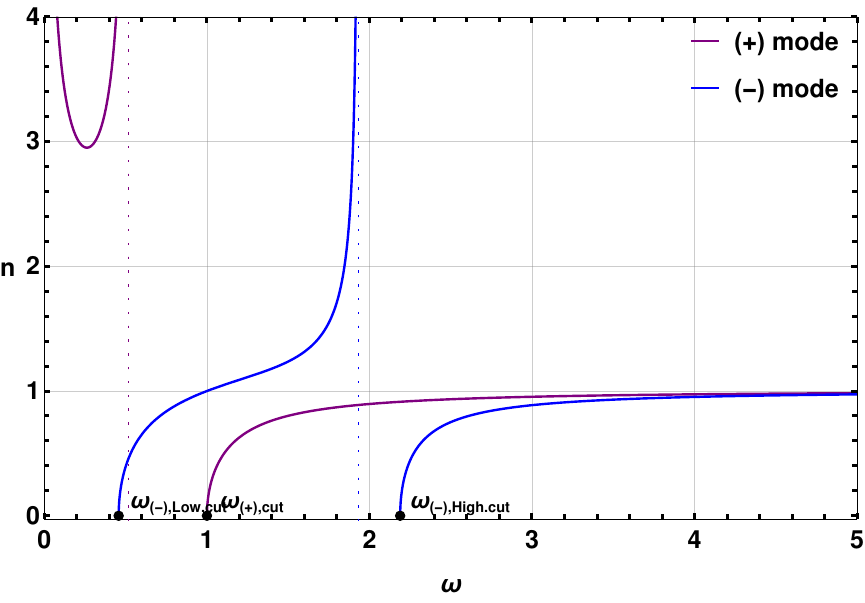}
 		\caption{Refractive indices (given by Eq.\;\eqref{eq:fr_umf_arb_dir_ref_ind}) as a function of $\omega$ in a magnetic, cold, inhomogeneous plasma medium. Red and blue curves represent $(+)$ and $(-)$ modes respectively. We have set $\omega_{\text{\tiny p}}=\tilde{\omega}_{\text{\tiny x} }=\tilde{\omega}_{\text{\tiny y} }=\tilde{\omega}_{\text{\tiny z} }= 1~\text{hz}$.  There are four  modes $(+)_\text{\tiny Low(High)}$ and $(-)_\text{\tiny Low(High)}$ separated by cut-offs and resonances. The resonances are represented by vertical lines.}
 		\label{fig:ref_ind_arb_mag_field}
 	\end{figure}
    %%%%%%%%%%%%%%%
    
    Let us now briefly discuss the cut-offs and the resonances of these modes.  The cut-offs are obtained by taking $n_{\pm}=0$. For a particular branch, the refractive index is imaginary below the cut-off frequencies, resulting in complete attenuation of the branch in the plasma at those frequencies.  For the $(+)$ mode, a cut-off exists at
    %%%%%%%%%%%%%%%
	\begin{equation}
		\omega_{\text{\tiny $(+)$,cut.}}\equiv \omega_{\text{\tiny p}}\;.
	\end{equation}
    %%%%%%%%%%%%%%%
	However, for the $(-)$ mode, there exist two cut-offs at
    %%%%%%%%%%%%%%%
	\begin{equation}\label{eq:fr_umf_arb_dir_cut}
		\omega_{\text{\tiny $(-)$,Low.cut.}}\equiv-\frac{\tilde{\omega} }{2}+\frac{1}{2} \sqrt{\tilde{\omega} ^2+4 \omega_{\text{\tiny p}}^2}\;, \quad \omega_{\text{\tiny $(-)$,High.cut.}}\equiv \frac{\tilde{\omega} }{2}+\frac{1}{2} \sqrt{\tilde{\omega} ^2+4 \omega_{\text{\tiny p}}^2}\;.
	\end{equation}
    %%%%%%%%%%%%%%%
    Apart from these cut-offs, there exist resonances when $n_{\pm}\to \infty$. They occur at characteristic frequencies
    %%%%%%%%%%%%%%%
	\begin{equation}\label{eq:fr_umf_arb_dir_res}
			\omega_{\text{\tiny ($\pm$),res.}}=\frac{\sqrt{\tilde{\omega} ^2+\omega_{\text{\tiny p}}^2\mp \sqrt{\left(\tilde{\omega} ^2+\omega_{\text{\tiny p}}^2\right)^2-4\tilde{\omega}^2_{\text{\tiny z}}\omega^2_{\text{\tiny p}}}}}{\sqrt{2}}\;.
	\end{equation} 
    %%%%%%%%%%%%%%%
	 At these resonances, electrons gyrate synchronously with the wave, allowing them to continuously extract energy and causing the waves to be absorbed by the plasma. As shown in Fig.\,\ref{fig:ref_ind_arb_mag_field}, these frequency cutoffs and resonances divide $(\pm)$ modes into lower and higher frequency branches. Therefore, the higher frequency branches exist for 
    %%%%%%%%%%%%%%%
	\begin{equation}
		\omega_{\text{\tiny $(+)$,High.}}>\omega_{\text{\tiny $(+)$,cut.}}\;, \hquad  	\omega_{\text{\tiny $(-)$,High.}}>\omega_{\text{\tiny $(-)$,High.cut.}}\;.
	\end{equation} 
    %%%%%%%%%%%%%%%
	while the lower frequency branches exist for 
    %%%%%%%%%%%%%%%
	\begin{equation}
		0<\omega_{\text{\tiny $(+)$,Low.}}<\omega_{\text{\tiny $(+)$,res.}}\;, \hquad  	\omega_{\text{\tiny $(-)$,Low.cut.}}<\omega_{\text{\tiny $(-)$,Low.}}<\omega_{\text{\tiny $(-)$,res.}}\;.
	\end{equation}
    %%%%%%%%%%%%%%%
	
The $(+)$ mode is non-propagating at frequencies $\omega_{\text{\tiny $(+)$,res.}}<\omega_{\text{\tiny$(+)$}}<\omega_{\text{\tiny $(+)$,cut.}}$, while the $(-)$ mode is non-propagating at frequencies $\omega_{\text{\tiny $(-)$,res.}}<\omega_{\text{\tiny$(-)$}}<\omega_{\text{\tiny $(-)$,High.cut.}}$. Furthermore, if a wave, which has frequency $\omega<\omega_{\text{\tiny ($\pm$),res.}}$, travels along the direction of decreasing magnetic field or equivalently decreasing $\tilde{\omega}$, then along its propagation, we see from Eq.\,\eqref{eq:fr_umf_arb_dir_res}, $\omega_{\text{\tiny ($\pm$),res.}}$ also decreases. Hence, when $\omega_{\text{\tiny ($\pm$),res.}}$ becomes equal to $\omega$, the resonance occurs, resulting in absorption of the corresponding mode.  Consequently, throughout propagation, the wave frequency must exceed $\omega_{\text{\tiny ($\pm$),res.}}$. As mentioned before, only higher frequency branches exist for $\omega>\omega_{\text{\tiny ($\pm$),res.}}$, and therefore solely these branches will propagate, which is pertinent to our context.  

Based on the above discussions, we will restrict ourselves to those frequencies where higher branches, for both $(\pm)$ modes, exist. This happens for frequencies satisfying
    %%%%%%%%%%%%%%%
	\begin{equation}\label{eq:fr_umf_arb_dir_freq_lim}
		 \omega>\omega_{\text{\tiny $(-)$,High. cut.}}= \frac{\tilde{\omega} }{2}+\frac{1}{2} \sqrt{\tilde{\omega} ^2+4 \omega_{\text{\tiny p}}^2}\;.
	\end{equation}
    %%%%%%%%%%%%%%%
    Notice, from Eq.\,\eqref{eq:fr_umf_arb_dir_cut} $\omega_{\text{\tiny $(-)$,High. cut.}}> \text{max}\left({ \omega_{\text{\tiny p}},\tilde{\omega}}\right)$, consequently the relevant frequencies for propagation will have  $\omega> \text{max}\left({ \omega_{\text{\tiny p}},\tilde{\omega}}\right)$.
    Additionally, from Eqs.\,\eqref{eq:fr_fru_arb_dir_dil_tens} and \eqref{eq:fr_umf_arb_dir_wave_eq_2}, one can obtain the relation between the transverse components of the characteristic modes as
  %%%%%%%%%%%%%%%
    \begin{equation}\label{eq:fr_umf_arb_dir_Ex_Ey}
        \left(\frac{E_{\text{x}}^{\,\text{(1)}}}{E_{\text{y}}^{\,\text{(1)}}}\right)_{{ (\pm)}}=\frac{i  \left(\omega\left(\tilde{\omega}_{\text{\tiny x}}^2-\tilde{\omega}_{\text{\tiny y}}^2\right)\pm\sqrt{{4 \tilde{\omega}_{\text{\tiny z}}^2 \left(\omega ^2-\omega_{\text{\tiny p}}^2\right)^2}+\omega^2\left(\tilde{\omega}_{\text{\tiny x}}^2+\tilde{\omega}_{\text{\tiny y}}^2\right)^2}\right)}{2\left( \left(\omega ^2- \omega_{\text{\tiny p}}^2\right) \tilde{\omega}_{\text{\tiny z}}+ i \omega  \tilde{\omega}_{\text{\tiny x}} \tilde{\omega}_{\text{\tiny y}} \right)}\;.
    \end{equation}
    %%%%%%%%%%%%%%%
    Since $\omega>\tilde{\omega}$, and $\tilde{\omega}$ decreases away from the magnetic field source, we may expand the above expression in powers of $\tilde{\omega}_{\text{\tiny (x,y,z)}}/\omega$, as given by Eq.\,\eqref{eq:app_fr_umf_arb_dir_Ex_Ey_approx}. Then, it is revealed that the characteristic waves approximately manifest as left and right circular polarizations, i.e. $E_{\text{x}}^{\,\text{(1)}}/E_{\text{y}}^{\,\text{(1)}}\simeq \pm i\,\text{sgn}(\tilde{\omega}_{\text{\tiny z}}) $, when 
  %%%%%%%%%%%%%%%
  \begin{equation}\label{eq:fr_umf_arb_dir_Ex_Ey_cond}
      \abs{\frac{ \omega\left( \tilde{\omega}_{\text{\tiny x}}^2 + \tilde{\omega}_{\text{\tiny y}}^2 \right)}{2 (\omega^2-\omega^2_{\text{\tiny p}})\tilde{\omega}_{\text{\tiny z}}}  }\ll1\;.
  \end{equation}
  %%%%%%%%%%%%%%%
    Therefore, in the above limit, we infer that for $\tilde{\omega}_{\text{\tiny z}}>0$, $(+)$ and $(-)$ modes are close to left and right circular modes, respectively. However, for $\tilde{\omega}_{\text{\tiny z}}<0$, $(+)$ and $(-)$ modes are close to right and left circular modes. 
  We can therefore write,
  %%%%%%%%%%%%%%%
  \begin{equation}\label{eq:fr_umf_arb_dir_ref_ind_fin}
      n_{\text{\tiny L(R)}}=\begin{cases}
          n_{{ +(-)}}\; ; &  \tilde{\omega}_{\text{\tiny z}}>0\;,\\
          n_{{ -(+)}}\; ; &  \tilde{\omega}_{\text{\tiny z}}<0\;.\\
      \end{cases}
  \end{equation}
  %%%%%%%%%%%%%%%
    Here, $n_{\text{\tiny L(R)}}$ are the refractive indices for left and right circular modes. It is evident from Eqs.\,\eqref{eq:fr_umf_arb_dir_ref_ind} and \eqref{eq:fr_umf_arb_dir_ref_ind_fin} that the refractive index for left and right circularly polarized waves are distinct, and for $\tilde{\omega}_{\text{\tiny z}}>0$, $n_{\text{\tiny L}}>n_{\text{\tiny R}}$ and for $\tilde{\omega}_{\text{\tiny z}}<0$, $n_{\text{\tiny L}}<n_{\text{\tiny R}}$. Again, this dependence specifically on the $z$ component of the magnetic field comes from our choice of the direction of propagation.

   With these results, let us calculate an expression for the expected change in the polarization angle. If the light source is linearly polarized, one can analyse the propagation of the left and right circularly polarised wave components in the medium and then compare the $x$ and $y$ components in the linear basis to obtain the polarization angle (See Appendix\,\ref{appendix:Far_rot} for details). Therefore, the change in polarization angle of the passing electromagnetic wave is given by
     \begin{equation}\label{eq:fr_fru_cp_pol_ang}
         \psi_{\text{\tiny pol.}}=\frac{\omega}{2 c}\int~dz \left( n_{\text{\tiny L}}(\vec{r})-n_{\text{\tiny R}}(\vec{r})\right)\;.
     \end{equation}
     %%%%%%%%%%%%%%%
     Here, the integration is over the path of the electromagnetic wave. We have neglected the influence of gravitational lensing on the trajectory of the electromagnetic wave. Our estimates suggest that for the lensed trajectory,  $\psi_{\text{\tiny pol.}}$ differs only by up to $\mathcal{O}(10^{-3})\,\%$ for the masses pertinent to our study. Now, since the left and right circular waves have distinct refractive indices, and moreover, may even each change along the path, the $\psi_{\text{\tiny pol.}}$ given by Eq.\,\eqref{eq:fr_fru_cp_pol_ang} will evolve along the wave's trajectory. If the initial polarization angle of the source is $\psi_{\text{\tiny pol,0}}$, the observer will detect the electromagnetic wave with polarization,
     %%%%%%%%%%%%%%%
     \begin{equation}\label{eq:fr_fru_cp_pol_ang_tot}
         \psi_{\text{\tiny obs.}}= \psi_{\text{\tiny pol,0}}+\psi_{\text{\tiny pol.}}\;.
     \end{equation}
    %%%%%%%%%%%%%%%
    
    The rotation measure (RM), which is a closely associated quantity, is defined as
    %%%%%%%%%%%%%%%
     \begin{equation}\label{eq:fr_fru_cp_RM_def}
         \text{RM}(\lambda)\equiv \frac{d\psi_{\text{\tiny pol.}}(\lambda)}{d\lambda^2}\;.
     \end{equation}
    %%%%%%%%%%%%%%%
    Here, $\lambda$ is the wavelength of light. The RM is useful for determining the properties of both the magnetic field and the plasma density along the wave's path. 
    
    Let us now calculate explicit expressions for these quantities of interest in our context. As mentioned before, the frequencies of interests are $\omega> \text{max}\left({ \omega_{\text{\tiny p}},\tilde{\omega}}\right)$  and in the limit when $\omega\gg \text{max}\left({ \omega_{\text{\tiny p}},\tilde{\omega}}\right)$, using Eqs.\,\eqref{eq:fr_umf_arb_dir_ref_ind} and \eqref{eq:fr_umf_arb_dir_ref_ind_fin}  we get
    \begin{equation}\label{eq:fr_umf_arb_ref_mul-mur_2}
		n_{\text{\tiny L}}(\vec{r})-n_{\text{\tiny R}}(\vec{r})\simeq	  \frac{\omega^2_{\text{\tiny p}}}{\omega^3}\tilde{\omega}_{\text{\tiny z}}+\frac{\omega^2_{\text{\tiny p}}\tilde{\omega}_{\text{\tiny z}}}{\omega^5}\left(\tilde{\omega}^2_{\text{\tiny x}}+\tilde{\omega}^2_{\text{\tiny y}}
		+\frac{\tilde{\omega}^2_{\text{\tiny x}}\tilde{\omega}^2_{\text{\tiny y}}}{4\tilde{\omega}^2_{\text{\tiny z}}}\right)+\mathcal{O}\left(\left(\frac{\omega^2_{\text{\tiny p}}}{\omega^2}\right)^2,\left(\frac{\tilde{\omega}}{\omega}\right)^4\right)
		\;.
	\end{equation}
    The change in polarization may hence be written, using Eq.\,\eqref{eq:fr_fru_cp_pol_ang}, as
    %%%%%%%%%%%%%%%
  \begin{equation}\label{eq:fr_fru_cp_pol_ang_approx_1}
      \psi_{\text{\tiny pol.}}\simeq \int dz~\left( \frac{\omega^2_{\text{\tiny p}}}{2\omega^2 c }\tilde{\omega}_{\text{\tiny z}}+\frac{\omega^2_{\text{\tiny p}}\tilde{\omega}_{\text{\tiny z}}}{2\omega^4 c}\left(\tilde{\omega}^2_{\text{\tiny x}}+\tilde{\omega}^2_{\text{\tiny y}}
		+\frac{\tilde{\omega}^2_{\text{\tiny x}}\tilde{\omega}^2_{\text{\tiny y}}}{4\tilde{\omega}^2_{\text{\tiny z}}}\right)\right)+\mathcal{O}\left(\left(\frac{\omega^2_{\text{\tiny p}}}{\omega^2}\right)^2,\left(\frac{\tilde{\omega}}{\omega}\right)^4\right)\;.
    \end{equation}
  %%%%%%%%%%%%%%%
  This may be further simplified in terms of the background magnetic field as 
     %%%%%%%%%%%%%%%
  \begin{eqnarray}\label{eq:fr_fru_cp_pol_ang_approx} 
     \psi_{\text{\tiny pol.}} &\simeq&\frac{e^3\lambda^2}{8 \pi^2 \epsilon_0m^2_{\text{\tiny e}} c^3}\int dz~  \mathcal{N}_{\text{\tiny e}}(r)B_{\text{\tiny z}}(\vec r)\\
		&&
		+\frac{e^5\lambda^4}{32 \pi^4 \epsilon_0m^4_{\text{\tiny e}} c^5}\int dz~ \mathcal{N}_{\text{\tiny e}}(r)B_{\text{\tiny z}}(\vec r) \left(B_{\text{\tiny x}}(\vec r)^2+B_{\text{\tiny y}}(\vec r)^2+\frac{B_{\text{\tiny x}}(\vec r)^2B_{\text{\tiny y}}(\vec r)^2}{4 B_{\text{\tiny z}}(\vec r)^2}\right)\nonumber\;.
  \end{eqnarray}
  %%%%%%%%%%%%%%%
  For the RM, using Eq.\,\eqref{eq:fr_fru_cp_RM_def}, we get
    %%%%%%%%%%%%%%%
  \begin{eqnarray}\label{eq:fr_fru_cp_RM_approx}
      \text{RM}(\lambda)&\simeq&\frac{e^3}{8 \pi^2 \epsilon_0m^2_{\text{\tiny e}} c^3}\int dz~  \mathcal{N}_{\text{\tiny e}}(r)B_{\text{\tiny z}}(\vec r)\\
		&&
		+\frac{e^5\lambda^2}{16 \pi^4 \epsilon_0m^4_{\text{\tiny e}} c^5}\int dz~ \mathcal{N}_{\text{\tiny e}}(r)B_{\text{\tiny z}}(\vec r) \left(B_{\text{\tiny x}}(\vec r)^2+B_{\text{\tiny y}}(\vec r)^2+\frac{B_{\text{\tiny x}}(\vec r)^2B_{\text{\tiny y}}(\vec r)^2}{4 B_{\text{\tiny z}}(\vec r)^2}\right)\nonumber\;.
  \end{eqnarray}
  %%%%%%%%%%%%%%%
  
   We note that in the limit $\omega\gg \text{max}\left({ \omega_{\text{\tiny p}},\tilde{\omega}}\right)$, one obtains the dominant contribution to the change in the polarisation angle and the RM from the component of the magnetic field along the wave's trajectory. However, for comprehensiveness in all the numerical analyses, we will always use Eqs.\,\eqref{eq:fr_umf_arb_dir_ref_ind}, \eqref{eq:fr_fru_cp_pol_ang} and \eqref{eq:fr_fru_cp_RM_def} directly which take into account all the components. Also, we will work in the limit where the modes are propagating and close to circular modes. These requirements are embodied in Eqs.\,\eqref{eq:fr_umf_arb_dir_freq_lim} and \eqref{eq:fr_umf_arb_dir_Ex_Ey_cond}. 
    
    We will now consider two sources for the background magnetic field---an NS and an MBH. In turn, we will compute the change of the polarization angle and the RM assuming these two sources and compare the spatial dependencies.  This spatial dependence may be useful in observationally distinguishing between an NS and an MBH. Apart from this, we will also establish a quantitative measure based on the spatial dependence of the polarisation angle to further discriminate them.

      %%%%%%%%%%%%%%%%%%%%%%%%%%%%%%%%%%%%%%%
	\subsection{Faraday rotations due to a neutron star}
         \label{sec:fr_ns}
\begin{figure}[h]
	 	\centering
	 	\includegraphics[scale=1]{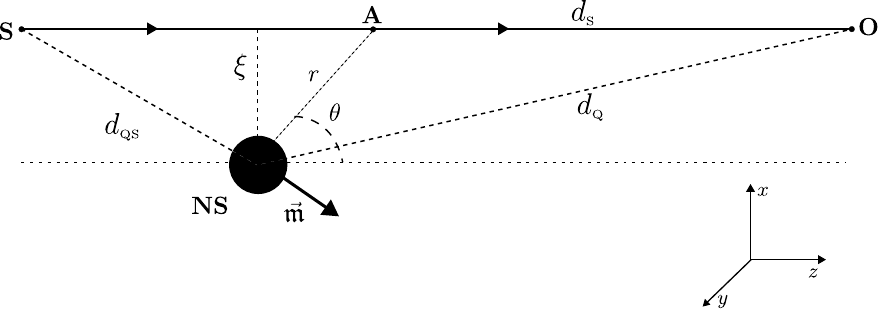}
	 	\caption[]{Schematic for Faraday rotation analysis in the case of a NS. An NS with magnetic moment $\vec{\mathfrak{m}}$ is at the origin. An observer (O) is located at a distance $d_{\text{\tiny Q}}$ from it, and a source (S) emitting linearly polarized light is located at a distance $d_{\text{\tiny S}}$ from the observer. The distance between the NS and the source is $d_{\text{\tiny QS}}$. The light rays are moving along the $z$ direction, with an impact parameter $\xi=\sqrt{x^2+y^2}$ with respect to the NS. }
	 	\label{fig:Faraday_effect_Schematics_2}
	 \end{figure}

     %%%%%%%%%%%%%%%%%%%%%%%%%%%%%%%%%%%%%%%
        Let us now compute the Faraday effect produced by an NS, which, for our purposes, can approximately be modelled as a magnetic dipole. Consider the schematic of Fig.\,\ref{fig:Faraday_effect_Schematics_2}---the origin is taken at the location of the NS, and an observer O is located at a distance $d_{\text{\tiny Q}}$ from it. The NS has a magnetic dipole moment $\vec{\mathfrak{m}}$ and mass $M_{\text{\tiny NS}}$. A source (S) located at a distance $d_{\text{\tiny QS}}$ from the NS is emitting plane-polarized light. We denote by $d_{\text{\tiny S}}$ the distance between the light source (S) and the observer, and by $\xi$ the distance of closest approach of light with the NS. In this setup, we assume that $d_{\text{\tiny S}}\simeq d_{\text{\tiny Q}}\gg\xi $, and we can write $d_{\text{\tiny QS}}\simeq d_{\text{\tiny S}}-d_{\text{\tiny Q}}$. Furthermore, we assume that there is a hydrogen plasma background with electron density $N_\text{\tiny e}(\Vec{r})$ permeating the region under consideration. 
        
        The orientation of the magnetic moment of the NS is determined by  the polar angle $\theta_{\text{\tiny NS}}$ and the azimuthal angle $\phi_{\text{\tiny NS}}$. We can, therefore, write
         
        %%%%%%%%%%%%%%%
 	\begin{equation}\label{eq:fr_GCDP_ns_dpmoment}
 		\vec{\mathfrak{m}}=\mathfrak{m} \left(\sin\theta_{\text{\tiny NS}}\cos\phi_{\text{\tiny NS}}\hat{x}+\sin\theta_{\text{\tiny NS}}\sin\phi_{\text{\tiny NS}}\hat{y}+\cos\theta_{\text{\tiny NS}}\hat{z}\right)\;,
 	\end{equation}
    %%%%%%%%%%%%%%%
 	where $\mathfrak{m}$ is the magnitude of the magnetic dipole moment. The magnetic field at an arbitrary point A with spherical coordinates $(r,\theta,\phi)$ is then given by  
     %%%%%%%%%%%%%%%
  	\begin{equation}\label{eq:fr_ns_mf_B}
		\vec{B}_{\text{\tiny NS}}(\vec{r})=\frac{\mu_0}{4 \pi}\left(\frac{3 (\vec{\mathfrak{m}}\cdot \vec{r})\vec{r}}{r^5}-\frac{\vec{\mathfrak{m}}}{r^3}\right)\;.
	\end{equation}
  %%%%%%%%%%%%%%%

    %%%%%%%%%%%%%%%
  %%%%%%%%%%%%%%%
  Notice that the magnetic field scales as 
    %%%%%%%%%%%%%%%
  	\begin{equation}\label{eq:fr_ns_mf_B_approx}
		\abs{\vec{B}_{\text{\tiny NS}}}\sim\frac{\mu_0}{4 \pi}\frac{\mathfrak{m}}{r^3}\;.
	\end{equation}
  %%%%%%%%%%%%%%%
  Furthermore, the cyclotron frequency of electrons at a radial distance $r$ due to the presence of an NS in the plasma is given by 
    %%%%%%%%%%%%%%%
    \begin{equation}\label{eq:fr_ns_cyc_freq}
    	\tilde{\omega}_{\text{\tiny NS}}(\vec{r})=\abs{\frac{\mu_0 e}{4 \pi m_{\text{\tiny e}}}\left(\frac{3 (\vec{\mathfrak{m}}\cdot \vec{r})\vec{r}}{r^5}-\frac{\vec{\mathfrak{m}}}{r^3}\right)}\;.
    \end{equation}
    %%%%%%%%%%%%%%%
    In this case, as appropriate, we take  $\{\tilde{\omega}_{\text{\tiny NS,x}},\tilde{\omega}_{\text{\tiny NS,y}},\tilde{\omega}_{\text{\tiny NS,z}}\}\equiv  \{e B_{\text{\tiny NS, x}}/m_{\text{\tiny e}},e B_{\text{\tiny NS,y}}/m_{\text{\tiny e}},e B_{\text{\tiny NS,z}}/m_{\text{\tiny e}}\}$, where $B_{\text{\tiny NS, (x,y,z)}}$ are the components of the magnetic field of the NS. Using Eqs.\,\eqref{eq:fr_umf_arb_dir_ref_ind}, \eqref{eq:fr_umf_arb_dir_ref_ind_fin} and \eqref{eq:fr_ns_cyc_freq}, we may calculate the refractive index for the left and right circularly polarized light at position $\vec{r}$
   %%%%%%%%%%%%%%%
  \begin{equation}\label{eq:fr_ns_ref_ind}
    \hspace{-0.5 cm} n_{\text{\tiny L,(R)}}^{\text{\tiny NS}}=\begin{cases}\left(1-\frac{\omega^2_{\text{\tiny p}} \left(\omega^2-\omega^2_{\text{\tiny p}}\right)}{\omega^2\left(\omega^2-\omega^2_{\text{\tiny p}}-\frac{1}{2}\left(\tilde{\omega}_{\text{\tiny NS,x}}^2+\tilde{\omega}_{\text{\tiny NS,y}}^2\right)\pm\left(\frac{1}{4}\left(\tilde{\omega}_{\text{\tiny NS,x}}^2+\tilde{\omega}_{\text{\tiny NS,y}}^2\right)^2+(\omega^2-\omega^2_{\text{\tiny p}})^2\frac{\tilde{\omega}_{\text{\tiny NS,z}}^2}{\omega^2}\right)^{1/2}\right)}\right)^{1/2}
          \;,\,\tilde{\omega}_{\text{\tiny NS,z}}>0\;,\\
          \left(1-\frac{\omega^2_{\text{\tiny p}} \left(\omega^2-\omega^2_{\text{\tiny p}}\right)}{\omega^2\left(\omega^2-\omega^2_{\text{\tiny p}}-\frac{1}{2}\left(\tilde{\omega}_{\text{\tiny NS,x}}^2+\tilde{\omega}_{\text{\tiny NS,y}}^2\right)\mp\left(\frac{1}{4}\left(\tilde{\omega}_{\text{\tiny NS,x}}^2+\tilde{\omega}_{\text{\tiny NS,y}}^2\right)^2+(\omega^2-\omega^2_{\text{\tiny p}})^2\frac{\tilde{\omega}_{\text{\tiny NS,z}}^2}{\omega^2}\right)^{1/2}\right)}\right)^{1/2}
          \;,\,\tilde{\omega}_{\text{\tiny NS,z}}<0\;.\\
      \end{cases}
  \end{equation}
  %%%%%%%%%%%%%%%
 Here, $\omega_{\text{\tiny p}}^2\equiv e^2\mathcal{N}_{\text{\tiny e}}(r)/(\epsilon_{0}m_{\text{\tiny e}})$ is the plasma frequency. 
    
    The spatial dependencies of these refractive indices are highly dependent upon the orientation of the magnetic dipole. For instance, we have found that when the magnetic moment is oriented along the direction of propagation (i.e. $(\theta_{\text{\tiny NS}},\phi_{\text{\tiny NS}})=(0,0)$), the refractive index of the left circular wave decreases while the refractive index of the right circular wave increases between the NS and observer. Moreover, in this case, the magnetic field is axially symmetric along the direction of propagation, and consequently, the refractive index variation also has this axial symmetry. When the magnetic moment of an NS is instead oriented perpendicular to the direction of propagation,  the refractive index of the left circular wave first increases and then decreases, between the NS and observer. In contrast, for this alignment, the refractive index of the right circular wave first decreases and then increases. At infinity, the refractive indices for both the left and right circular components match and acquire the value $\left(1-\omega^2_{\text{\tiny p}}/\omega^2\right)^{1/2}$.
    
    Now, for the NS case, we can calculate the change in polarization using Eq.\,\eqref{eq:fr_fru_cp_pol_ang}
     %%%%%%%%%%%%%%%
    \begin{equation}\label{eq:fr_ns_cp_pol_ang}
    	\psi_{\text{\tiny pol.}}^{\text{\tiny NS}}=\frac{\omega}{2 c}\int_{-d_{\text{\tiny QS}}}^{d_{\text{\tiny Q}}}~dz \left( n_{\text{\tiny L}}^{\text{\tiny NS}}(\vec{r})-n_{\text{\tiny R}}^{\text{\tiny NS}}(\vec{r})\right)\;.
    \end{equation}
    %%%%%%%%%%%%%%%
     Similarly, the RM value accrued due to an NS along the light's path will be given by 
      %%%%%%%%%%%%%%%
    \begin{equation}\label{eq:fr_ns_cp_RM_def}
    	\text{RM}^{\text{\tiny NS}}(\lambda)=\frac{d}{d\lambda^2}\left(\psi_{\text{\tiny pol.}}^{\text{\tiny NS}}\right)\;.
    \end{equation} 
    %%%%%%%%%%%%%%%
    For the reasons discussed earlier, we are working in a regime where Eq.\,\eqref{eq:fr_umf_arb_dir_Ex_Ey_cond} is satisfied for the NS parameters, i.e.
    %%%%%%%%%%%%%%%
  \begin{equation}\label{eq:fr_ns_Ex_Ey_cond}
      \abs{\frac{ \omega\left( \tilde{\omega}_{\text{\tiny NS,x}}^2 + \tilde{\omega}_{\text{\tiny NS,y}}^2 \right)}{2 (\omega^2-\omega^2_{\text{\tiny p}})\tilde{\omega}_{\text{\tiny NS,z}}}  }\ll1\;.
  \end{equation}
  %%%%%%%%%%%%%%%
    For a fixed $\mathfrak{m},\,\omega$ and $\omega_{\text{\tiny p}}$, this will help us identify the regions where we can consider the characteristic modes as left and right circularly polarized.  Hence, we are implicitly assuming the applicability of Eq.\,\eqref{eq:fr_ns_cp_pol_ang} in the regime permitted by the above equation in conjunction with  Eq.\,\eqref{eq:fr_umf_arb_dir_freq_lim}.
    
    If we assume a constant plasma density and use  Eq.\,\eqref{eq:fr_ns_mf_B_approx}, then Eq.\,\eqref{eq:fr_umf_arb_dir_freq_lim} gives us a radius cut-off, above which the modes are propagating. The result is 
    %%%%%%%%%%%%%%%
      \begin{equation}\label{eq:fr_ns_cut}
          r^{\text{\tiny NS}}_{\text{\tiny cut}} \sim \left(\frac{\mu_{0}   e \,\mathfrak{m}}{4 \pi  m_{\text{\tiny e}}}\frac{\omega}{ \left(\omega ^2-\omega_{\text{\tiny p}}^2\right)}\right)^{1/3}\;.
      \end{equation}
     %%%%%%%%%%%%%%%
     For a fixed $\mathfrak{m},\,\omega$ and $\omega_{\text{\tiny p}}$, we may calculate this distance explicitly. We may then utilize Eqs.\,\eqref{eq:fr_ns_ref_ind} and \,\eqref{eq:fr_ns_cp_pol_ang} in a bonafide way for the evaluation of the change in polarization. 
     
   Intriguingly, we will have potentially interesting observational signatures in the region $r\lesssim r^{\text{\tiny NS}}_{\text{\tiny cut}}$. Only the $(+)$ characteristic mode would be observed inside $r\lesssim r^{\text{\tiny NS}}_{\text{\tiny cut}}$, since the $(-)$ modes will be optically blocked. For the polarization angle and RM maps we will focus mainly on the larger $r \gtrsim r^{\text{\tiny NS}}_{\text{\tiny cut}}$ regions.
   
   Now, we will consider two semi-realistic plasma profiles to compute the Faraday rotations by an NS. We will examine both a constant plasma density and a more realistic galactic plasma density profile to evaluate this effect.
    %%%%%%%%%%%%%%%%
\begin{figure}
	\centering
		\includegraphics[scale=0.6]{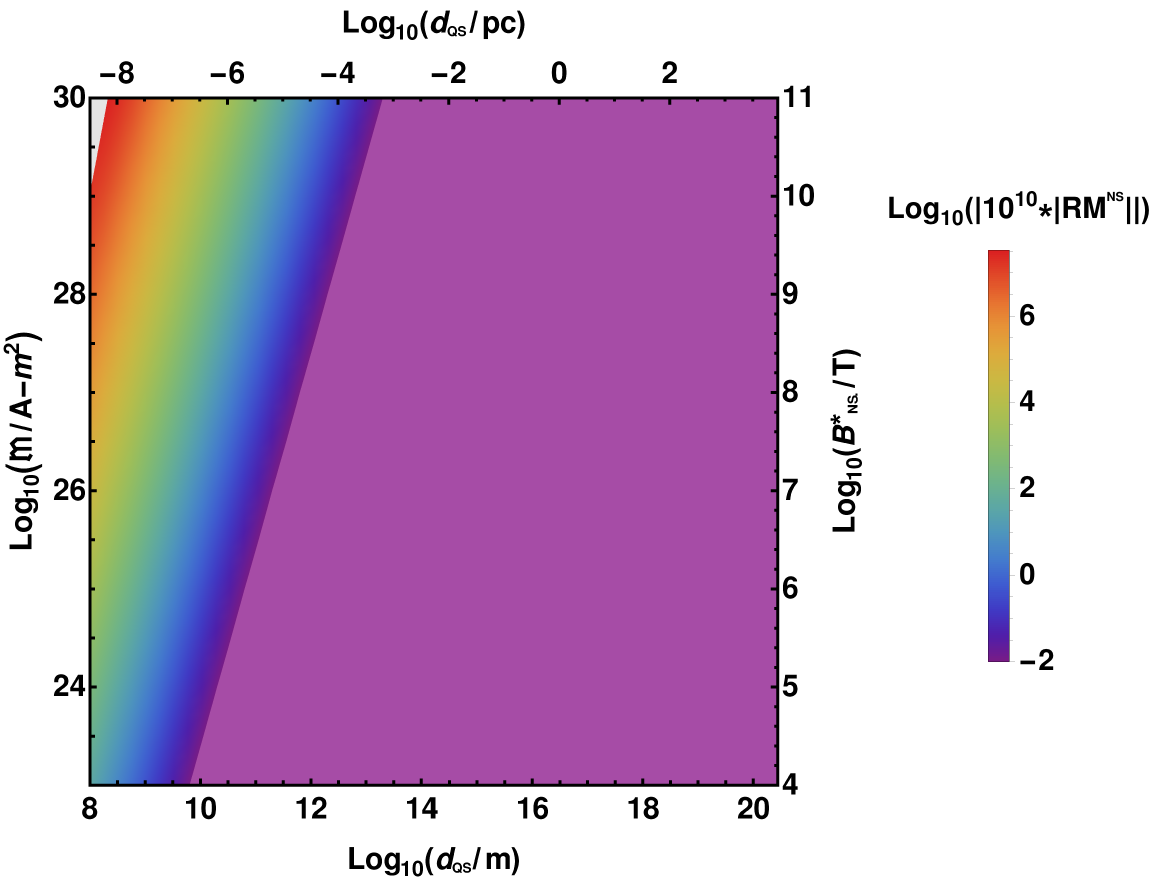}
  \includegraphics[scale=0.6]{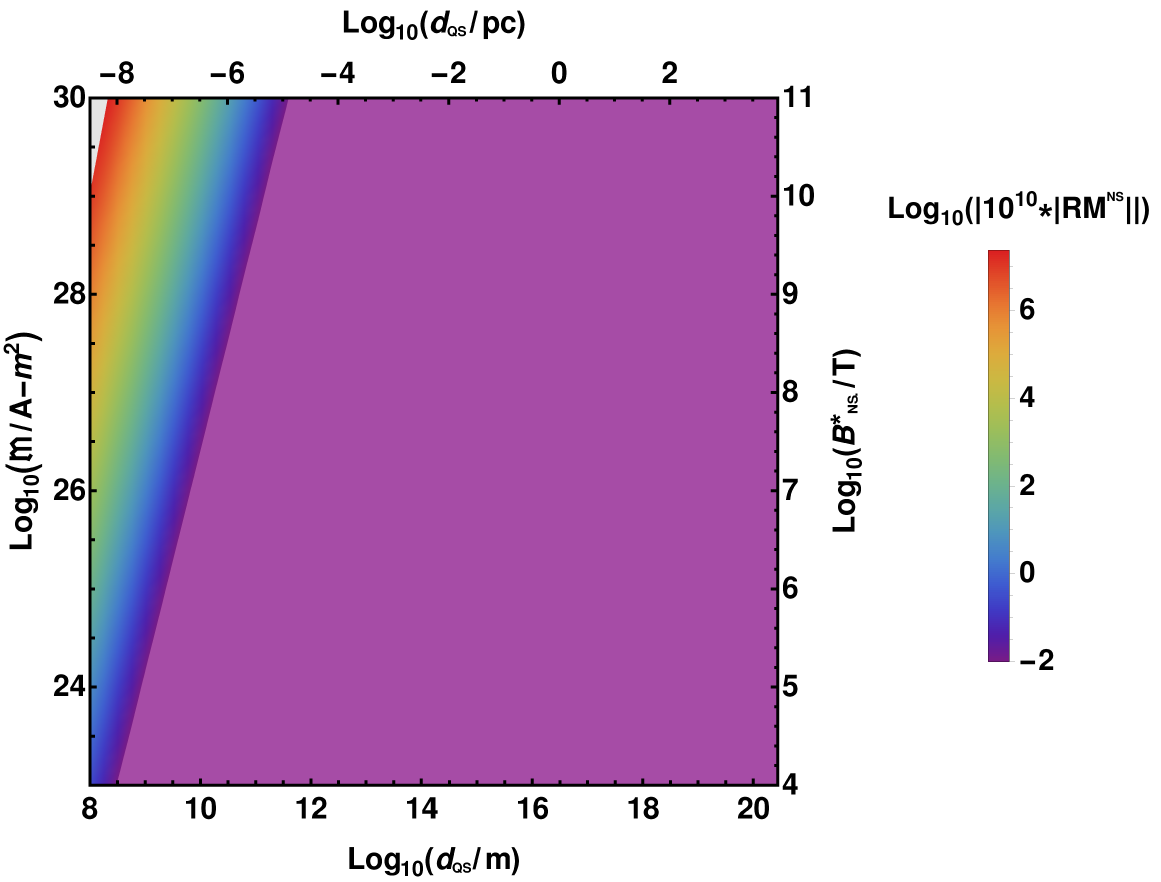}
	\caption{Rotation measure (RM) values due to an NS located inside the Milky Way galaxy as a function of the magnetic moment $\mathfrak{m}$ and the distance between the NS and the source ($d_{\text{\tiny QS}}$). The plasma density is assumed to be constant. The plots are for when the magnetic moment is parallel i.e., $(\theta_{\text{\tiny NS}},\phi_{\text{\tiny NS}})=(0,0)$ (top) and perpendicular i.e., $(\theta_{\text{\tiny NS}},\phi_{\text{\tiny NS}})=(\pi/2,0)$ (bottom). The alignments are with respect to the direction of wave propagation, i.e. $z$-axis.  The surface magnetic field ($B^*_{\text{\tiny NS}}$) of the NS is in the range $10^{4}-10^{11}$ T. Here, we have evaluated RM values at the impact parameter $\xi=r^{\text{\tiny NS}}_{\text{\tiny cut}}$ (See Eq.\,\eqref{eq:fr_ns_cut}), and taken the distance between the source and observer to be $d_{\text{\tiny S }} = 8.5 ~\text{kpc}$. The wavelength of light is taken to be $\lambda=1~ \text{m}$ and the MW electron number density is taken to be $N_{\text{\tiny e},0}^{\text{\tiny MW}}$ = 0.015 cm$^{-3}$\,\cite{Ocker:2020tnt}. The gray region depicts the parameter space for which $(-)$ modes are blocked. Notice that in the parameter space regions shown, the RM values are all too small for observation i.e. $\text{RM}^{\text{\tiny NS}}\ll0.01$.}
	\label{fig:gal_MW_RM_ex_QM_plot_ns}
\end{figure}
%%%%%%%%%%%%%%5

%%%%%%%%%%%%%%%%
\begin{figure}
	\centering
	\includegraphics[scale=0.5]{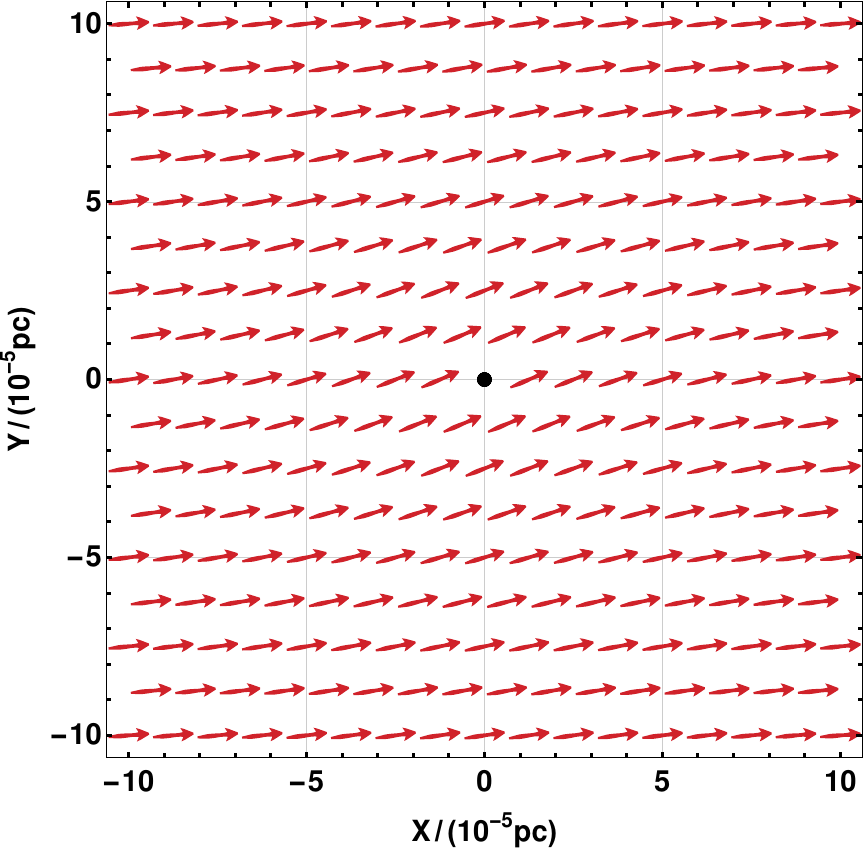}
	\includegraphics[scale=0.5]{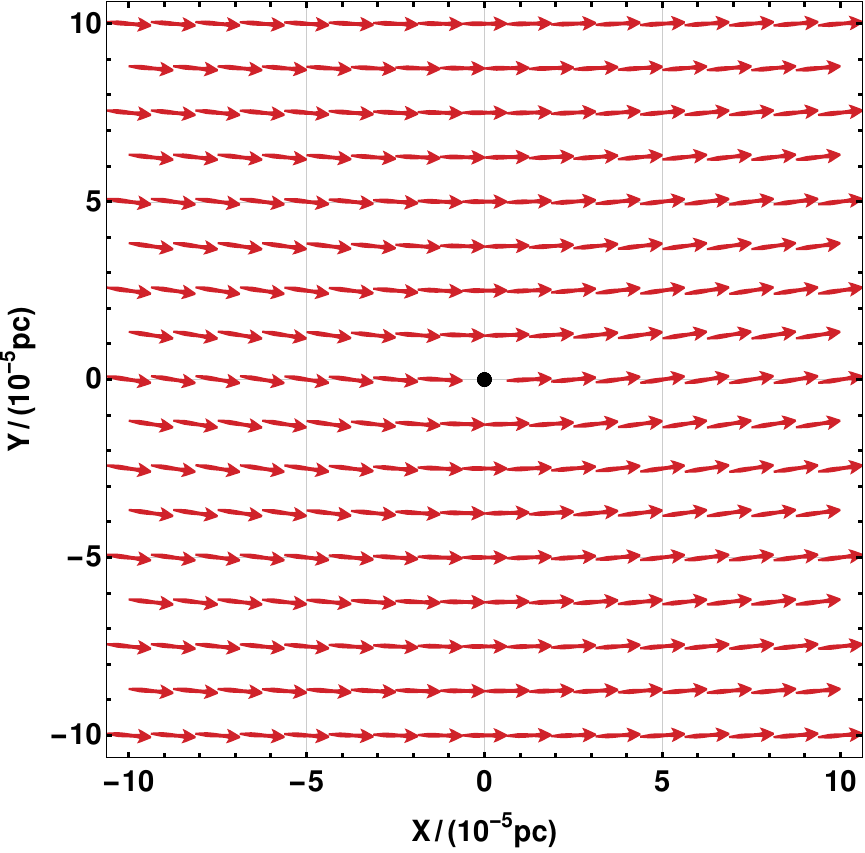}
	\caption{Polarization angle maps ($\psi_{\text{\tiny pol.}}^{\text{\tiny NS}}\times 10^{10}$) induced by an NS located inside the Milky Way galaxy. It is illustrated for linearly polarized electromagnetic waves from a uniform background source. The initial polarization is along the $\hat{x}$-axis. The plasma density is assumed to be constant. The plots are shown when the NS magnetic moments are parallel (left) and perpendicular (right) to the direction of wave propagation ($z$-axis; coming out of the plane in the figures above). Here, we have taken, $\mathfrak{m}=10^{30}~\text{A-m$^2$}$, or equivalently $B^*_{\text{\tiny NS}}=10^{11}~\text{T}$, $d_{\text{\tiny QS}}= 10^{-4}~ \text{pc}$, $d_{\text{\tiny S }} = 10~ \text{kpc}$, $\lambda=1 ~\text{m}$  and  $N_{\text{\tiny e},0}^{\text{\tiny MW}}$ = 0.015 cm$^{-3}$\,\cite{Ocker:2020tnt}. The black circle denotes an NS (exaggerated for visibility).}
	\label{fig:gal_MW_PA_ex_QM_plot_ns_1}
\end{figure}
%%%%%%%%%%%%%%%%

%%%%%%%%%%%%%%%%%%%%%%%%%%%%%%%%%%%%%%%%%%%%%%%%%%%%%%%%%%%%%%%%%%%%%
    \subsubsection{Constant density plasma}\label{subsec:fr_ns_cdp}
    As a simplistic model, we will first consider that the plasma is uniformly distributed inside the galaxy. This will help us explore the qualitative and quantitative features of the Faraday effect. 
    
    We assume the NS under consideration to be situated within our Milky Way galaxy (MW) and take the plasma density as the average Milky Way galactic plasma density. Thus, the distribution under consideration is
     %%%%%%%%%%%%%%%
     \begin{equation}\label{eq:fr_ns_plas_dens}
         N^{\text{\tiny MW}}_{\text{\tiny e}}(r)=
         \begin{cases}
             0\;; &r<R_{\text{\tiny NS}}\;, \\
              N_{\text{\tiny e},0}^{\text{\tiny MW}}\;;&r>R_{\text{\tiny NS}}\;.
         \end{cases}
     \end{equation}
    %%%%%%%%%%%%%%%
    Here, $R_{\text{\tiny NS}}$ is the radius of the NS and $N_{\text{\tiny e},0}^{\text{\tiny MW}}$ is the average galactic plasma density of the Milky Way. The plasma frequency in this scenario may be expressed as
    %%%%%%%%%%%%%%%
     \begin{equation}\label{eq:fr_ns_MW_plas_freq}
        \omega^2_{\text{\tiny p}}(r)=
         \begin{cases}
             0\;; & r<R_{\text{\tiny NS}}\;,\\
              \frac{e^2 N_{\text{\tiny e},0}^{\text{\tiny MW}}}{\epsilon_0m_{\text{\tiny e}}}\;; &r>R_{\text{\tiny NS}}\;.
         \end{cases}
     \end{equation}
    %%%%%%%%%%%%%%%
    
    A general, analytic expression for the change in polarization and RM cannot be obtained, even in the constant plasma density case. However, when  $\omega\gg \text{max}\left({ \omega_{\text{\tiny p}},\tilde{\omega}}\right)$, using Eqs.\,\eqref{eq:fr_ns_ref_ind}, and \eqref{eq:fr_ns_cp_pol_ang}, we get    
      %%%%%%%%%%%%%%%
  \begin{equation}\label{eq:fr_fru_ns_cp_pol_ang_approx_1}
      \psi_{\text{\tiny pol.}}^{\text{\tiny NS}}\simeq \int_{-d_\text{\tiny QS}}^{d_\text{\tiny Q}} dz~ \left(\frac{\omega^2_{\text{\tiny p}}}{2\omega^2 c }\tilde{\omega}_{\text{\tiny NS,z}}\right)+\mathcal{O}\left(\left(\frac{\omega^2_{\text{\tiny p}}}{\omega^2}\right)^2,\left(\frac{\tilde{\omega}_{\text{\tiny NS}}}{\omega}\right)^2\right)\;.
    \end{equation}
  %%%%%%%%%%%%%%%
  
    Using the  Eqs.\,\eqref{eq:fr_ns_cyc_freq} and \eqref{eq:fr_ns_MW_plas_freq}, the change in polarization angle can be written as
     %%%%%%%%%%%%%%%
    \begin{eqnarray}\label{eq:fr_ns_cp_pol_ang_approx} 
      \psi_{\text{\tiny pol.}}^{\text{\tiny NS}} &\simeq&-\frac{e^3\mathfrak{m} N_{\text{\tiny e},0}^{\text{\tiny MW}}\lambda^2}{32 \pi^3 \epsilon_0m^2_{\text{\tiny e}} c^3} \int_{-d_\text{\tiny QS}}^{d_\text{\tiny Q}} dz~ \left(\frac{3 \left(x\sin\theta_{\text{\tiny NS}}\cos\phi_{\text{\tiny NS}}+y\sin\theta_{\text{\tiny NS}}\sin\phi_{\text{\tiny NS}}+z\cos\theta_{\text{\tiny NS}}\right)z}{\left(\xi^2 +z^2\right)^{5/2}}\right.\nonumber \\
      &&\hquad \hquad \left.-\frac{\cos\theta_{\text{\tiny NS}}}{\left(\xi^2 +z^2\right)^{\frac{3}{2}}}\right)\;.,\nonumber \\
      &\simeq&-\frac{e^3\mathfrak{m} N_{\text{\tiny e},0}^{\text{\tiny MW}}\lambda^2}{32 \pi^3 \epsilon_0m^2_{\text{\tiny e}} c^3}\left(\frac{\sin \theta_{\text{\tiny NS}} (x \cos \phi_{\text{\tiny NS}}+y \sin \phi_{\text{\tiny NS}})+d_{\text{\tiny Q}} \cos \theta_{\text{\tiny NS}}}{\left(\xi ^2+d_{\text{\tiny Q}}^2\right)^{\frac{3}{2}}}\right.\nonumber\\
 		&&\hquad \hquad -\left.\frac{\sin \theta_{\text{\tiny NS}} (x \cos \phi_{\text{\tiny NS}}+y \sin \phi_{\text{\tiny NS}})+d_{\text{\tiny QS}} \cos \theta_{\text{\tiny NS}}}{\left(\xi ^2+d_{\text{\tiny QS}}^2\right)^{\frac{3}{2}}}\right)\;.
  \end{eqnarray}
  %%%%%%%%%%%%%%%
  
Using Eq.\,\eqref{eq:fr_fru_cp_RM_def}, we get the corresponding RM as 
    %%%%%%%%%%%%%%%
  \begin{eqnarray}\label{eq:fr_ns_cp_RM_approx}
      \text{RM}(\lambda)&\simeq&-\frac{e^3\mathfrak{m} N_{\text{\tiny e},0}^{\text{\tiny MW}}}{32 \pi^3 \epsilon_0m^2_{\text{\tiny e}} c^3}\left(\frac{\sin \theta_{\text{\tiny NS}} (x \cos \phi_{\text{\tiny NS}}+y \sin \phi_{\text{\tiny NS}})+d_{\text{\tiny Q}} \cos \theta_{\text{\tiny NS}}}{\left(\xi ^2+d_{\text{\tiny Q}}^2\right)^{\frac{3}{2}}}\right.\nonumber\\
 		&&\hquad \hquad -\left.\frac{\sin \theta_{\text{\tiny NS}} (x \cos \phi_{\text{\tiny NS}}+y \sin \phi_{\text{\tiny NS}})+d_{\text{\tiny QS}} \cos \theta_{\text{\tiny NS}}}{\left(\xi ^2+d_{\text{\tiny QS}}^2\right)^{\frac{3}{2}}}\right)\;.
  \end{eqnarray}
  %%%%%%%%%%%%%%%

   We note that in the high-frequency limit $\omega\gg \text{max}\left({ \omega_{\text{\tiny p}},\tilde{\omega}}\right)$, the change in polarization angle and RM exhibits a linear proportionality to the NS magnetic moment. One also notes in passing that since the effects are all proportional to  $1/m_e^{2}$, the impact of ions in the plasma is subdominant. Additionally, it's interesting that the RM is independent of the wavelength of light, at leading order. In the next-to-leading order term, $\lambda^2$ contributes, as elucidated in equation Eq.\,\eqref{eq:fr_fru_cp_RM_approx}. 
   
   It is also evident that the change in polarization and RM explicitly depend upon coordinates $x$ and $y$ generically. However, when the magnetic moment of the NS aligns with the direction of wave propagation, i.e. $ \theta_{\text{\tiny NS }}=0$, this specific dependency is eliminated. For this alignment, quantities depend upon the impact parameter $\xi$ as expected from the axial symmetry. Additionally, it is worth mentioning that the change in polarization angle and RM undergoes a sign reversal when we exchange $d_{\text{\tiny QS}}$ with $d_{\text{\tiny Q}}$. Consequently, when $d_{\text{\tiny Q}}\simeq d_{\text{\tiny QS}}$ (i.e. when the neutrons star is midway between source and observer), both the change in the polarization angle and the RM will become zero.
    
    We will employ Eqs.\,\eqref{eq:fr_GCDP_ns_dpmoment}, \eqref{eq:fr_ns_mf_B}, \eqref{eq:fr_ns_ref_ind}-\eqref{eq:fr_ns_cp_RM_def}, and \eqref{eq:fr_ns_MW_plas_freq} to numerically calculate the change in polarization and the RM due to an NS. 
    We have assumed the radius of the NS ($R_{\text{\tiny NS}}$) to be $10~\text{Km}$, and then evaluated the RM and the change in the polarisation angles for surface magnetic fields ($B^*_{\text{NS}}$) in the range $10^{4}-10^{11}~\text{T}$. The relation between the dipole moment and magnetic field is obtained using Eq.\,\eqref{eq:fr_ns_mf_B}. 
    
    In Fig.\,\ref{fig:gal_MW_RM_ex_QM_plot_ns}, we have presented a density plot illustrating the magnitude of the RM as a function of the NS magnetic dipole moment and the NS-source distance. The magnetic moment of the NS is aligned parallel (top) and perpendicular (bottom) to the wave propagation direction. The source-observer distance is taken to be $d_{\text{\tiny S }} = 8.5 ~\text{kpc}$ and the galactic electron number density as $N_{\text{\tiny e},0}^{\text{\tiny MW}}$ = 0.015 cm$^{-3}$\,\cite{Ocker:2020tnt}. We have calculated the RM at an impact parameter $\xi=r^{\text{\tiny NS}}_{\text{\tiny cut}}$ (using Eq.\,\eqref{eq:fr_ns_cut}), at which the RM value is maximum. From the figure, we conclude that the RM values for NS are typically small and do not generically reach observationally relevant values at present (i.e. $\text{RM}^{\text{\tiny NS}}\ll0.01$). 
     
    For both parallel and perpendicular orientations, it is evident that the RM values increase as we increase the dipole moment, as should be expected. Furthermore, increasing  $d_{\text{\tiny QS}}$ leads to a drastic decrease in the RM values because of the $1/r^3$ dependence of the magnetic field. Also, notice that one gets relatively higher RM values for the parallel case since the contribution of the $z$-component of the magnetic field is larger in this case.  It is also important to mention that inside the grey regions in Fig.\,\ref{fig:gal_MW_RM_ex_QM_plot_ns}, the cut-off condition specified by Eq.\,\eqref{eq:fr_ns_cut} is not satisfied and they are therefore excluded. 

    In Fig.\,\ref{fig:gal_MW_PA_ex_QM_plot_ns_1}, we present the polarization angle map (PA map). The uniform, extended, linearly polarised light source is in the plane perpendicular to the direction of wave propagation. It is located at a distance $d_{\text{\tiny QS}}= 10^{-4}~ \text{pc}$ from the NS. The source produces linearly polarized light, which is initially aligned along the $ x$-direction. The NS is assumed to have a magnetic dipole moment $\mathfrak{m}=10^{30}~\text{A-m$^2$}$ or equivalently a surface magnetic field  $B^*_{\text{\tiny NS}}=10^{11}~\text{T}$. The PA map shows a patch $10^{-5}\,\text{pc} \times 10^{-5}\,\text{pc}$ as seen by the observer---for both the parallel (left) and perpendicular (right) orientations of the NS. The actual values for the change in polarization angle are exceedingly small to be detected in current observations. In the case of parallel orientation, it is notable in the PA map that for fixed impact parameters $\xi$, the electric fields are unidirectional, rendering a uniform polarisation angle. This is because of the axial symmetry of the NS's magnetic field in this orientation.  However, in perpendicular orientation, it's significant to note that the PA map has mirror symmetry about the $x$-axis. This distinctive pattern is a consequence of the specific alignment of the magnetic field in this configuration. Additionally, as expected, we note that as we increase the impact parameter $\xi$, the change in the polarization angle diminishes due to the weakening of the magnetic field strength.

    We observe that the change in polarization values within the PA map lacks axial symmetry in the case of a general orientation of the NS. Hence, if we choose a circular contour of any given radius on the PA map and proceed to integrate the change in polarization values along this contour, the resulting value for the integral will satisfy a normalized inequality given by
    %%%%%%%%%%%%%%%%
    \begin{equation}\label{eq:fr_ns_cp_cont_ineq}
      \mathcal{M}^{\text{\tiny NS}} \equiv \abs{\frac{1}{ 2\pi \psi_{\text{\tiny pol.}}^{\text{\tiny NS}}(\phi_{0})}\left[\left(\oint_C d \phi\,\psi_{\text{\tiny pol.}}^{\text{\tiny NS}}(\phi)\right)- 2\pi \psi_{\text{\tiny pol.}}^{\text{\tiny NS}}(\phi_{0})\right]}\ge 0\;.
    \end{equation}
    %%%%%%%%%%%%%%%%
    Here, $C$ represents any closed circular contour centred around the NS and $\phi_{0}$ is an arbitrary azimuthal angle. The above inequality should be approximately satisfied in many realistic scenarios for the plasma density.
    
    As may be easily surmised, and as we shall see later, the right-hand side of the above equation evaluates to zero for a monopolar magnetic field. This simple observation is noteworthy, as it offers a potentially robust method to differentiate in PA maps conventional astrophysical magnetic field configurations---typically dipoles or higher multipoles---from MBHs, which have unique monopolar fields.

%%%%%%%%%%%%%%%%
\begin{figure}
	\centering
	\includegraphics[scale=0.6]{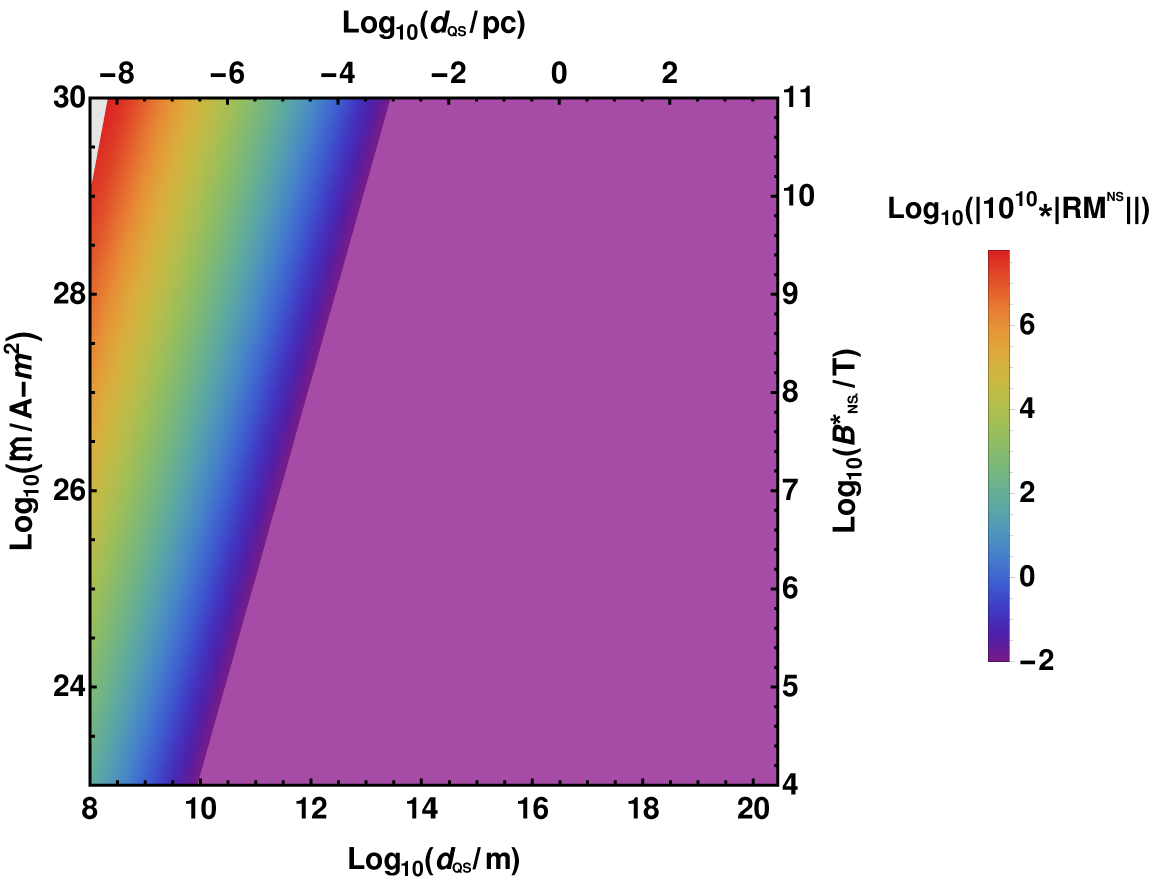}
	\includegraphics[scale=0.6]{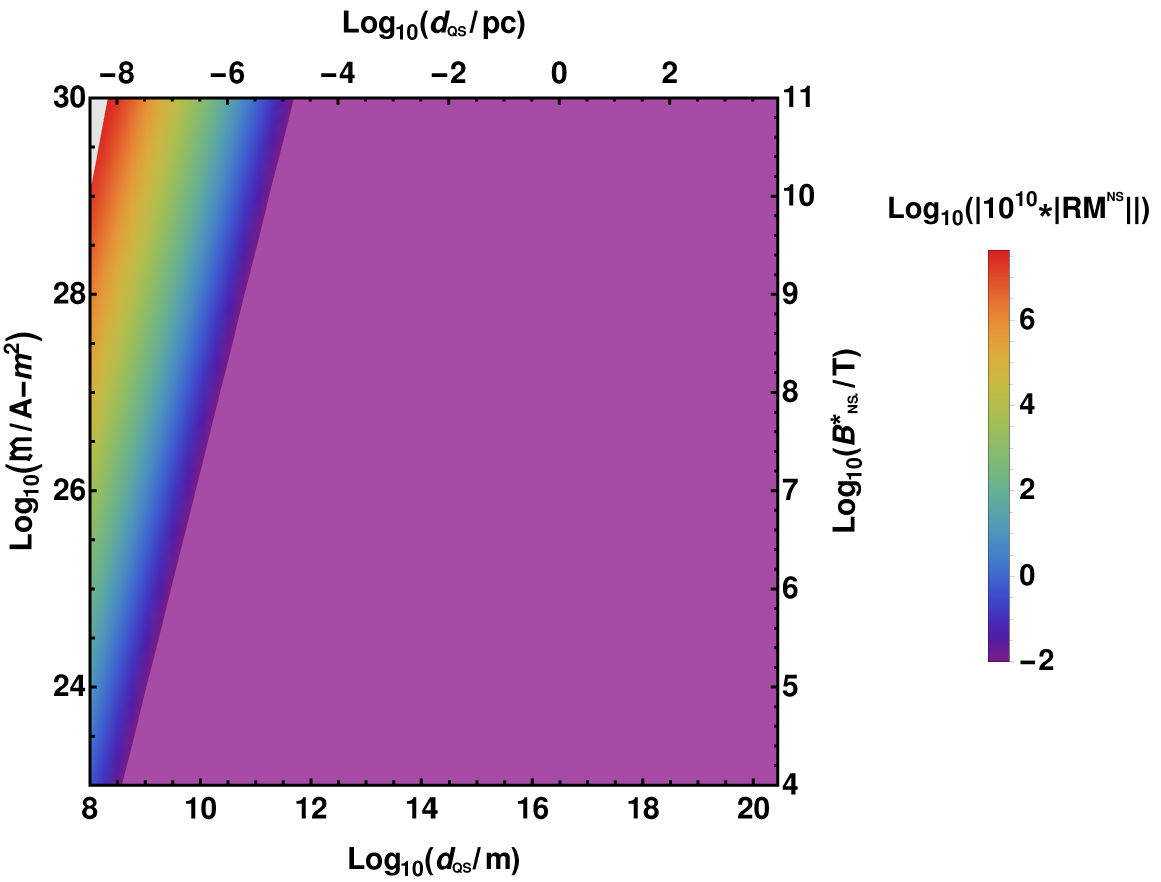}
	\caption{Rotation measure values due to an NS located inside the Milky Way galaxy as a function of $\mathfrak{m}$ and distance $d_{\text{\tiny QS}}$, for a galactic plasma distribution. Again, the plots are shown when the magnetic moment is parallel (top) and perpendicular (bottom) to the $z$-axis. As before, the NS surface magnetic fields are in the range $10^{4}-10^{11}$ T . We have taken $\xi=r^{\text{\tiny NS}}_{\text{\tiny cut}}$, $d_{\text{\tiny S }} = 8.5 ~\text{kpc}$, and $\lambda=1~ \text{m}$. For the galactic plasma density model parameters we take the values $N_{\text{\tiny e},1}^{\text{\tiny MW}}$ = 0.025 cm$^{-3}$, $N_{\text{\tiny e},2}^{\text{\tiny MW}}$ = 0.2 cm$^{-3}$, $A_{1}^{\text{\tiny MW}}$ = 20 kpc, and $A_{2}^{\text{\tiny MW}}$ = 2 kpc\,\cite{Cordes_1991}. The grey region again depicts the parameter space for which the $(-)$ modes are blocked. $\text{RM}^{\text{\tiny NS}}\ll0.01$ in all the regions, and hence observationally lackluster.}
	\label{fig:gal_MW_RM_gauss_ex_QM_plot_ns}
\end{figure}
%%%%%%%%%%%%%%5

%%%%%%%%%%%%%%%%
\begin{figure}
	\centering
	\includegraphics[scale=0.5]{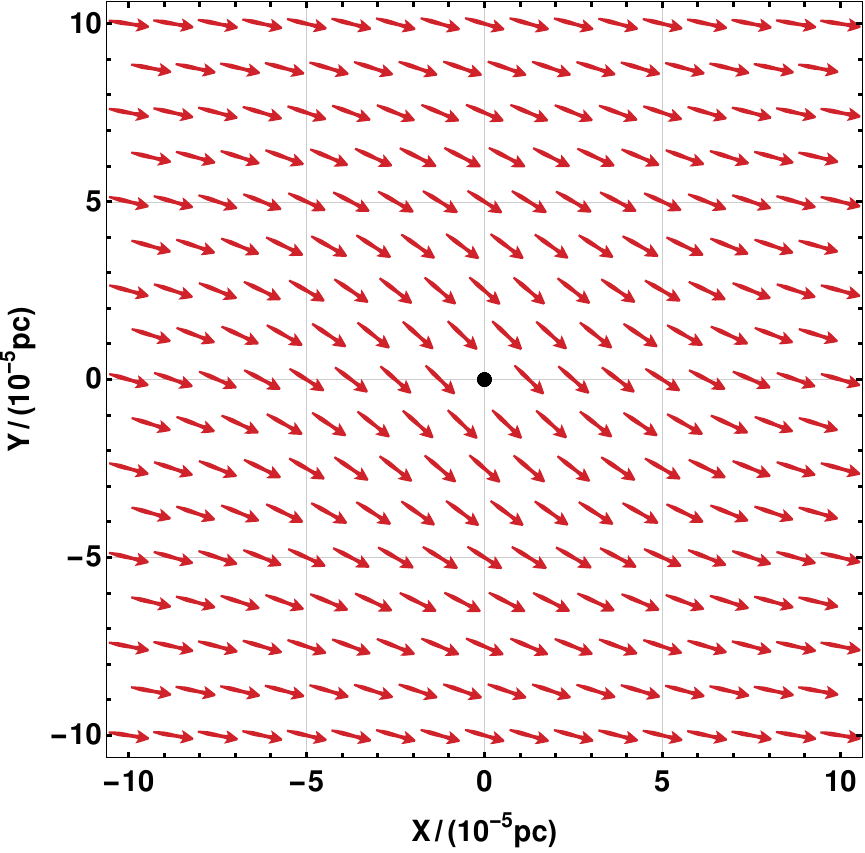}
	\includegraphics[scale=0.5]{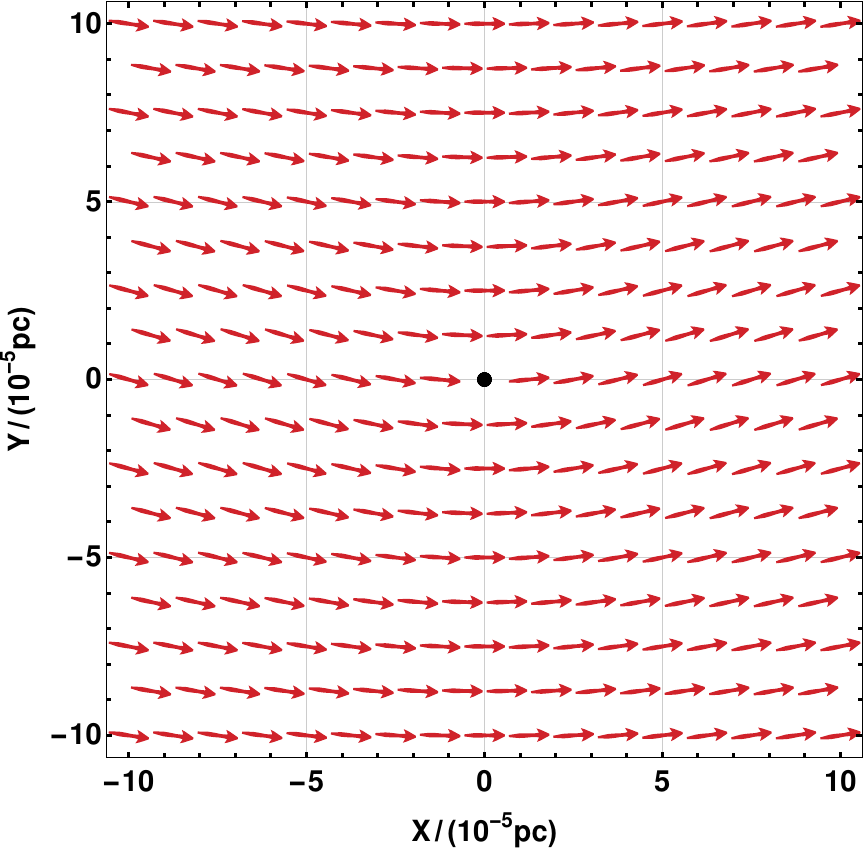}
	\caption{ NS polarization map for a galactic plasma distribution. Parallel (left) and perpendicular (right) NS orientations are shown. The parametric values chosen are similar to those in  Fig.\,\ref{fig:gal_MW_PA_ex_QM_plot_ns_1}.}
	\label{fig:gal_MW_PA_gauss_ex_QM_plot_ns_1}
\end{figure}
%%%%%%%%%%%%%%%%

%%%%%%%%%%%%%%%%%%%%%%%%%%%%%%%%%%%%%%%%%%%%%%
\subsubsection{Galactic distribution of plasma}
\label{subsec:fr_ns_gal}
%%%%%%%%%%%%%%%%
   After exploring the general characteristics of Faraday rotation by an NS, in the presence of a constant plasma density, we will shift to a slightly more realistic scenario. 
   
   We will adopt the galactic plasma distribution functions from\,\cite{Cordes_1991}, and again consider that the NS is situated within our Milky Way galaxy. We align the direction of light propagation with the galactic plane and assume that the source $S$  is located at the galactic centre. In this scenario, along the direction of wave propagation, the plasma density is assumed to follow the sum of two Gaussian profiles\,\cite{Cordes_1991}
     %%%%%%%%%%%%%%%
     \begin{equation}\label{eq:fr_ns_gal_plas_dens}
         \mathcal{N}^{\text{\tiny MW}}_{\text{\tiny e}}(r)=
         \begin{cases}
             0\;; & r<R_{\text{\tiny NS}}\;, \\
              N_{\text{\tiny e},1}^{\text{\tiny MW}}e^{-\left(\left(z-d_{\text{\tiny QS}}\right)/A_1^{\text{\tiny MW}}\right)^2}+N_{\text{\tiny e},2}^{\text{\tiny MW}}e^{-\left(\left(z-d_{\text{\tiny QS}}-2A_2^{\text{\tiny MW}}\right)/A_2^{\text{\tiny MW}}\right)^2}\;; & r>R_{\text{\tiny NS}}\;.
         \end{cases}
     \end{equation}
    %%%%%%%%%%%%%%%
    Here, $R_{\text{\tiny NS}}$ is the radius of the NS and $N_{\text{\tiny e},1}^{\text{\tiny MW}}$, $N_{\text{\tiny e},1}^{\text{\tiny MW}}$, $A_1^{\text{\tiny MW}}$, and $A_2^{\text{\tiny MW}}$ are model parameters. We will appropriate values for these parameters from\,\cite{Cordes_1991}. As before, we will again assume that $d_{\text{\tiny S}}\simeq d_{\text{\tiny Q}}\gg\xi $. In the above equation, the first term represents the thick disk component of the galactic distribution of electrons, while the second term represents the inner thin disk component. The associated plasma frequency is
    %%%%%%%%%%%%%%%
     \begin{equation}\label{eq:fr_ns_gal_MW_plas_freq}
        \omega^2_{\text{\tiny p}}(r)=
         \begin{cases}
             0\;; & r<R_{\text{\tiny NS}}\;, \\
              \frac{e^2 }{\epsilon_0m_{\text{\tiny e}}}\left(N_{\text{\tiny e},1}^{\text{\tiny MW}}e^{-\left(\left(z-d_{\text{\tiny QS}}\right)/A_1^{\text{\tiny MW}}\right)^2}+N_{\text{\tiny e},2}^{\text{\tiny MW}}e^{-\left(\left(z-d_{\text{\tiny QS}}-2A_2^{\text{\tiny MW}}\right)/A_2^{\text{\tiny MW}}\right)^2}\right)\;; & r>R_{\text{\tiny NS}}\;.
         \end{cases}
     \end{equation}
    %%%%%%%%%%%%%%%
        
     We will once again use  Eqs.\,\eqref{eq:fr_GCDP_ns_dpmoment}, \eqref{eq:fr_ns_mf_B}, \eqref{eq:fr_ns_ref_ind}-\eqref{eq:fr_ns_cp_RM_def}, and \eqref{eq:fr_ns_gal_MW_plas_freq} to numerically calculate the change in the polarization angle and the RM value in this case.

    We again assume $R_{\text{\tiny NS}}=10~\text{Km}$, and that the surface field $B^*_{\text{NS}}$ is in the range $10^{4}-10^{11}~\text{T}$. The RM density plot is presented in Fig.\,\ref{fig:gal_MW_RM_gauss_ex_QM_plot_ns}.  The magnetic moment of the NS is again assumed to be oriented parallel (top) or perpendicular (bottom) to wave propagation. As before, we have taken $d_{\text{\tiny S }} = 8.5 ~\text{kpc}$ and  $\xi=r^{\text{\tiny NS}}_{\text{\tiny cut}}$. The values of the galactic plasma density model parameters are taken to be $N_{\text{\tiny e},1}^{\text{\tiny MW}}$ = 0.025 cm$^{-3}$, $N_{\text{\tiny e},2}^{\text{\tiny MW}}$ = 0.2 cm$^{-3}$, $A_{1}^{\text{\tiny MW}}$ = 20 kpc, and $A_{2}^{\text{\tiny MW}}$ = 2 kpc\,\cite{Cordes_1991}. The RM values for the galactic plasma distribution case are slightly larger than the constant density case. Additionally, it is also found that the RM values exhibit very little variation, even when the source is located farther from the galactic centre. The RM values are exceedingly small, $\text{RM}^{\text{\tiny NS}}\ll0.01$, throughout.

    Fig.\,\ref{fig:gal_MW_PA_gauss_ex_QM_plot_ns_1} depicts the PA map in this scenario for $B^*_{\text{\tiny NS}}=10^{11}~\text{T}$ and $d_{\text{\tiny QS}}=10^{-4} ~\text{pc}$. The initial polarization direction for the wave is assumed to be along the $x$-direction. Similar to the PA maps in Sec.\,\ref{subsec:fr_ns_cdp}, we have displayed  $10^{-5}\,\text{pc} \times 10^{-5}\,\text{pc}$  patches as seen by the observer; for the parallel (left) and perpendicular (right) orientation of the magnetic moment of an NS. Although slightly higher than the values in the constant density case, the polarization angle values still remain exceedingly small for current observational capabilities. 
             
   Again, in the case of parallel orientation, we observe that the electric fields are consistently aligned for fixed impact parameter $\xi$. This leads to a uniform polarization angle on any circle centered at NS. In the case of a perpendicular orientation, the polarization angle map once more exhibits mirror symmetry with respect to the $x$-axis. Again, the inequality in Eq.\,\eqref{eq:fr_ns_cp_cont_ineq} is satisfied for the galactic plasma density profile.

   It is noteworthy that, for the impact parameters of interest, i.e., $\xi \lesssim \mathcal{O}(1)\,\text{pc}$, variation in the plasma density perpendicular to the galactic plane are expected to be minimal\,\cite{Cordes_1991}. However, any significant non-uniformity in the plasma density may disrupt the aforementioned characteristics of the PA map as well as the integral measure $\mathcal{M}^{\text{\tiny NS}}$. 
    
    Having gained some insight into the Faraday rotation characteristics of NS, let us now explore the Faraday effects induced by an MBH. This will help us contrast the two cases.

  %%%%%%%%%%%%%%%%%%%%%%%%%%%%%%%%%%%%%%%%%%%%%%%%%%%%%%%%%%%%%%%%%%%%
	\subsection{Faraday rotations due to a primordial magnetic black hole}
        \label{sec:fr_bh}
 %%%%%%%%%%%%%%%%%%%%%%%%%%%%%%%%%%%%%%%
 In this section, we will explore the Faraday effect produced by MBHs and make a comparative analysis with the NS case. 
 
\begin{figure}
	 	\centering
	 	\includegraphics[scale=1]{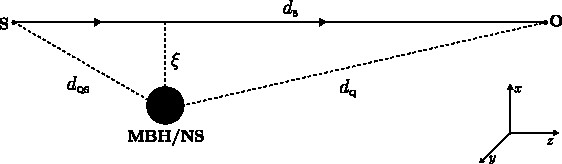}
	 	\caption[]{Schematic for the MBH Faraday rotation analysis. The MBH with magnetic monopole charge $Q_{\text{\tiny BH}}$ is taken as the origin. It is located at a distance $d_{\text{\tiny Q}}$ from an observer (O). The distance between the MBH and the source (S) is taken as $d_{\text{\tiny QS}}$. $d_{\text{\tiny S}}$ is the source-observer distance. The light rays are moving along the $z$ direction, with impact parameter $\xi=\sqrt{x^2+y^2}$. }
	 	\label{fig:Faraday_effect_Schematics}
	 \end{figure}

 Consider the schematics presented in Fig.\,\ref{fig:Faraday_effect_Schematics}. The origin is taken at the MBH, and an observer (O) is located at a distance $d_{\text{\tiny Q}}$ from it. The MBH has a magnetic charge $Q_{\text{\tiny BH}}$ and mass $M_{\text{\tiny BH}}$. Similar to the case of an NS in Sec.\,\ref{sec:fr_ns}, a source (S) located at a distance $d_{\text{\tiny QS}}$ from the MBH is emitting plane-polarized electromagnetic waves. Furthermore, $d_{\text{\tiny S}}$ denotes the distance between the source (S) and the observer, and $\xi$ denotes the distance of closest approach of the waves with the MBH. In this setup, as before, we assume $d_{\text{\tiny S}}\simeq d_{\text{\tiny Q}}\gg\xi$ and write $d_{\text{\tiny QS}}\simeq d_{\text{\tiny S}}-d_{\text{\tiny Q}}$. The electron density profile due to the hydrogen plasma background is denoted $N_\text{\tiny e}(\Vec{r})$.
  %%%%%%%%%%%%%%%
  %%%%%%%%%%%%%%%
  
  The magnetic field due to  an MBH, with a magnetic charge  $Q_{\text{\tiny BH}}$, at an arbitrary point A with spherical coordinates $(r,\theta,\phi)$ is given by
     %%%%%%%%%%%%%%%
  	\begin{equation}\label{eq:fr_mbh_mf_B}
		\vec{B}_{\text{\tiny BH}}(\vec{r})\equiv\frac{\mu_0}{4 \pi}\frac{Q_{\text{\tiny BH}}}{r^2}\hat{r}=\frac{\mu_0}{4 \pi}\frac{Q_{\text{\tiny BH}}}{r^2}\left(\sin\theta\cos\phi~\hat{x}+\sin\theta\sin\phi~\hat{x}+\cos\theta~\hat{z}\right)\;.
	\end{equation}
  %%%%%%%%%%%%%%%
  The cyclotron frequency of electrons at radial distance $r$ is given by 
    %%%%%%%%%%%%%%%
  	\begin{equation}\label{eq:fr_mbh_cyc_freq}
		\tilde{\omega}_{\text{\tiny BH}}(r)\equiv \frac{e}{m_{\text{\tiny e}}}\abs{\vec{B}_{\text{\tiny BH}}(r)}=\frac{\mu_0}{4 \pi}\frac{eQ_{\text{\tiny BH}}}{m_{\text{\tiny e}}r^2}\;.
	\end{equation}
  %%%%%%%%%%%%%%%
  We also denote $\{\tilde{\omega}_{\text{\tiny BH,x}},\tilde{\omega}_{\text{\tiny BH,y}},\tilde{\omega}_{\text{\tiny BH,z}}\}\equiv \tilde{\omega}_{\text{\tiny BH}}(r)\{\sin\theta\cos\phi,\sin\theta\sin\phi,\cos\theta\}$. 
  
  Now, as before, the refractive indices for the left and the right circularly polarized light at position $\vec{r}$ are obtained by using Eqs.\,\eqref{eq:fr_umf_arb_dir_ref_ind}, \eqref{eq:fr_umf_arb_dir_ref_ind_fin} and \eqref{eq:fr_mbh_cyc_freq}. They have the form
   %%%%%%%%%%%%%%%
  \begin{equation}\label{eq:fr_mbh_ref_ind}
     n_{\text{\tiny L,(R)}}^{\text{\tiny BH}}=\begin{cases}
          \left(1-\frac{\omega^2_{\text{\tiny p}} \left(\omega^2-\omega^2_{\text{\tiny p}}\right)}{\omega^2\left(\omega^2-\omega^2_{\text{\tiny p}}-\frac{1}{2}\tilde{\omega}_{\text{\tiny BH}}^2\sin^2\theta\pm\left(\frac{1}{4}\tilde{\omega}_{\text{\tiny BH}}^4\sin^4\theta+(\omega^2-\omega^2_{\text{\tiny p}})^2\frac{\tilde{\omega}_{\text{\tiny BH}}^2}{\omega^2}\cos^2\theta\right)^{1/2}\right)}\right)^{1/2}\;, \quad0<\theta<\pi/2\;,\\
          \left(1-\frac{\omega^2_{\text{\tiny p}} \left(\omega^2-\omega^2_{\text{\tiny p}}\right)}{\omega^2\left(\omega^2-\omega^2_{\text{\tiny p}}-\frac{1}{2}\tilde{\omega}_{\text{\tiny BH}}^2\sin^2\theta\mp\left(\frac{1}{4}\tilde{\omega}_{\text{\tiny BH}}^4\sin^4\theta+(\omega^2-\omega^2_{\text{\tiny p}})^2\frac{\tilde{\omega}_{\text{\tiny BH}}^2}{\omega^2}\cos^2\theta\right)^{1/2}\right)}\right)^{1/2}\;, \quad \pi/2 <\theta<\pi \;.\\
      \end{cases}
  \end{equation}
  %%%%%%%%%%%%%%%
    Here, $\omega_{\text{\tiny p}}^2\equiv e^2\mathcal{N}_{\text{\tiny e}}(r)/(\epsilon_{0}m_{\text{\tiny e}})$ is the plasma frequency. Notice that the refractive indices are independent of the angle $\phi$ and, therefore, have axial symmetry about $z$-axis. As we shall see later, this symmetry will be retained in the polarization angles and the RM values. 
    
    For $0<\theta<\pi/2$ the refractive index of the left circular wave is greater than the right circular wave, while for $\pi/2<\theta<\pi$ the refractive index of the left circular wave is less than the right circular waves. Furthermore, as a wave propagates away from the black hole, the refractive index of the left circular wave decreases, whereas the refractive index of the right circular wave increases.    
   
    The change in polarization due to the MBH is
    %%%%%%%%%%%%%%%
     \begin{equation}\label{eq:fr_mbh_cp_pol_ang}
         \psi_{\text{\tiny pol.}}^{\text{\tiny BH}}=\frac{\omega}{2 c}\int_{-d_{\text{\tiny QS}}}^{d_{\text{\tiny Q}}}~dz \left( n_{\text{\tiny L}}^{\text{\tiny BH}}(\vec{r})-n_{\text{\tiny R}}^{\text{\tiny BH}}(\vec{r})\right)\;,
     \end{equation}
     %%%%%%%%%%%%%%%
     and the RM value is given by
      %%%%%%%%%%%%%%%
     \begin{equation}\label{eq:fr_mbh_cp_RM_def}
         \text{RM}^{\text{\tiny BH}}(\lambda)=\frac{d}{d\lambda^2}\left(\psi_{\text{\tiny pol.}}^{\text{\tiny BH}}\right)\;.
     \end{equation}
    %%%%%%%%%%%%%%%
    
    As discussed, Eqs.\,\eqref{eq:fr_mbh_cp_pol_ang} and \eqref{eq:fr_mbh_cp_RM_def} are valid only when Eqs.\,\eqref{eq:fr_umf_arb_dir_freq_lim} and \eqref{eq:fr_umf_arb_dir_Ex_Ey_cond}  are satisfied. Assuming a constant plasma density and using the expression of the cyclotron frequency given by Eq.\,\eqref{eq:fr_mbh_cyc_freq}, Eq.\,\eqref{eq:fr_umf_arb_dir_Ex_Ey_cond} leads to
    %%%%%%%%%%%%%%%
      \begin{equation}\label{eq:fr_rcut}
          r\gtrsim\sqrt{\frac{\mu_{0}   e \abs{Q_{\text{\tiny BH}}}}{4 \pi  m_{\text{\tiny e}}}\frac{\omega\sin\theta \tan\theta}{ 2\left(\omega ^2-\omega_{\text{\tiny p}}^2\right)} }\;.
      \end{equation}
     %%%%%%%%%%%%%%%
     Again, for a fixed $Q_{\text{\tiny BH}},\,\omega$ and $\omega_{\text{\tiny p}}$, this will determine the region in which the characteristic modes are approximately the left and the right circular modes.
    Additionally, using Eqs.\,\eqref{eq:fr_umf_arb_dir_freq_lim} and \eqref{eq:fr_mbh_cyc_freq}, we get a cut-off
    %%%%%%%%%%%%%%%
      \begin{equation}\label{eq:fr_mbh_cut}
        r^{\text{\tiny BH}}_{\text{\tiny cut}} \simeq \sqrt{\frac{\mu_{0}   e Q_{\text{\tiny BH}}}{4 \pi  m_{\text{\tiny e}}}\frac{\omega}{ \left(\omega ^2-\omega_{\text{\tiny p}}^2\right)}}\; ,
      \end{equation}
     %%%%%%%%%%%%%%%
      where both modes are propagating. This will help us readily use Eqs.\,\eqref{eq:fr_mbh_ref_ind} and \,\eqref{eq:fr_mbh_cp_pol_ang} to calculate the change in polarization. 
      
      It is interesting to note that for a given $Q_{\text{\tiny BH}},\,\omega$ and $\omega_{\text{\tiny p}}$, the region $r< r^{\text{\tiny BH}}_{\text{\tiny cut}}$ is opaque to the observer. This is because a left circularly polarised light component entering this region is completely blocked, while after the light crosses an MBH, the right circular component is also absorbed; the latter is due to the reversal in modes given by Eq.\,\eqref{eq:fr_umf_arb_dir_ref_ind_fin}. Therefore, we will only calculate the change in polarization angle and RM at distances $r \gtrsim r^{\text{\tiny BH}}_{\text{\tiny cut}}$. 
      
      Let us now proceed to evaluate the RM and the polarization angle change for the same density profiles discussed in Sec.\,\ref{sec:fr_ns}, but now for an MBH.
      
     %%%%%%%%%%%%%%%			
     \begin{figure}
     	\centering
     	\includegraphics[scale=0.65]{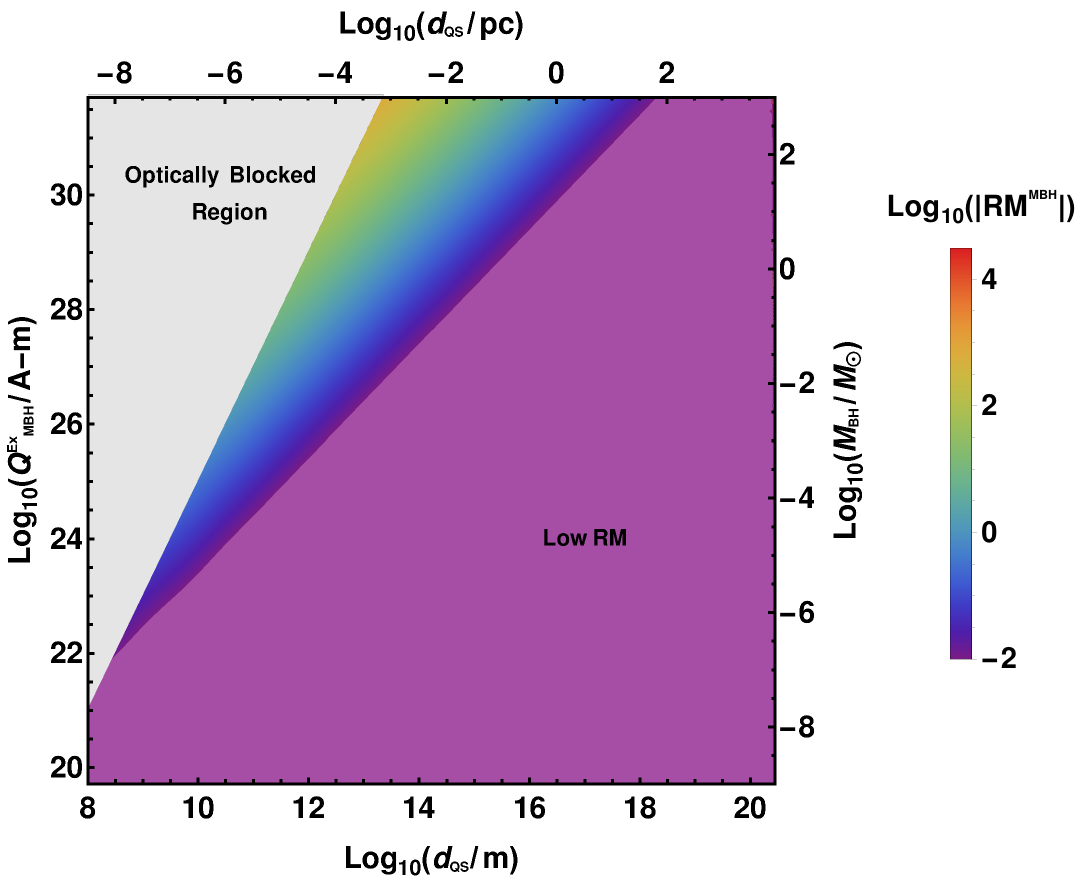}
     	\caption{Rotation measure (RM) is shown for an extremal MBH as a function of the magnetic charge of the black hole ($Q{\text{\tiny BH}}$) and the distance between the black hole and the source ($d_{\text{\tiny QS}}$). This density plot is for the case of constant plasma density. Here, we have taken the impact parameter $\xi=r^{\text{\tiny BH}}_{\text{\tiny cut}}$ (See Eq.\eqref{eq:fr_mbh_cut}), the distance between the source and observer $d_{\text{\tiny S }} = 8.5 ~\text{kpc}$, the wavelength of light $\lambda=1~ \text{m}$ and the electron number density $N_{\text{\tiny e},0}^{\text{\tiny MW}}$ = 0.015 cm$^{-3}$\,\cite{Ocker:2020tnt}. Notice that for $Q^{\text{\tiny Ex.}}_{\text{\tiny BH}}\gtrsim10^{22}~ \text{A-m}$, we get observable RM values i.e. $ \text{RM}^{\text{\tiny BH}}\gtrsim 0.01 $. In the above plot, the grey region depicts the parameter space for which the waves are optically blocked, and the light violet region depicts the region for which the RM values are too small for current observations.    }
     	\label{fig:gal_MW_RM_ex_QM_plot_mp}
     \end{figure}
      %%%%%%%%%%%%%%%
    %%%%%%%%%%%%%%%	
	 \begin{figure}
		\centering
            \includegraphics[scale=0.5]{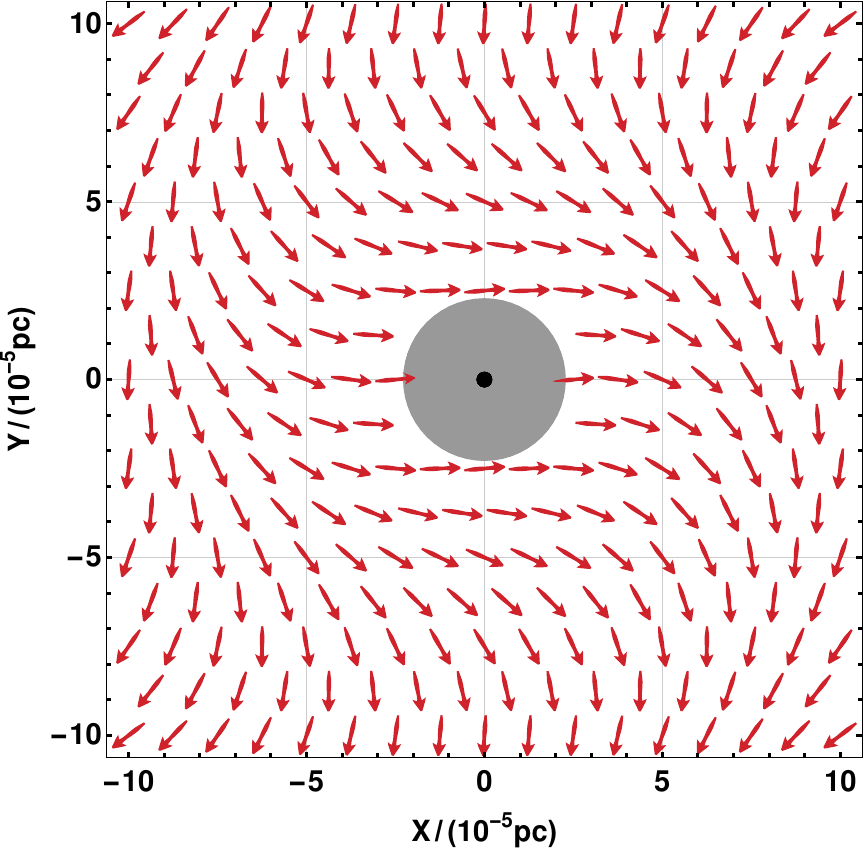}
		\includegraphics[scale=0.5]{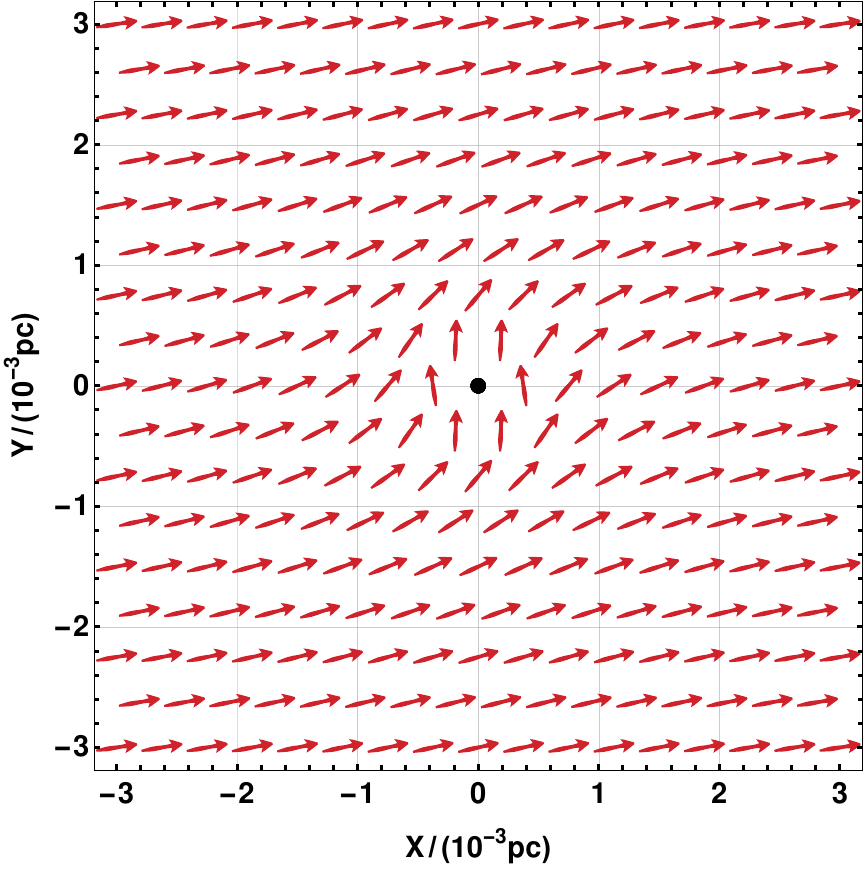}
		\caption{Polarization maps ($\psi_{\text{\tiny pol.}}^{\text{\tiny BH}}$) of linearly polarized electromagnetic waves from a uniform, extended background source, due to an extremal MBH. Constant plasma density is being assumed. As before, the initial polarization is along the $\hat{x}$-axis. The patches are $10^{-5}\,\text{pc} \times 10^{-5}\,\text{pc}$ (left) and $10^{-3}\,\text{pc} \times 10^{-3} \,\text{pc}$ (right). We have taken, $M_{\text{\tiny BH}}= 1~\solarmass$ or equivalently $Q_{\text{\tiny BH}}\simeq 0.5 \times 10^{29}~\text{A-m}$, $d_{\text{\tiny QS}}= 10^{-4}~ \text{pc}$, $d_{\text{\tiny S }} = 10~ \text{kpc}$, $\lambda=1 ~\text{m}$  and  $N_{\text{\tiny e},0}^{\text{\tiny MW}}$ = 0.015 cm$^{-3}$\,\cite{Ocker:2020tnt}. The black circle denotes an MBH (not to scale), and the grey region depicts the parameter space for which the waves are optically blocked.  }
			\label{fig:gal_MW_PA_ex_QM_plot_mp_1}
	\end{figure}
	%%%%%%%%%%%%%%%	

     %%%%%%%%%%%%%%%%%%%%%%%%%%%%%%%%%%%%%%%%%%%%%%%%%%%%%%%
    \subsubsection{Constant density plasma }
    \label{subsec:fr_bh_cdp}
     %%%%%%%%%%%%%%%
    As considered in Sec.\,\ref{subsec:fr_ns_cdp}, let us begin with a uniform plasma distribution within the galaxy. The MBH is located inside the Milky Way (MW) galaxy. The plasma density and the plasma frequency are taken from Eqs.\,\eqref{eq:fr_ns_plas_dens} and \eqref{eq:fr_ns_MW_plas_freq}, with $R_{\text{\tiny NS}}$ now replaced by $R_{\text{\tiny hor}}$---where, $R_{\text{\tiny hor}}\equiv G/c^2(M_{\text{\tiny BH}}+\sqrt{M_{\text{\tiny BH}}^2+\mu_0Q_{\text{\tiny BH}}^2/(4 \pi G)})$ is the outer horizon of the MBH. 

    A general theoretical expression for the change in polarization and RM cannot be obtained. Nevertheless, in the limit when  $\omega\gg \text{max}\left({ \omega_{\text{\tiny p}},\tilde{\omega}}\right)$, using Eqs.\,\eqref{eq:fr_ns_ref_ind} and \eqref{eq:fr_mbh_cp_pol_ang}, we obtain    
      %%%%%%%%%%%%%%%
  \begin{equation}\label{eq:fr_fru_cp_pol_ang_approx_2}
      \psi_{\text{\tiny pol.}}^{\text{\tiny BH}}\simeq \int_{-d_\text{\tiny QS}}^{d_\text{\tiny Q}} dz~ \left(\frac{\omega^2_{\text{\tiny p}}}{2\omega^2 c }\tilde{\omega}_{\text{\tiny BH}}\cos\theta\right)+\mathcal{O}\left(\left(\frac{\omega^2_{\text{\tiny p}}}{\omega^2}\right)^2,\left(\frac{\tilde{\omega}_{\text{\tiny BH}}}{\omega}\right)^2\right)\;.
    \end{equation}
  %%%%%%%%%%%%%%%
    Using Eqs.\,\eqref{eq:fr_ns_MW_plas_freq} and \eqref{eq:fr_mbh_cyc_freq}, the change in polarization angle can be written as
     %%%%%%%%%%%%%%%
    \begin{eqnarray}\label{eq:fr_mbh_cp_pol_ang_approx} 
      \psi_{\text{\tiny pol.}}^{\text{\tiny BH}} &\simeq&\frac{e^3Q_{\text{\tiny BH}} N_{\text{\tiny e},0}^{\text{\tiny MW}}\lambda^2}{32 \pi^3 \epsilon_0m^2_{\text{\tiny e}} c^3} \int_{-d_\text{\tiny QS}}^{d_\text{\tiny Q}} dz~ \frac{z}{\left(\xi^2 +z^2\right)^{\frac{3}{2}}}\;,\nonumber \\
      &\simeq&-\frac{e^3Q_{\text{\tiny BH}} N_{\text{\tiny e},0}^{\text{\tiny MW}}\lambda^2}{32 \pi^3 \epsilon_0m^2_{\text{\tiny e}} c^3}\left(\frac{1}{\left(\xi^2 +d_{\text{\tiny Q}}^2\right)^{1/2}}-\frac{1}{\left(\xi^2 +d_{\text{\tiny QS}}^2\right)^{1/2}}\right)\;.
  \end{eqnarray}
  %%%%%%%%%%%%%%%
  Finally, using Eq.\,\eqref{eq:fr_fru_cp_RM_def}, we get for the RM 
    %%%%%%%%%%%%%%%
  \begin{eqnarray}\label{eq:fr_mbh_cp_RM_approx}
      \text{RM}(\lambda)&\simeq&-\frac{e^3Q_{\text{\tiny BH}} N_{\text{\tiny e},0}^{\text{\tiny MW}}}{32 \pi^3 \epsilon_0m^2_{\text{\tiny e}} c^3}\left(\frac{1}{\left(\xi^2 +d_{\text{\tiny Q}}^2\right)^{1/2}}-\frac{1}{\left(\xi^2 +d_{\text{\tiny QS}}^2\right)^{1/2}}\right)\;.
  \end{eqnarray}
  %%%%%%%%%%%%%%%
  
    The above expressions represent the leading order contributions to the MBH Faraday rotation. In this limit, the change in polarization angles and RM are directly proportional to the charge of the black hole and independent of its mass. Furthermore, the RM value to leading order is independent of $\lambda$. The next-to-leading order term will include $\lambda^2$ as shown in Eq.\,\eqref{eq:fr_fru_cp_RM_approx}. Notice again that the change in polarization angle and RM value flips sign when we exchange $d_{\text{\tiny QS}}$ with $d_{\text{\tiny Q}}$. This happens because the magnetic field of the MBH is spherically symmetric, as a result of which when $d_{\text{\tiny Q}}\simeq d_{\text{\tiny QS}}$, the change in polarization angle and RM value will become zero.
    
    Now, using Eqs.\,\eqref{eq:fr_ns_MW_plas_freq}, \eqref{eq:fr_mbh_mf_B}, and \eqref{eq:fr_mbh_ref_ind}-\eqref{eq:fr_mbh_cp_RM_def}, let us numerically calculate the change in the polarization angle and the RM value due to an MBH, assuming a constant plasma density background. 
    
    In Fig.\,\ref{fig:gal_MW_RM_ex_QM_plot_mp}, we show the density plot for the RM as a function of the magnetic charge of the extremal MBH and the distance between the black hole and the source. The black hole is assumed to be located inside the Milky Way galaxy, and we have taken $d_{\text{\tiny S }} = 8.5 ~\text{kpc}$, $N_{\text{\tiny e},0}^{\text{\tiny MW}}$ = 0.015 cm$^{-3}$\,\cite{Ocker:2020tnt}, $\xi=r^{\text{\tiny BH}}_{\text{\tiny cut}}$, and the wavelength of light is taken to be $1~ \text{m}$. It is found that for $Q_{\text{\tiny BH}}\gtrsim10^{24}~\text{A-m}$ ($M_{\text{\tiny BH}}\gtrsim10^{-5} ~\solarmass$), the values of the RM are potentially sufficient for current observations, i.e. $ \text{RM}^{\text{\tiny BH}}\gtrsim 0.01 $. As expected, it can be seen that the RM values get larger with increase in the charge of the black hole. Furthermore, for a fixed source position, the value of the RM decreases as the distance between the source and the black hole ($d_{\text{\tiny QS}}$) increases. This happens till $d_{\text{\tiny QS}}\simeq d_{\text{\tiny Q}}$ when the RM vanishes, and then it increases but with an opposite sign as $d_{\text{\tiny QS}}$ gets larger. Inside the grey region of the density plots, the cut-off condition specified by \eqref{eq:fr_umf_arb_dir_freq_lim} is not satisfied, and as discussed before, the left and right circular polarizations are completely blocked. RM values are below the current observational capabilities for the light-purple shaded regions. 
 			
 	Having explored the RM values by an MBH, let us now focus on Fig.\,\ref{fig:gal_MW_PA_ex_QM_plot_mp_1}, where we have shown a polarization angle map (PA map) for linearly polarized light with initial polarization along $x$-direction. The waves are assumed to be passing through an extremal MBH with $M_{\text{\tiny BH}}= 1~\solarmass$ or $Q_{\text{\tiny BH}}\simeq 0.5 \times 10^{29}~\text{A-m}$. The MBH is located at a distance $10^{-4} ~\text{pc}$  from the source, located inside the Milky Way.  We show $10^{-5}\,\text{pc} \times 10^{-5}\,\text{pc}$ (right) and $10^{-3}\,\text{pc} \times 10^{-3} \,\text{pc}$ (left) patches as seen by the observer. It is notable that the electric fields in the PA map maintain a consistent direction for a fixed impact parameter $\xi$, leading to a constant polarization angle. This is a manifestation of the fact that the refractive index distribution, given by Eq.\,\eqref{eq:fr_mbh_ref_ind}, is independent of the azimuthal angle $\phi$. Furthermore, as expected, when we move away from the center of the MBH, the change in polarization angle decreases because the strength of the magnetic field wanes.

    As mentioned before, the PA map shows uniform polarisation angle values for fixed $\xi$. Therefore, if we select a circular contour of any specific radius on the PA map and integrate the change in polarization values along this contour, the resulting quantity will satisfy
    %%%%%%%%%%%%%%%%
    \begin{equation}\label{eq:fr_mbh_cp_cont_ineq}
         \mathcal{M}^{\text{\tiny BH}} \equiv \abs{\frac{1}{ 2\pi \psi_{\text{\tiny pol.}}^{\text{\tiny BH.}}}\left[\left(\oint_C d \phi\,\psi_{\text{\tiny pol.}}^{\text{\tiny BH}}(\phi)\right)- 2\pi \psi_{\text{\tiny pol.}}^{\text{\tiny BH}}\right]}= 0\;.
    \end{equation}
    %%%%%%%%%%%%%%%%

    $C$ represents, as before, any closed circular contour centered around the MBH. This simple global measure holds observational significance as it provides a way to distinguish between an MBH with a unique monopolar field and other astrophysical magnetic field configurations that are always non-monopolar in nature. This condition should approximately hold true in many realistic plasma density distributions around MBHs.

      %%%%%%%%%%%%%%%			
     \begin{figure}
     	\centering
     	\includegraphics[scale=0.65]{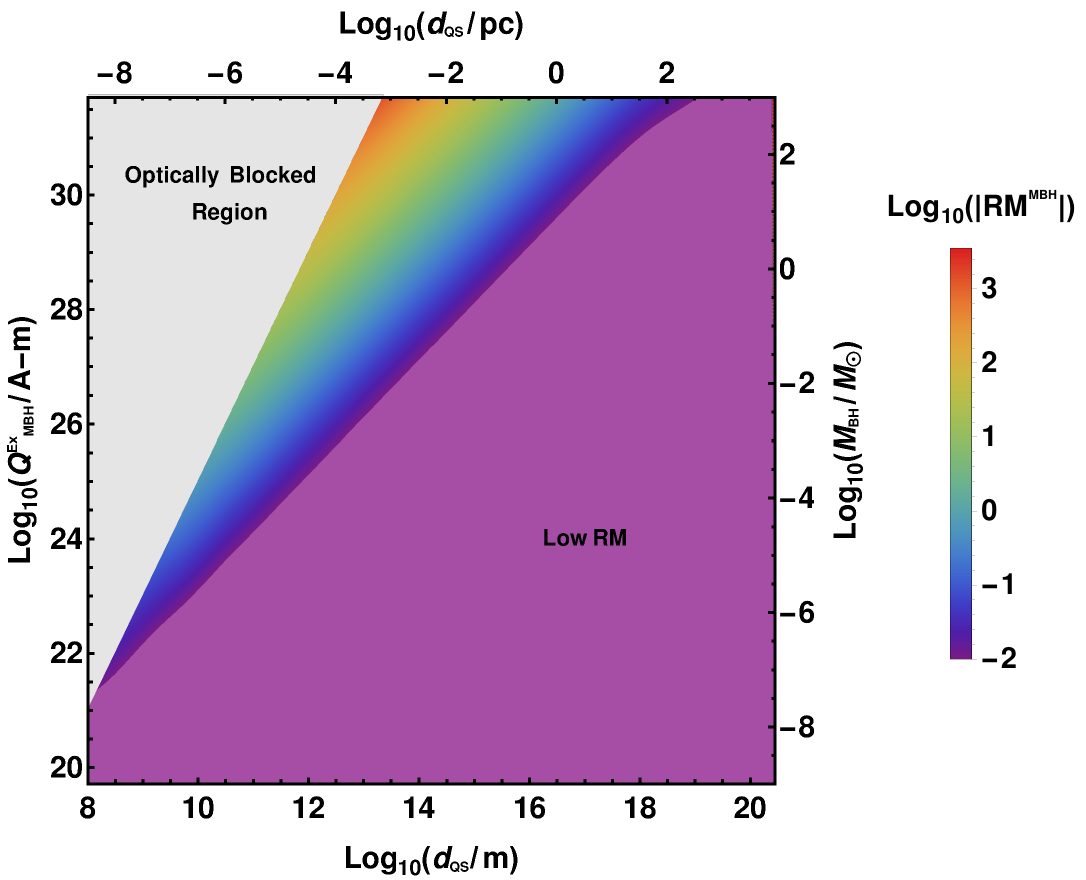}
     	\caption{Rotation measure (RM) density plot for an extremal MBH located inside the Milky Way galaxy for the galactic plasma distribution scenario. The parameters adopted are described in the text.}
     	\label{fig:gal_MW_RM_gauss_ex_QM_plot_mp}
     \end{figure}
      %%%%%%%%%%%%%%%
    %%%%%%%%%%%%%%%	
	 \begin{figure}
		\centering
            \includegraphics[scale=0.5]{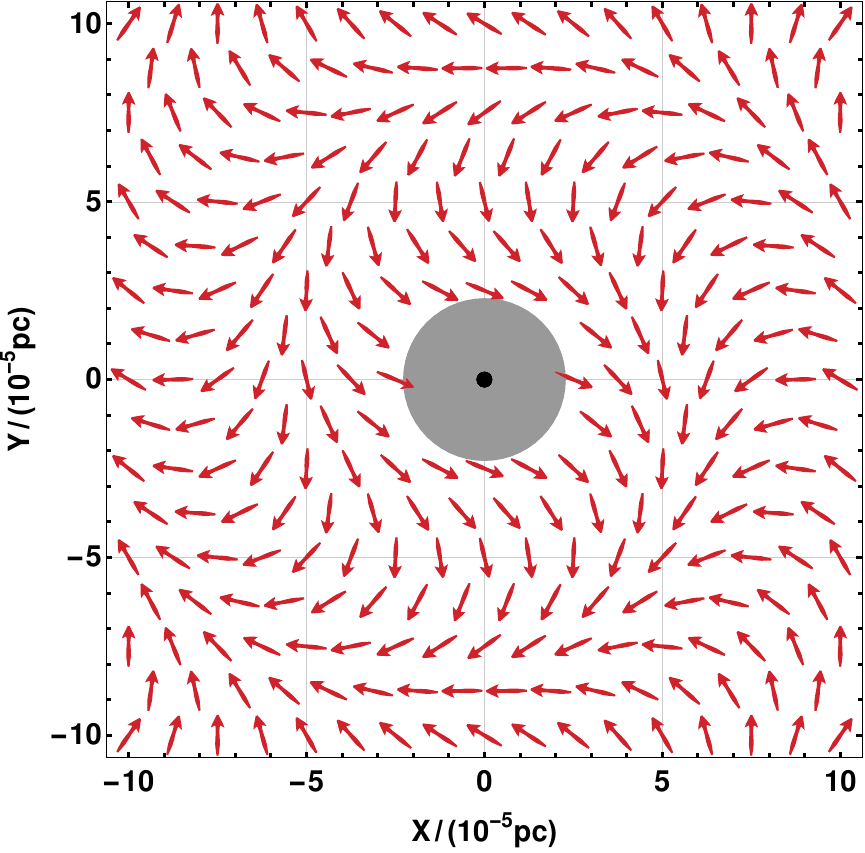}
		\includegraphics[scale=0.5]{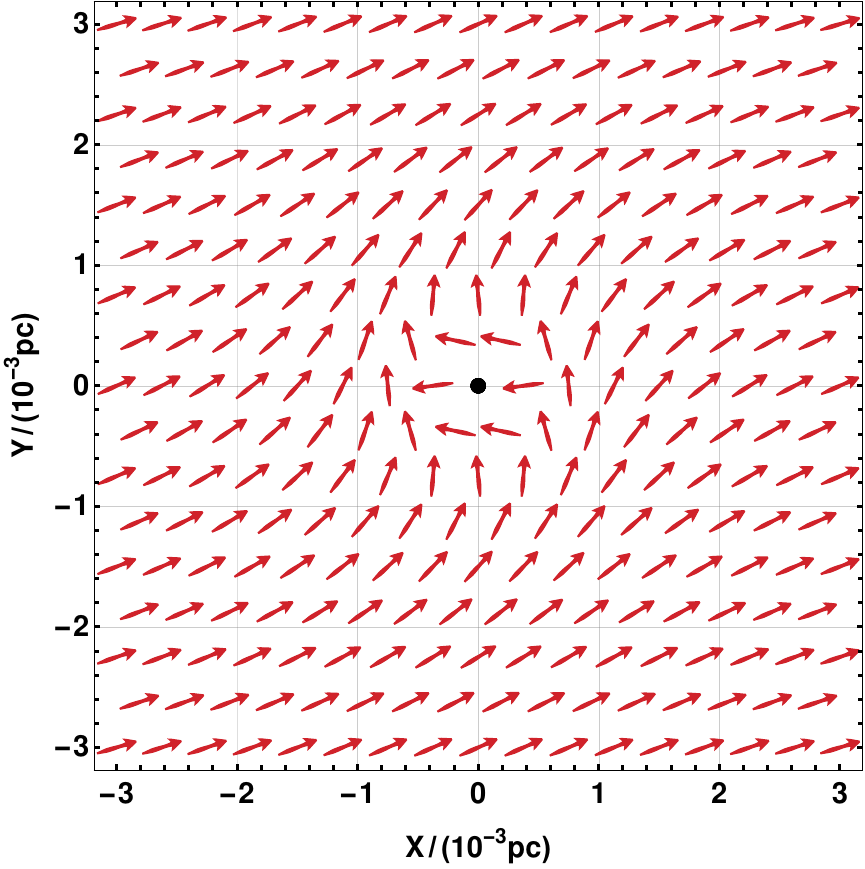}
		\caption{ The PA maps for an MBH located inside the MW galaxy assuming a galactic plasma distribution function. Distinctive features are visible in the PA map, absent from the corresponding NS case.}
			\label{fig:gal_MW_PA_gauss_ex_QM_plot_mp_1}\label{fig:gal_M31_PA_gauss_ex_QM_plot_mp_1}
	\end{figure}
	%%%%%%%%%%%%%%%	

     %%%%%%%%%%%%%%%%%%%%%%%%%%%%%%%%%%%%%%%%%%%%%%%%%%%%%%%%%%%%%%%%%%%%%%%%
    \subsubsection{Galactic distribution for plasma}
    \label{subsec:fr_bh_gal}
     %%%%%%%%%%%%%%%
    In this section, as before, we make a transition to a possibly more realistic scenario by considering a specific galactic plasma distribution that has been proposed\,\cite{Cordes_1991}.
    
    As in the previous sections, the MBH is assumed to be situated within our Milky Way galaxy. The plasma density and plasma frequency are derived from equations\,\eqref{eq:fr_ns_gal_plas_dens}, and  \eqref{eq:fr_ns_gal_MW_plas_freq}, with $R_{\text{\tiny NS}}$ replaced by $R_{\text{\tiny hor}}\equiv G/c^2(M_{\text{\tiny BH}}+\sqrt{M_{\text{\tiny BH}}^2+\mu_0Q_{\text{\tiny BH}}^2/(4 \pi G)})$, the outer horizon radius of the MBH. 
      
    Using Eqs.\,\eqref{eq:fr_ns_gal_MW_plas_freq}-\eqref{eq:fr_mbh_cp_RM_def}, we numerically calculate the change in the polarization angle and the RM value due to an MBH. We show the results in Fig.\,\ref{fig:gal_MW_RM_gauss_ex_QM_plot_mp}. The parametric values are fixed similar to the earlier analyses. 
    
    Similar to the constant density case, the RM for $Q_{\text{\tiny BH}}\gtrsim10^{22}~\text{A-m}$ ($M_{\text{\tiny BH}}\gtrsim10^{-5} ~\solarmass$) are in a range $ \text{RM}^{\text{\tiny BH}}\gtrsim 0.01 $ and potentially sufficient for observational evidence, if MBHs exist in the universe. Furthermore, for $d_{\text{\tiny QS}}\ll 1\,\text{kpc}$, the contribution from the thick disk results in qualitatively similar behaviour but with slightly elevated RM values compared to the constant density case. Additionally, the contribution of the inner disk near $d_{\text{\tiny QS}}\sim 1\,\text{kpc}$ leads to further enhancement in the RM values. Grey and light-purple denote regions where the cut-off condition is not satisfied and where the RM values are below observational thresholds. 
    
    %%%%%%%%%%%%%%%
     \begin{figure}
    	\centering
                
    		\includegraphics[scale=0.5]{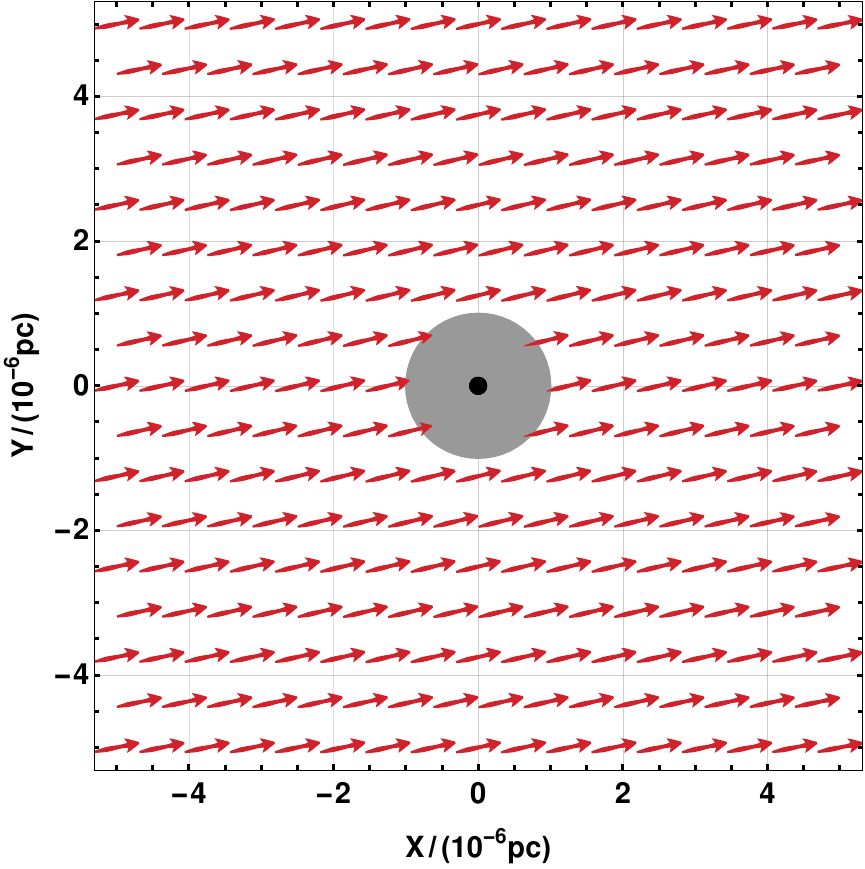}
                \includegraphics[scale=0.5]{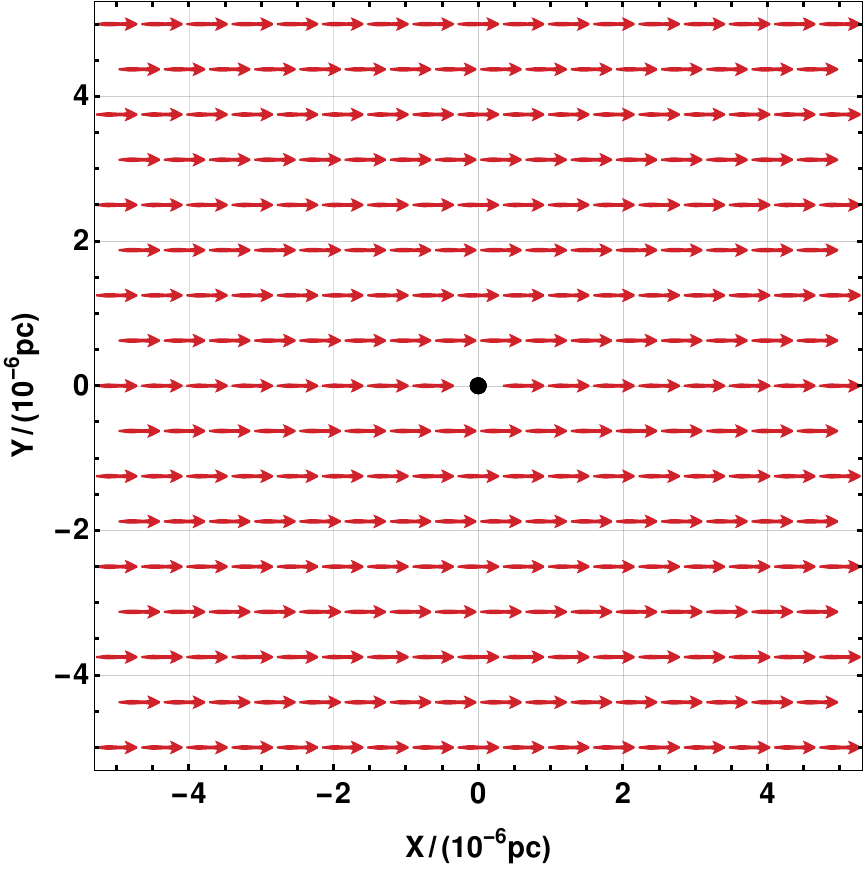}
    	\caption{Comparison of the change in polarization angle ($\psi_{\text{\tiny pol.}}^{\text{\tiny BH,(NS)}}$) between an extremal MBH (Left) and a matched NS (Right). The initial polarization is along the $\hat{x}$-axis, as before. The magnetic fields at $10\,\mathrm{Km}$ are kept the same for both the astrophysical compact objects. Specifically, we have taken, $Q_{\text{\tiny BH}}=10^{26} ~\text{A-m}$, (for MBH) and $\mathfrak{m}=10^{30}~\text{A-m$^2$} $ (for NS). This corresponds to a surface magnetic field of $10^{11}\,\mathrm{T}$ for the NS, corresponding to expected Magnetar surface fields. The plasma density is assumed to follow the proposed galactic plasma density distribution in\,\cite{Cordes_1991}. Again, the MBH produces a significant change in polarisation compared to an NS, with a distinctive pattern to boot. Moreover, a major take-home message is that even other astrophysical magnetic configurations, with their typical non-monopolar magnetic fields, will not be able to effectively mimic Faraday rotations due to an MBH. }
    	\label{fig:gal_MW_PA_gauss_NS_MBH_comp_v2}
    \end{figure}
    %%%%%%%%%%%%%%%
    
 	Fig.\,\ref{fig:gal_MW_PA_gauss_ex_QM_plot_mp_1} shows the PA map in the MBH case with a galactic plasma distribution. The parameters chosen are similar to the earlier sections. A distinctive pattern is observed again in this case, and the equality described in Eq.\,\eqref{eq:fr_mbh_cp_cont_ineq} is once again satisfied to good approximation.

  Once again, it is noteworthy that the variation in plasma density perpendicular to the galactic plane is anticipated to be negligible\,\cite{Cordes_1991}, especially when $\xi\lesssim \mathcal{O}(1)\,\text{pc}$. However, the mentioned features of the PA map, as well as the integral measure $\mathcal{M}^{\text{\tiny BH}}$, might not be present in the event of significant non-uniformity in plasma density.

    In Fig.\,\ref{fig:gal_MW_PA_gauss_NS_MBH_comp_v2}, we compare the change in the polarisation angles due to an extremal MBH (left) and an NS (right) for the galactic distribution of plasma. In order to make this comparison, we consider the magnetic field at $r=10\,\text{km}$ to be the same for both the NS and the MBH; for context, an extremal MBH with mass $M_{\text{\tiny BH}}= 1\,\solarmass$ has outer horizon at $r\simeq1.5\,\text{km}$. We assume the NS magnetic moment is oriented parallel to the direction of wave propagation (i.e.$\{\theta_{\text{\tiny dp}},\phi_{\text{\tiny dp}}\}=\{0,0\}$). For an NS with a surface magnetic field of $10^{11}\,\mathrm{T}$, this matching stipulates an MBH monopole charge $Q_{\text{\tiny BH}}=10^{26} ~\text{A-m}$. The corresponding NS dipole moment is $\mathfrak{m}=10^{30} ~\text{A-m$^2$}$. We also take $d_{\text{\tiny QS}}= 10^{-5}\,\text{pc}$ and a $10^{-6}\times 10^{-6}\,\mathrm{pc}$ patch in the observer's sky to construct the figure. For the MBH, a polarization pattern is clearly visible relative to the initial source polarization along the $\hat{x}$-direction. In contrast, the NS is impotent to effect such a polarisation change. Moreover, as we commented earlier, the polarization pattern furnished by the uncommon monopole field of an MBH will be distinctive from any other astrophysical magnetic configuration, which will be non-monopolar in nature.
    
    In Table\,\,\ref{tab:fr_NS_MBH}, we compare the 
    change in the polarisation angle ($\psi_{\text{\tiny pol.}}(r=r_{\text{\tiny cut}}^{\text{\tiny BH}})$) and the PA map integral measure ($\mathcal{M}^{\text{\tiny}}$) due to an extremal MBH and an NS for different values of $B^*_{\text{\tiny BH(NS)}}$ at $r=10~\text{km}$. We tabulate for different orientations of the NS's dipole magnetic moment. Clearly, extremal MBHs induce around $10^{8}$ times larger $\psi_{\text{\tiny pol.}}$ as compared to an NS.
    Besides, as commented earlier, the value of the integral measure $\mathcal{M}^{\text{\tiny BH}}$ is exactly zero for MBHs. For an NS - apart from the parallel orientation case - this is always positive and non-zero. As $\theta_{dp}$ changes from $0$ to $\pi/2$ for the NS orientaton, the value of $\mathcal{M}^{\text{\tiny NS}}$ changes from $0$ to $1$.  Again, we stress that the above qualitative and quantitative characteristics are unique to an MBH, owing to its distinctive monopolar field. Non-monopolar fields, due to other astrophysical magnetic configurations, are incapable of producing such characteristics.

    %%%%%%%%%%%%%%%%%%%%%%%%%%%%%%%%%%%%%%%%%%%%%%%%%%%%%%%%%%%%%%
    \begin{table}[H]
   	\center
   	\renewcommand{\arraystretch}{1.25}
   	\begin{tabular}{|c|c|c|c|c|c| } 
   		\hline 
   		~$B^*_{\text{\tiny BH(NS)}}(T)$~&~$\text{NS or BH}~$~&~$Q_{\text{\tiny BH}} (A-m) $~& $\theta_{\text{\tiny}dp}$~& ~$\psi_{\text{\tiny pol.}}(r=r_{\text{\tiny cut}}^{\text{\tiny BH(NS)}})  $~& $\mathcal{M}^{\text{\tiny BH(NS)}}$~\\
     
        ~~&~~&~$\text{ or }\mathfrak{m} (A-m^2)$~& ~~& ~~&~\\
   		\hline

     \multirow{4}{*}{~$10^{5}$}~&~$\text{BH}$~&~$10^{20}$~&~$-$~&~$2.4\times10^{-7}$~&~$0$~\\\cline{2-6}
     
         ~~&\multirow{3}{*}{~$\text{NS}$~}&\multirow{3}{*}{~$10^{24}$~}&~$0$~&~$ 3.7 \times 10^{-15}$~&~$ 0 $~\\
        ~~&~~&~~&~$\pi/4$~&~$2.6 \times 10^{-15}$~&~$ 0.9 \times  10^{-4}$~\\
         ~~&~~&~~&~$\pi/2$~&~$ 3.5 \times  10^{-19}$~&~$1$~\\
   		\hline

      \multirow{4}{*}{~$10^{8}$}~&~$\text{BH}$~&~$10^{23}$~&~$-$~&~$2.4 \times10^{-4}$~&~$0$~\\\cline{2-6}
     
         ~~&\multirow{3}{*}{~$\text{NS}$~}&\multirow{3}{*}{~$10^{27}$~}&~$0$~&~$ 3.7 \times 10^{-12}$~&~$0 $~\\
         ~~&~~&~~&~$\pi/4$~&~$2.6 \times 10^{-12}$~&~$ 3.0 \times  10^{-3}$~\\
         ~~&~~&~~&~$\pi/2$~&~$ 3.3 \times  10^{-14}$~&~$1$~\\
   		\hline

     \multirow{4}{*}{~$10^{11}$}~&~$\text{BH}$~&~$10^{26}$~&~$-$~&~$0.24$~&~$0$~\\\cline{2-6}
     
         ~~&\multirow{3}{*}{~$\text{NS}$~}&\multirow{3}{*}{~$10^{30}$~}&~$0$~&~$ 3.5 \times 10^{-9}$~&~$ 0 $~\\
         ~~&~~&~~&~$\pi/4$~&~$2.7 \times 10^{-9}$~&~$ 0.86\times 10^{-1}$~\\
        ~~&~~&~~&~$\pi/2$~&~$ 3.3 \times  10^{-10}$~&~$1$~\\
   		\hline

   	\end{tabular}
 	\caption{Table displaying the change in polarization angle ($\psi_{\text{\tiny pol.}}^{\text{\tiny BH,(NS)}}$) and integral measure $\mathcal{M}^{\text{\tiny BH(NS)}}$ for linearly polarized electromagnetic waves, due to an extremal MBH and NS. A galactic plasma density distribution is assumed. For bonafide comparisons, we have taken the magnetic fields of the extremal MBH and the NS to be the same at $r=10~\text{Km}$. The MBH effects are very pronounced compared to the NS and display characteristic patterns that are manifested in their integral measure values. }
   	\label{tab:fr_NS_MBH}
   \end{table}
%%%%%%%%%%%%%%%%%%%%%%%%%%%%%%%%%%%%%%%%%%%%%%%%%%%%%%%%%%%%%%

   %%%%%%

     %%%%%%
  
 %%%%%%   

 %%%%%%%%%%%%%%%%%%%%%%%%%%%%%%%%%%%%%%%
\section{Summary and conclusions}
%%%%%%%%%%%%%%%%%%%%%%%%%%%%%%%%%%%%%%%
\label{sec:sum_and_conc}
    In this study, we explored a few notable constraints and signatures associated with magnetically charged primordial black holes (MBHs). Such astrophysical entities are particularly intriguing as they may evade the usual Hawking radiation constraints on primordial black holes even for masses $M_{\text{\tiny BH}} \lesssim 10^{15}$g (see\,\cite{Carr_2021} and references therein, for instance). This may lead to a significant population for them in the current epoch even for masses $M_{\text{\tiny BH}} \lesssim 10^{15}$g, and thereby contribute to a fraction of the dark matter in the universe today. This is especially pertinent as their electrically charged or spinning extremal counterparts may be more readily neutralised or degraded in angular momenta in typical astrophysical environments.
    
    In Sec.\,\ref{sec:park_bound_pmbh}, we investigated Parker-type bounds on MBHs from galactic and extra-galactic magnetic configurations. Although Parker-type bounds on MBHs have previously been explored ---for instance, from galactic magnetic fields\,\cite{Ghosh_2020,Bai_2020}, yielding a limit on their dark matter fraction $f_{\text{\tiny DM}} \lesssim \mathcal{O}(10^{-3})$--- we find that constraints from cosmic void and cosmic filament magnetic fields provide novel and more stringent bounds on the dark matter fraction than those extant in the literature\,\cite{Ghosh_2020,Bai_2020, Kobayashi_2022,Kobayashi_2023}. Apart from the detailed theoretical analyses of the various scenarios that could arise while leveraging intergalactic magnetic field systems, the main results we derive are given in Eqs.\,\eqref{eq:fDM_primordial_IGMF}, \eqref{eq:FMBH_primordial_IGMF}, \eqref{eq:fDM_cos_fil}, and \eqref{eq:bound_NGC4501_1}. The most exacting constraints on the population of extremal MBHs arise from intergalactic magnetic fields in cosmic voids $\left(f_{\text{\tiny DM}} \lesssim 10^{-8}\right)$ and cosmic web filaments $\left(f_{\text{\tiny DM}} \lesssim 10^{-4}\right)$. These bounds comes from the unclustered MBHs exhibiting typical velocities of $v \sim 10^{-3}~ c$ in cosmic voids and filaments\,\cite{Padilla:2005ea,rost_2021}. For non-extremal MBHs, these constraints are influenced by the ratio of their magnetic charge to their mass. These constraints are detailed  in  Eqs.\;\eqref{eq:fDM_primordial_IGMF_non} and \eqref{eq:fDM_cos_fil_non}. We have also pointed out a few theoretical features in the analyses---for instance, that for slow extremal MBHs or extremal MBHs whose velocities are isotropically distributed, the $f_{\text{\tiny DM}}$ bounds are completely independent of the MBH mass. It is essential to mention that these boundaries depend on the presence of magnetic fields in cosmic structures; specifically, there should be no cosmic voids or web filaments characterized by depleted magnetic fields.

    Cognizant of the stringent Parker-type bounds we derived, suggesting a small MBH population if they exist in the universe, we then speculated on whether these rare exotic compact objects may have any unique observational signatures. This led us to a detailed investigation of the Faraday rotation signatures produced by MBHs in Sec.\,\ref{sec:farad_effec}. Faraday rotation refers to the rotation in the polarization angle of a linearly polarized wave as it passes through a magnetized plasma. We comprehensively analysed the rotation measure (RM) and the polarisation angle change induced by an NS and an MBH. These were done assuming two different density distributions for the hydrogen plasma---a simplistic constant density profile and a more realistic galactic profile\,\cite{Cordes_1991}.
    
    Along with some of the detailed theoretical and numerical analyses presented in Sec.\,\ref{sec:farad_effec}, a few of the main results are displayed in Figs.\,\ref{fig:gal_MW_RM_ex_QM_plot_ns}-\ref{fig:gal_MW_PA_gauss_ex_QM_plot_ns_1}, Figs.\,\ref{fig:gal_MW_RM_ex_QM_plot_mp}-\ref{fig:gal_MW_PA_gauss_NS_MBH_comp_v2} and Table\,\,\ref{tab:fr_NS_MBH}. We noted that for extremal MBHs with a magnetic charge $Q^{\text{\tiny Ex.}}_{\text{\tiny BH}}\gtrsim10^{22}~ \text{A-m}$ or equivalently mass $M^{\text{\tiny Ex.}}_{\text{\tiny BH}}\gtrsim10^{-6}~ \solarmass$, the RM values are significant enough to be detected by earth-based radio telescopes. The corresponding values for an NS are well below observational thresholds. We also pointed out that the unique monopolar fields of MBHs lead to unique patterns in the polarization maps, which are very distinct from those induced by other astrophysical sources, which all have non-monopolar fields. We illustrated these uncommon MBH characteristics by making detailed comparisons to an NS with a dipolar magnetic field. In this context, we defined a simple integral measure to quantify the symmetry in the polarization maps and established an inequality for them in Eqs.\,\eqref{eq:fr_ns_cp_cont_ineq} and \eqref{eq:fr_mbh_cp_cont_ineq}; which may potentially help us discriminate between astrophysical sources. 
    
    Observing uncommon spatial dependences in polarization angle and RM maps may be a smoking gun for MBHs. The pertinent observations are probably within the capabilities of earth-based telescopic arrays with a resolution of $ \mathcal{O}(10^{-6})$ arcseconds. Two of the existing telescopic arrays that are likely promising in this regard are the Global mm-VLBI Array\footnote{https://www3.mpifr-bonn.mpg.de/div/vlbi/globalmm/} and the Event Horizon Telescope\,\cite{akiyama2019first}, both of which utilise very long baseline interferometry.

%%%%%%%%%%%%%%%%%%%%%%%%%%%%%%%%%%%%%%%%%%%%%%%%%%%%%%%%%%%%%%%%%%%%%%%%%%
\acknowledgments
LB acknowledges support from a Senior Research
Fellowship, granted by the Human Resource Development Group, Council of Scientific and
Industrial Research, Government of India. AB acknowledges support from Science and Engineering Research Board (SERB) India via the Startup Research Grant SRG/2023/000378.

%%%%%%%%%%%%%%%%%%%%%%%%%%%%%%%%%%%%%%%%%%%%%%%%%%%%%%%%%%%%%%%%%%%%%%%%%%
\appendix

%%%%%%%%%%%%%%%%%%%%%%%%%%%%%%%%%%%%%%%%%%%%%%%%%%%%
\section{Derivation of Faraday rotation in a magnetic field}
%%%%%%%%%%%%%%%%%%%%%%%%%%%%%%%%%%%%%%%%%%%%%%%%%%%
\label{appendix:Far_rot}
 Here, we will present a formal derivation of Faraday rotation in a magnetic field. We consider an electromagnetic wave propagating along the $z$-axis through an inhomogeneous, weakly-ionised hydrogen plasma. A spatially non-uniform background curl-less magnetic field is present. We assume that the background magnetic field is dominant compared to the magnetic field of the wave. It takes the form      
       %%%%%%%%%%%%%%%
       \begin{equation}\label{eq:app_fr_fru_arb_mf}
	 	\vec{\mathcal{B}}_{\text{\tiny ext.}}(\vec r)=
	 	B_{\text{\tiny ext,x}}(\vec r)~\hat{x}+	B_{\text{\tiny ext,y}}(\vec r)~\hat{y}+	B_{\text{\tiny ext,z}}(\vec r)~\hat{z}\;,
        \end{equation}
        %%%%%%%%%%%%%%%        
       with, $B_{\text{\tiny ext.}}(\vec r)=\sqrt{B_{\text{\tiny ext,x}}^2+B_{\text{\tiny ext,y}}^2+B_{\text{\tiny ext,z}}^2}$. As the electromagnetic wave passes through the magnetized plasma, the electrons in the plasma will experience a Lorentz force given by, 
       %%%%%%%%%%%%%%%
        \begin{equation}\label{eq:app_fr_fru_arb_eom}
	 	m\text{\tiny e}\frac{d^2\vec{r}_{\text{\tiny e}}}{dt^2}=-e \left(\vec{E}+\frac{d \vec{r}}{d t}\times \vec{B}\right)\;.
	 \end{equation}
 	%%%%%%%%%%%%%%%  
    Here, $\vec{r}_{\text{\tiny e}}$, $m_{\text{\tiny e}}$ and $-e$ are the position, mass, and charge of the electron, and $\vec{E}$ and $\vec{B}$ are the total electric and magnetic fields. 
    
    We can solve the Lorentz force equation (Eq.\,\eqref{eq:app_fr_fru_arb_eom}) by considering a perturbation about a background solution. In the steady state, these
    perturbation can be written in the time domain as    
       %%%%%%%%%%%%%%%
	 \begin{eqnarray}\label{eq:app_fr_fru_arb_pert}
	 	\vec{r}_{\text{\tiny e}}&=&\vec{r}_{\text{\tiny e}}^{\,\text{(0)}}+\vec{r}_{\text{\tiny e}}^{\,\text{(1)}}e^{-i\omega t}\;, \nonumber\\
	 	\vec{E}(\vec{r})&=&0+\vec{E}^{\,\text{(1)}}(\vec{r})e^{-i\omega t}\;, \\
	 	\vec{B}(\vec{r})&=&\vec{\mathcal{B}}_{\text{\tiny ext.}}(\vec r)+\vec{B}^{\,\text{(1)}}(\vec{r})e^{-i\omega t}\;. \nonumber
	 \end{eqnarray}	 
	 %%%%%%%%%%%%%%%
    Here, $\vec{r}_{\text{\tiny e}}^{\,\text{(0)}}$ and $\vec{r}_{\text{\tiny e}}^{\,\text{(1)}}$ are the mean position and amplitude of displacement of the electron, and $\omega$, $\vec{E}^{\,\text{(1)}}$ and $\vec{B}^{\,\text{(1)}}$, are respectively the frequency, electric field and magnetic field of the wave. 
    
    Now, using Eqs.\,\eqref{eq:app_fr_fru_arb_eom} and \eqref{eq:app_fr_fru_arb_pert}, and taking terms upto linear order we get
      %%%%%%%%%%%%%%%
	 \begin{equation}\label{eq:app_fr_fru_arb_eom_1}
	 	\frac{e \vec{E}^{\,\text{(1)}}}{m_{\text{\tiny e}}}=\omega^2\vec{r}_{\text{\tiny e}}^{\,\text{(1)}} 	+\frac{ie\omega}{m_{\text{\tiny e}}}\left(\vec{r}_{\text{\tiny e}}^{\,\text{(1)}}\times \vec{\mathcal{B}}_{\text{\tiny ext.}}(\vec r)\right)\;,
	 \end{equation}
     %%%%%%%%%%%%%%%
    which can be written in the component form as
     %%%%%%%%%%%%%%%
	\begin{equation}\label{eq:app_fr_fru_arb_E_r_mat}
	 \frac{e}{m_{\text{\tiny e}}}	\begin{pmatrix} 
	 		E_{\text{x}}^{\,\text{(1)}}\\E_{\text{y}}^{\,\text{(1)}} \\E_{\text{z}}^{\,\text{(1)}} \\
	 	\end{pmatrix}=\begin{pmatrix} 
	 			\omega^2 & i \omega \tilde{\omega}_{\text{\tiny z}} & -i \omega \tilde{\omega}_{\text{\tiny y}}   \\
	 		-i \omega \tilde{\omega}_{\text{\tiny z}} & \omega^2  & i\omega \tilde{\omega}_{\text{\tiny x}}   \\ 
	 		 i \omega \tilde{\omega}_{\text{\tiny y}} &- i\omega  \tilde{\omega}_{\text{\tiny x}} & \omega^2  \\
	 	\end{pmatrix}
 	\begin{pmatrix} 
 		x^{\,\text{(1)}}\\y^{\,\text{(1)}} \\z^{\,\text{(1)}}\\
 	\end{pmatrix}\;. 
	 \end{equation}
     %%%%%%%%%%%%%%%
	 Here, we have denoted $\{\tilde{\omega}_{\text{\tiny x}},\tilde{\omega}_{\text{\tiny y}},\tilde{\omega}_{\text{\tiny z}}\}\equiv \{e B_{\text{\tiny ext,x}}(\vec r)/m_{\text{\tiny e}},e B_{\text{\tiny ext,y}}(\vec r)/m_{\text{\tiny e}},e B_{\text{\tiny ext,z}}(\vec r)/m_{\text{\tiny e}}\}$. $\tilde{\omega}\equiv e B_{\text{\tiny ext.}}(\vec r)/m_{\text{\tiny e}}$ is the cyclotron frequency of the electron and $\vec{E}^{\,\text{(1)}}=\left(E_{\text{x}}^{\,\text{(1)}}, E_{\text{y}}^{\,\text{(1)}}, E_{\text{z}}^{\,\text{(1)}}\right)$, $\vec{r}^{\,\text{(1)}}_{\text{\tiny e}}=\left(x^{\,\text{(1)}}_{\text{\tiny e}},y^{\,\text{(1)}}_{\text{\tiny e}},z^{\,\text{(1)}}_{\text{\tiny e}}\right)$ are the amplitudes of the electric field and the electron oscillations in the $(x,y,z)$ directions respectively. 
  
    The oscillation amplitudes can be determined by inverting Eq.\,\eqref{eq:app_fr_fru_arb_E_r_mat} and is given by
     %%%%%%%%%%%%%%%
      \begin{eqnarray}\label{eq:app_fr_fru_arb_r_E_rel_mat}
	 \begin{pmatrix} 
	 		x_{\text{\tiny 1}}\\y_{\text{\tiny 1}} \\z_{\text{\tiny 1}} \\
	 	\end{pmatrix}= 
	 	\frac{e}{m_{\text{\tiny e}} \left(\omega^2-\tilde{\omega}^2\right)}\begin{pmatrix} 
	 		1-\frac{\tilde{\omega}_{\text{\tiny x}}^2}{\omega^2} & -\frac{\tilde{\omega}_{\text{\tiny x}} \tilde{\omega}_{\text{\tiny y}}+i \omega  \tilde{\omega}_{\text{\tiny z}}}{\omega ^2} & \frac{-\tilde{\omega}_{\text{\tiny x}} \tilde{\omega}_{\text{\tiny z}}+i \omega  \tilde{\omega}_{\text{\tiny y}}}{\omega ^2}    \\
	 		\frac{-\tilde{\omega}_{\text{\tiny x}} \tilde{\omega}_{\text{\tiny y}}+i \omega  \tilde{\omega}_{\text{\tiny z}}}{\omega ^2} & 	1-\frac{\tilde{\omega}_{\text{\tiny y}}^2}{\omega^2} & -\frac{\tilde{\omega}_{\text{\tiny y}} \tilde{\omega}_{\text{\tiny z}}+i \omega  \tilde{\omega}_{\text{\tiny x}}}{\omega ^2}  \\ 
	 		-\frac{\tilde{\omega}_{\text{\tiny x}} \tilde{\omega}_{\text{\tiny z}}+i \omega  \tilde{\omega}_{\text{\tiny y}}}{\omega ^2}   & \frac{-\tilde{\omega}_{\text{\tiny y}} \tilde{\omega}_{\text{\tiny z}}+i \omega  \tilde{\omega}_{\text{\tiny x}}}{\omega ^2} & 	1-\frac{\tilde{\omega}_{\text{\tiny z}}^2}{\omega^2} \\
	 	\end{pmatrix}
	 		\begin{pmatrix} 
	 		E_{\text{\tiny 1,x}}\\E_{\text{\tiny 1,y}} \\E_{\text{\tiny 1,z}} 
	 	\end{pmatrix}\;.
	 \end{eqnarray}
     %%%%%%%%%%%%%%%
    Eq.\,\eqref{eq:app_fr_fru_arb_r_E_rel_mat} relates the electrons' oscillation amplitude with the wave's electric field and the background magnetic field.  Due to the background magnetic field, the oscillation in each direction is coupled to the total electric field. Furthermore, these electron oscillations will induce a time-dependent dipole moment with amplitude, 
      %%%%%%%%%%%%%%%
      \begin{eqnarray}\label{eq:app_fr_fru_arb_P_E_rel_mat}
	\vec{P}\equiv- \mathcal{N}_{\text{\tiny e}} e \vec{r}_{\text{\tiny e}}^{\,\text{(1)}}= -
 	  	\frac{\epsilon_{0}\omega^2_{\text{\tiny p}}}{ \left(\omega^2-\tilde{\omega}^2\right)}\begin{pmatrix} 
 	  		1-\frac{\tilde{\omega}_{\text{\tiny x}}^2}{\omega^2} & -\frac{\tilde{\omega}_{\text{\tiny x}} \tilde{\omega}_{\text{\tiny y}}+i \omega  \tilde{\omega}_{\text{\tiny z}}}{\omega ^2} & \frac{-\tilde{\omega}_{\text{\tiny x}} \tilde{\omega}_{\text{\tiny z}}+i \omega  \tilde{\omega}_{\text{\tiny y}}}{\omega ^2}    \\
 	  		\frac{-\tilde{\omega}_{\text{\tiny x}} \tilde{\omega}_{\text{\tiny y}}+i \omega  \tilde{\omega}_{\text{\tiny z}}}{\omega ^2} & 	1-\frac{\tilde{\omega}_{\text{\tiny y}}^2}{\omega^2} & -\frac{\tilde{\omega}_{\text{\tiny y}} \tilde{\omega}_{\text{\tiny z}}+i \omega  \tilde{\omega}_{\text{\tiny x}}}{\omega ^2}  \\ 
 	  		-\frac{\tilde{\omega}_{\text{\tiny x}} \tilde{\omega}_{\text{\tiny z}}+i \omega  \tilde{\omega}_{\text{\tiny y}}}{\omega ^2}   & \frac{-\tilde{\omega}_{\text{\tiny y}} \tilde{\omega}_{\text{\tiny z}}+i \omega  \tilde{\omega}_{\text{\tiny x}}}{\omega ^2} & 	1-\frac{\tilde{\omega}_{\text{\tiny z}}^2}{\omega^2} \\
 	  	\end{pmatrix}
 	  \begin{pmatrix} 
 	  	E_{\text{\tiny 1,x}}\\E_{\text{\tiny 1,y}} \\E_{\text{\tiny 1,z}} 
 	  \end{pmatrix}\;.\nonumber \\
	 \end{eqnarray}
     %%%%%%%%%%%%%%%
       Here, $\mathcal{N}_{\text{\tiny e}}$ and $\omega_{\text{\tiny p}}^2\equiv e^2\mathcal{N}_{\text{\tiny e}}/(\epsilon_{0}m_{\text{\tiny e}})$ are the number density and plasma frequency respectively. 
       
       Using the relation between induced dipole moment and dielectric tensor, i.e., $\vec{P}=\epsilon_{0}({\bm \epsilon_{\text{\tiny r}}}-{\bm I})\cdot\vec{E}^{\,\text{(1)}}$, where ${\bm I}$ is the identity matrix, we obtain the dielectric tensor as
     %%%%%%%%%%%%%%%
      \begin{equation}\label{eq:app_fr_fru_arb_dir_dil_tens}
  		{\bm \epsilon_{\text{\tiny r}}}=\begin{pmatrix} 
  			1-\frac{\omega^2_{\text{\tiny p}}\left(\omega^2-\tilde{\omega}^2_{\text{\tiny x}} \right)}{\omega^2\left(\omega^2-\tilde{\omega}^2\right)} &  \frac{\omega^2_{\text{\tiny p}} \left(\tilde{\omega}_{\text{\tiny x}} \tilde{\omega}_{\text{\tiny y}}+i \omega  \tilde{\omega}_{\text{\tiny z}}\right)}{\omega^2\left(\omega^2-\tilde{\omega}^2\right)} & \frac{\omega^2_{\text{\tiny p}} \left(\tilde{\omega}_{\text{\tiny x}} \tilde{\omega}_{\text{\tiny z}}-i \omega  \tilde{\omega}_{\text{\tiny y}}\right)}{\omega^2\left(\omega^2-\tilde{\omega}^2\right)}   \\
  			\frac{\omega^2_{\text{\tiny p}} \left(\tilde{\omega}_{\text{\tiny x}} \tilde{\omega}_{\text{\tiny y}}-i \omega  \tilde{\omega}_{\text{\tiny z}}\right)}{\omega^2\left(\omega^2-\tilde{\omega}^2\right)} & 1-\frac{\omega^2_{\text{\tiny p}}\left(\omega^2-\tilde{\omega}^2_{\text{\tiny y}} \right)}{\omega^2\left(\omega^2-\tilde{\omega}^2\right)}&\frac{\omega^2_{\text{\tiny p}} \left(\tilde{\omega}_{\text{\tiny y}} \tilde{\omega}_{\text{\tiny z}}+i \omega  \tilde{\omega}_{\text{\tiny x}}\right)}{\omega^2\left(\omega^2-\tilde{\omega}^2\right)} \\ 
  			\frac{\omega^2_{\text{\tiny p}} \left(\tilde{\omega}_{\text{\tiny x}} \tilde{\omega}_{\text{\tiny z}}+i \omega  \tilde{\omega}_{\text{\tiny y}}\right)}{\omega^2\left(\omega^2-\tilde{\omega}^2\right)}    & \frac{\omega^2_{\text{\tiny p}} \left(\tilde{\omega}_{\text{\tiny y}} \tilde{\omega}_{\text{\tiny z}}-i \omega  \tilde{\omega}_{\text{\tiny x}}\right)}{\omega^2\left(\omega^2-\tilde{\omega}^2\right)} & 1-\frac{\omega^2_{\text{\tiny p}}\left(\omega^2-\tilde{\omega}^2_{\text{\tiny z}} \right)}{\omega^2\left(\omega^2-\tilde{\omega}^2\right)} \\
  		\end{pmatrix}\;.
  	\end{equation} 
   %%%%%%%%%%%%%%%
  
    We will now solve Maxwell's equations\,\cite{Constantinidis:2016wjo} with magnetic monopole to obtain the characteristic modes and their refractive indices. The Maxwell's equations for the case of interest to us are given by
   %%%%%%%%%%%%%%%
	  \begin{eqnarray}
	 	\label{eq:app_fr_umf_arb_dir_ME_1}
	 	\nabla\cdot \vec{E}&=& \frac{e}{\epsilon_{0}}(\mathcal{N}_{\text{\tiny p}}-\mathcal{N}_{\text{\tiny e}})\;, \\
	 	\label{eq:app_fr_umf_arb_dir_ME_2}
	 	\nabla\cdot \vec{B}&=&\mu_{0}Q_{\text{\tiny BH}}\delta^3(\vec{r})\;, \\
	 	\label{eq:app_fr_umf_arb_dir_ME_3}
	 	\nabla\times\vec{E}+\frac{\partial \vec{B}}{\partial t}
	 	&=& 0\;, \\
	 	\label{eq:app_fr_umf_arb_dir_ME_4}
	 	\nabla\times\vec{B}-\frac{1}{c^2}\frac{\partial \vec{E}}{\partial t}&=&\mu_{0}\vec{J}=\mu_{0}e(\mathcal{N}_{\text{\tiny p}}\vec{v}_{\text{\tiny p}}-\mathcal{N}_{\text{\tiny e}}\vec{v}_{\text{\tiny e}})\;.
	 \end{eqnarray}
  %%%%%%%%%%%%%%%
 	$Q_{\text{\tiny BH}}$ is the monopole magnetic charge, which is at rest and may source a background magnetic field. $\mathcal{N}_{\text{\tiny p}}$ and $\mathcal{N}_{\text{\tiny e}}$ are the number densities of protons and electrons, and  $\vec{v}_{\text{\tiny p}}$ and $\vec{v}_{\text{\tiny e}}$ are the velocities of the protons and electrons respectively. As protons are much heavier than electrons, we may neglect $\vec{v}_{\text{\tiny p}}$, since $\vec{v}_{\text{\tiny p}}\ll\vec{v}_{\text{\tiny e}}$.

    Using Eq.\,\eqref{eq:app_fr_fru_arb_pert}, Eqs.\,\eqref{eq:app_fr_umf_arb_dir_ME_3}	 and \eqref{eq:app_fr_umf_arb_dir_ME_4} can be linearized as	
    %%%%%%%%%%%%%%%
 	\begin{eqnarray}\label{eq:app_fr_umf_arb_dir_lin_cont_eq}
 		\label{eq:app_fr_umf_arb_dir_lin_ME_3_0}
 		\nabla\times\vec{E}^{\,\text{(1)}}-i\omega\vec{B}^{\,\text{(1)}}
 		&=& 0\;, \\
 		\label{eq:app_fr_umf_arb_dir_lin_ME_4_0}
 		\nabla\times\vec{B}^{\,\text{(1)}}+\frac{i\omega}{c^2}\vec{E}^{\,\text{(1)}}&= &i\omega n_{0}e\mathcal{N}_{\text{\tiny e}}\vec{r}_{\text{\tiny e}}^{\,\text{(1)}}\;.
 	\end{eqnarray}
    %%%%%%%%%%%%%%% 
    We have used the fact that the curl of the background magnetic field is zero. Using $\vec{P}\equiv- \mathcal{N}_{\text{\tiny e}} e \vec{r}_{\text{\tiny e}}^{\,\text{(1)}}=\epsilon_{0}({\bm \epsilon_{\text{\tiny r}}}-{\bm I})\cdot\vec{E}^{\,\text{(1)}}$, we get
    %%%%%%%%%%%%%%%
 	\begin{eqnarray}\label{eq:app_fr_umf_arb_dir_lin_cont_eq_1}
 		\label{eq:app_fr_umf_arb_dir_lin_ME_3}
 		\nabla\times\vec{E}^{\,\text{(1)}}-i\omega\vec{B}^{\,\text{(1)}}
 		&=& 0\;, \\
 		\label{eq:app_fr_umf_arb_dir_lin_ME_4}
 		\nabla\times\vec{B}^{\,\text{(1)}}+i\frac{\omega}{c^2}{\bm \epsilon_{\text{\tiny r}}}\cdot\vec{E}^{\,\text{(1)}}&= &0\;.
 	\end{eqnarray}
    %%%%%%%%%%%%%%% 
   We have used $c=(\mu_0\epsilon_0)^{-1/2}$. 
   
   Now, substituting Eq.\,\eqref{eq:app_fr_umf_arb_dir_lin_ME_3} in Eq.\,\eqref{eq:app_fr_umf_arb_dir_lin_ME_4} and simplifying it, leads to,
   
        %%%%%%%%%%%%%%% 
	 \begin{equation}\label{eq:app_fr_umf_arb_dir_wave_eq}
	   \nabla\left(\nabla\cdot\vec{E}^{\,\text{(1)}}\right)-\nabla^2\vec{E}^{\,\text{(1)}}-\frac{\omega^2}{c^2}{\bm \epsilon_{\text{\tiny r}}}\cdot \vec{E}^{\,\text{(1)}}=0\;.
	 \end{equation}
	 %%%%%%%%%%%%%%%
    For plane electromagnetic waves, we can write $\vec{E}^{\,\text{(1)}}(\vec{r})\propto e^{i\psi_{\text{\tiny ph.}}(\vec{r})}$, where $\psi_{\text{\tiny ph.}}$ is the phase of the electric field and the wave vector is $\vec{k}=\nabla \psi_{\text{\tiny ph.}}(\vec{r})$.
    Then, Eq.\,\eqref{eq:app_fr_umf_arb_dir_wave_eq} may be re-written as
    %%%%%%%%%%%%%%% 
	 \begin{equation}\label{eq:app_fr_umf_arb_dir_wave_eq_1}
	   \left(\vec{k}\cdot\vec{E}^{\,\text{(1)}}\right)\vec{k}-k^2\vec{E}^{\,\text{(1)}}+\frac{\omega^2}{c^2}{\bm \epsilon_{\text{\tiny r}}}\cdot \vec{E}^{\,\text{(1)}}=i\nabla \left(\vec{k}\cdot\vert{\vec{E}^{\,\text{(1)}}}\vert\right)-i(\nabla\cdot\vec{k})\vec{E}^{\,\text{(1)}}\;,
	 \end{equation}
	 %%%%%%%%%%%%%%%
	where $k=\vert\vec{k}\vert$. In the eikonal limit, i.e., when $\vert\nabla{k}\vert/k^2\ll 1$, the right-hand side of Eq.\,\eqref{eq:app_fr_umf_arb_dir_wave_eq_1} is negligible. Therefore, we can write
	  %%%%%%%%%%%%%%% 
	 \begin{equation}\label{eq:app_fr_umf_arb_dir_wave_eq_2}
	   \left(\vec{k}\cdot\vec{E}^{\,\text{(1)}}\right)\vec{k}-k^2\vec{E}^{\,\text{(1)}}+\frac{\omega^2}{c^2}{\bm \epsilon_{\text{\tiny r}}}\cdot \vec{E}^{\,\text{(1)}}\approx0\;.
	 \end{equation}
	 %%%%%%%%%%%%%%%
  
    For an electromagnetic wave traveling along the $z$ direction, the wave vector takes the form $\vec{k}=(0,0,d \psi_{\text{\tiny ph.}}(\vec{r})/dz)$ and the phase can be written as 
     %%%%%%%%%%%%%%% 
	 \begin{equation}\label{eq:app_fr_umf_arb_dir_phase}
	  \psi_{\text{\tiny ph.}}(\vec{r})=\int dz~ k(\vec{r})=\frac{c}{\omega}\int dz~ n(\vec{r})\;,
	 \end{equation}
	 %%%%%%%%%%%%%%%
  where $n(\vec{r})$ is the refractive index. Solving Eqs.\,\eqref{eq:app_fr_fru_arb_dir_dil_tens} and \eqref{eq:app_fr_umf_arb_dir_wave_eq_2} for the refractive indices (i.e. $n =c k/\omega $), we get 
     %%%%%%%%%%%%%%%
    \begin{equation}\label{eq:app_fr_umf_arb_dir_ref_ind}
 		n_{{ (\pm)}}(\vec{r})=\left(1-\frac{\omega^2_{\text{\tiny p}} \left(\omega^2-\omega^2_{\text{\tiny p}}\right)}{\omega^2\left(\omega^2-\omega^2_{\text{\tiny p}}-\frac{1}{2}\left(\tilde{\omega}^2_{\text{\tiny x}}+\tilde{\omega}^2_{\text{\tiny y}}\right)\pm\left(\frac{1}{4}\left(\tilde{\omega}^2_{\text{\tiny x}}+\tilde{\omega}^2_{\text{\tiny y}}\right)^2+(\omega^2-\omega^2_{\text{\tiny p}})^2\frac{\tilde{\omega}^2_{\text{\tiny z}}}{\omega^2}\right)^{1/2}\right)}\right)^{1/2}\;.
 	\end{equation}
    %%%%%%%%%%%%%%%
    
Furthermore, the transverse components of the characteristic modes are related by
  %%%%%%%%%%%%%%%
    \begin{equation}\label{eq:app_fr_umf_arb_dir_Ex_Ey}
        \left(\frac{E_{\text{x}}^{\,\text{(1)}}}{E_{\text{y}}^{\,\text{(1)}}}\right)_{{ (\pm)}}=\frac{i  \left(\omega\left(\tilde{\omega}_{\text{\tiny x}}^2-\tilde{\omega}_{\text{\tiny y}}^2\right)\pm\sqrt{{4 \tilde{\omega}_{\text{\tiny z}}^2 \left(\omega ^2-\omega_{\text{\tiny p}}^2\right)^2}+\omega^2\left(\tilde{\omega}_{\text{\tiny x}}^2+\tilde{\omega}_{\text{\tiny y}}^2\right)^2}\right)}{2\left( \left(\omega ^2- \omega_{\text{\tiny p}}^2\right) \tilde{\omega}_{\text{\tiny z}}+ i \omega  \tilde{\omega}_{\text{\tiny x}} \tilde{\omega}_{\text{\tiny y}} \right)}\;.
    \end{equation}
    %%%%%%%%%%%%%%%
   For $\omega \gg \tilde{\omega}_a$, the above equation may be expanded as 
    \begin{equation}\label{eq:app_fr_umf_arb_dir_Ex_Ey_approx}
 		\left(\frac{E_{\text{x}}^{\,\text{(1)}}}{E_{\text{y}}^{\,\text{(1)}}}\right)_{{ (\pm)}}\simeq \pm i \,\text{sgn}(\tilde{\omega}_{\text{\tiny z}})+i\frac{ \omega\left( \tilde{\omega}_{\text{\tiny x}} \mp i\,\text{sgn}(\tilde{\omega}_{\text{\tiny z}})\tilde{\omega}_{\text{\tiny y}} \right)^2}{2 (\omega^2-\omega^2_{\text{\tiny p}})\tilde{\omega}_{\text{\tiny z}}}  \;,
 	\end{equation}
      where $\text{sgn}$ is signum function. From above we note that the characteristic modes approximately map to left and right circular polarizations, i.e. $E_{\text{x}}^{\,\text{(1)}}/E_{\text{y}}^{\,\text{(1)}}\simeq \pm i\,\text{sgn}(\tilde{\omega}_{\text{\tiny z}})$, when 
  %%%%%%%%%%%%%%%
  \begin{equation}\label{eq:app_fr_umf_arb_dir_Ex_Ey_cond}
      \abs{\frac{ \omega\left( \tilde{\omega}_{\text{\tiny x}}^2 + \tilde{\omega}_{\text{\tiny y}}^2 \right)}{2 (\omega^2-\omega^2_{\text{\tiny p}})\tilde{\omega}_{\text{\tiny z}}}  }\ll1\;.
  \end{equation}
  %%%%%%%%%%%%%%%
  
  Thus, when the above limit is satisfied, from Eq.\,\eqref{eq:app_fr_umf_arb_dir_Ex_Ey_approx}, for $\tilde{\omega}_{\text{\tiny z}}>0$, we conclude that the $(+)$ and the $(-)$ modes are very close to the left and the right circular modes, respectively. However, for $\tilde{\omega}_{\text{\tiny z}}<0$, the $(+)$ and $(-)$ modes are approximately mapped to the right and left circular modes. We may hence write,
  %%%%%%%%%%%%%%%
  \begin{equation}\label{eq:app_fr_umf_arb_dir_ref_ind_fin}
      n_{\text{\tiny L(R)}}=\begin{cases}
          n_{{ (\pm)}}\; ; &  \tilde{\omega}_{\text{\tiny z}}>0\;,\\
          n_{{ (\mp)}}\; ; &  \tilde{\omega}_{\text{\tiny z}}<0\;.\\
      \end{cases}
  \end{equation}
  %%%%%%%%%%%%%%%
     
      Now, let us calculate the polarization angle of the wave. The total electric field can be written as the sum of the left and right circular waves
  %%%%%%%%%%%%%%%
    	\begin{equation}\label{eq:app_fr_fru_cp_ref_elec}
		\vec{E}^{\,\text{(1)}}_{\text{\tiny tot.}}(t,\vec{r})=\vec{E}^{\,\text{(1)}}_{\text{\tiny L}}e^{i(\psi_{\text{\tiny ph.(L)}}(\vec{r}) -\omega t)}\hat{e}_{\text{\tiny L}}+\vec{E}^{\,\text{(1)}}_{\text{\tiny R}}e^{i(\psi_{\text{\tiny ph. (R)}}(\vec{r}) -\omega t)}\hat{e}_{\text{\tiny R}}\; .
	\end{equation}
 %%%%%%%%%%%%%%%
     $\psi_{\text{\tiny ph.(L,(R))}}$ and $\vec{E}^{\,\text{(1)}}_{\text{\tiny L,(R)}}$, are the phases and the amplitudes of the left and right circular waves respectively. $\hat{e}_{\text{\tiny L,(R)}}=\left(\hat{x}\mp i\hat{y}\right)/2 $ are the corresponding direction vectors. Assuming that the light source is linearly polarized, i.e. $\vec{E}^{\,\text{(1)}}_{\text{\tiny L}}=\vec{E}^{\,\text{(1)}}_{\text{\tiny R}}\equiv\vec{E}^{\,\text{(1)}}_{\text{\tiny in.}}$, one obtains from  Eq.\,\eqref{eq:app_fr_fru_cp_ref_elec}
     %%%%%%%%%%%%%%%
    \begin{equation}\label{eq:app_fr_fru_cp_ref_elec_1}
		\vec{E}^{\,\text{(1)}}_{\text{\tiny tot.}}(t,\vec{r})=\vec{E}^{\,\text{(1)}}_{\text{\tiny in.}}e^{i\left(\psi_{+}-\omega t\right)}\left(\cos \psi_{-}\hat{x}+\sin  \psi_{-}\hat{y}\right) \; .
	\end{equation}
 %%%%%%%%%%%%%%%
     Here, $\psi_{-}\equiv\left(\psi_{\text{\tiny ph. (L)}}-\psi_{\text{\tiny ph. (R)}}\right)/2$ and $\psi_{\text{\tiny ph.(L
      +R)}}\equiv\left(\psi_{\text{\tiny ph. (L)}}+\psi_{\text{\tiny ph. (R)}}\right)/2$. Now, taking the ratio of the y-component and the x-component of the total electric field, we get the polarization angle of the wave
    %%%%%%%%%%%%%%%
     \begin{equation}\label{eq:app_fr_fru_cp_pol_ang}
         \psi_{\text{\tiny pol.}}\equiv\tan^{-1} \left(\vec{E}^{\,\text{(1)}}_{\text{\tiny tot.,y}}/\vec{E}^{\,\text{(1)}}_{\text{\tiny tot.,x}}\right)=\frac{1}{2 } \left(\psi_{\text{\tiny ph. (L)}}(\vec{r})-\psi_{\text{\tiny ph. (R)}}(\vec{r})\right)=\frac{\omega}{2 c}\int~dz \left( n_{\text{\tiny L}}(\vec{r})-n_{\text{\tiny R}}(\vec{r})\right)\;.
     \end{equation}
     %%%%%%%%%%%%%%%
     Here, we have used Eq.\,\eqref{eq:app_fr_umf_arb_dir_phase}. 
     
    The rotation measure RM is defined as the rate of change of polarization angle with $\lambda^2$
    %%%%%%%%%%%%%%%
     \begin{equation}\label{eq:app_fr_fru_cp_RM_def}
         \text{RM}(\lambda)\equiv \frac{d\psi_{\text{\tiny pol.}}(\lambda)}{d\lambda^2}\;,
     \end{equation}
    %%%%%%%%%%%%%%%
    where $\lambda$ is the wavelength of the light.
    %%%%%%%%%%%%%%%%%%%%%%%%%%%%%%%%%%%%%%%%%%%%%%%%%%%%%%%%%%%%%
%%%%%%%%%%%%%%%%%%%%%%%%%%%%%%%%%%%%%%%%%%%%%%%%%%%%%%%%%%%%%
\bibliographystyle{JHEP.bst}
\bibliography{Primordial_Magnetic_Relics_and_Signature}

%%%%%%%%%%%%%%%%%%%%%%%%%%%%%%%%%%%%%%%%%%%
\end{document}